\documentclass[universe,article,accept,pdftex,moreauthors]{Definitions/mdpi} 

\firstpage{1} 
\pubvolume{1}
\issuenum{1}
\articlenumber{0}
\pubyear{2023}
\copyrightyear{2023}
\datereceived{ } 
\daterevised{ } %
\dateaccepted{ } 
\datepublished{ } 
\hreflink{https://doi.org/} 
\pdfoutput=1 

\Title{Photodynamical Modeling of the Compact, Multiply Eclipsing
Systems KIC 5255552, KIC 7668648, KIC 10319590, \mbox{and EPIC 220204960}}

\TitleCitation{Photodynamical Modeling of the Compact, Multiply
Eclipsing Systems KIC 5255552, KIC 7668648, KIC 10319590, and EPIC 220204960}

\Author{Jerome A. Orosz \orcidA{}}

\AuthorNames{Jerome A. Orosz}

\AuthorCitation{Orosz, J.A.}

\address[1]{Department of Astronomy, San Diego State University,
5500 Campanile Drive, San Diego, CA 92182-1221; jorosz@sdsu.edu}

\abstract{We present photodynamical models of four eclipsing binary
systems that show strong evidence of being members of higher-order
multiple systems via their strong eclipse timing variations and/or
via the presence of extra eclipse events.  Three of these systems
are from the main {\em Kepler} mission, and the other is from the K2
mission.  We provide some ground-based radial velocities
measurements for the three {\em Kepler} systems and make use of
recent light curves from the {\em TESS} mission.  Our sample
consists of two 2 + 1 systems and two 2 + 2 systems.  The first 2 +
1 system, KIC 7668648, consists of an eclipsing binary ($P_{\rm
bin}$ = 27.8 days) with late-type stars (\mbox{$M_1=0.8403\pm
0.0090$ $M_{\odot}$}, $R_1=1.0066\pm 0.0036\,R_{\odot}$ and
$M_2=0.8000\pm 0.0085\,M_{\odot}$, $R_2=0.8779\pm
0.0032\,R_{\odot}$) with a low-mass star ($M_3=0.2750\pm
0.0029\,M_{\odot}$, $R_3=0.2874\pm 0.0010\,R_{\odot}$) on a roughly
coplanar outer orbit ($P_3=208$ days).  There are several eclipse
events involving the third star that allow for the precise
determination of the system parameters.  The second 2 + 1 system,
KIC 10319590, consists of a binary ($P_{\rm bin}=21.3$ days) with
late-type stars ($M_1=1.108\pm 0.043\,M_{\odot}$, $R_1=1.590\pm
0.019\,R_{\odot}$ and $M_2=0.743\pm 0.023\,M_{\odot}$,
$R_2=0.7180\pm 0.0086\,R_{\odot}$) that stopped eclipsing about a
third of the way into the nominal {\em Kepler} mission.  We show
here that the third star in this system is a Sun-like star
($M_3=1.049\pm 0.038\,M_{\odot}$, $R_3=1.39\pm 0.11\,R_{\odot}$) on
an inclined outer orbit ($P_3=456$ days).  In this case, there are
no extra eclipse events.  We present the first comprehensive
solution for KIC 5255552 and demonstrate that it is a 2 + 2 system
consisting of an eclipsing binary ($P_{\rm bin,1}=32.5$ days) with
late-type stars ($M_1=0.950\pm 0.018\,M_{\odot}$, $R_1=0.9284\pm
0.0063\,R_{\odot}$ and $M_2=0.745\pm 0.014\,M_{\odot}$,
$R_2=0.6891\pm 0.0051\,R_{\odot}$) paired with a non-eclipsing
binary ($P_{\rm bin,2}=33.7$ days) with somewhat lower-mass stars
($M_3=0.483\pm 0.010\,M_{\odot}$, $R_3=0.4640\pm 0.0036\,R_{\odot}$
and $M_4=0.507\pm 0.010\,M_{\odot}$, $R_4=0.4749\pm
0.0031\,R_{\odot}$).  The two binaries, which have nearly coplanar
orbits, orbit their common barycenter on a roughly aligned outer
orbit ($P_{\rm out}=878$ days).  There are extra eclipse events
involving the component stars of the non-eclipsing binary, which
leads to relatively small uncertainties in the system parameters.
The second 2 + 2 system, EPIC 220204960, consists of a pair of
eclipsing binaries ($P_{\rm bin,2}=13.3$ days, $P_{\rm bin,2}=14.4$
days) that both consist of two low-mass stars
($M_1=0.54\,M_{\odot}$, $R_1=0.46\,R_{\odot}$,
$M_2=0.46\,M_{\odot}$, $R_2=0.37\,R_{\odot}$ and
$M_3=0.38\,M_{\odot}$, $R_3=0.40\,R_{\odot}$, $M_4=0.38\,M_{\odot}$,
$R_4=0.37\,R_{\odot}$) that orbit their common barycenter on a
poorly determined outer orbit.  Because of the relatively short time
span of the observations ($\approx$80 days for the photometry and
$\approx$70 days for the radial velocity measurements), the masses
and radii of the four stars in EPIC 220204960 can only be determined
with accuracies of $\approx$10\% and $\approx$5\%, respectively.  We
show that the most likely period of the outer orbit is 957 days,
with a $1\sigma$ range of 595 to 1674 days.  We can only place weak
constraints on the mutual inclinations of the orbital planes, and
additional radial velocity measurements and/or additional eclipse
observations would allow for much tighter constraints on the
properties of the outer orbit.}


\begin{document}

\section{Introduction}

Roughly half of nearby Sun-like stars are in binary systems (e.g.,
\cite{Raghaven_2010}).  Furthermore, perhaps 10\% of those binary
systems are themselves members of triple or quadruple systems
(e.g.,\ see~\cite{Tokovinin_2018}).  Large space-based photometric
surveys such as {\em Kepler}
\cite{Borucki_2010a,Borucki_2010b,Koch_2010} and \mbox{{\em TESS}
\cite{Ricker_2015}} have discovered thousands of eclipsing binaries
(e.g.,\ \cite{Kirk_2016,Prsa_2011,Slawson_2011}).  Owing to the often
long, continuous photometric coverage with high signal-to-noise, many
of these eclipsing binaries have been shown to have nearby companions
found from eclipse timing variations (e.g.,
\cite{Borkovits_2015,Borkovits_2016,Rappaport_2013}) or from having
multiple eclipse events with distinct periodicities
(e.g.,\ \cite{Kostov_2022}).  It is well known that binary stars are
the foundation upon which stellar astrophysics is built because~ it is
through eclipsing binaries, we can measure stellar masses and radii
with high precision.  Triple stars offer even more, especially the
low-mass, short-period systems with extra eclipses, none of which were
known prior to the {\it Kepler} mission.  Very succinctly, triple
systems can often provide more information, and this allows a better
determination of the system properties.  This additional precision
allows us to probe deeper physics than could be done with simple
binary stars. Furthermore, the~sample of triple-star architectures can
be used to statistically test the predictions of tidal theory and
theories of the formation of multiple-star~systems.

The addition of a third body companion with a non-zero mass certainly
complicates the orbital motions of the binary star components:
generally, the orbits are no longer purely Keplerian.  But this
complexity provides additional information that we can
exploit. In~particular, the~orbits are quite sensitive to the exact
positions and momenta of the three bodies and the observed light curve
is very unforgiving: if the orbits, masses, and~radii are not known
{\em very} precisely (typically with relatives of less than a few
percent) the model light curve will not match the observations.  This
sensitivity is a curse in~that it can be exceedingly difficult to
match the data (e.g.,\ the title of a 2014 paper by
Marsh~et~al.\ \cite{Marsh_2014} says this nicely: ``KIC 2856960: the
impossible triple star'').  However, when the correct model is found,
it is a blessing since the stringent observational constraints impose
very tight limits on system parameters and~would perhaps allow us to
measure the effects to which we were previously insensitive. But~from
where exactly does the extra information in a triple system come?

In an isolated eclipsing binary system, there are two locations per
orbit that the relative positions and velocities of the stars are
directly measured with photometry, namely at the conjunction phases
when the eclipses occur (the velocity constraint comes from the
duration of the eclipses).  But in an eclipsing triple system, when
additional eclipses/transits occur they tell us the locations and
motions of the bodies at those tertiary eclipse times. Because~the
orbital periods of the binary and the outer orbit are generally not
related, tertiary eclipses can occur at any binary phase.
Consequently, an~eclipsing 3-body system offers qualitatively new
information, stroboscopically probing more of the binary orbit.  It is
this additional information that allows the extremely precise mass and
radii~determinations.

In this work we present comprehensive photodynamical models of KIC
5255552 and KIC 7668648, both of which show extra eclipse events.  KIC
5255552 turns out to be a \mbox{2 + 2} system (e.g.,\ a pair of close
binaries that orbit the system barycenter), and~our model here is the
first one that can account for all the extra eclipse events and the
large eclipse timing variations.  KIC 7668648 is a compact 2 + 1
system (e.g.,\ a binary with a circumbinary companion that orbits both
components of the binary).  Owing to the numerous extra eclipse
events, along with the inclusion of ground-based radial velocity
measurements and additional data from {\em TESS}, we can determine the
masses and radii of all three stars with $\approx$$1\%$ relative
uncertainties.  KIC 10319590 is a 2 + 1 system that shows dramatic
changes in the eclipse depths over time (the eclipses disappeared
completely after the first $\approx$$400$ days of the {\em Kepler}
mission), but~without any observed extra eclipse events above the
noise level of the data.  EPIC 220204960 is a 2 + 2 system that shows
strong evidence of tidal interactions between the two binaries, again
without any observed extra eclipse events above the noise level of the
data.

The layout of this article is as follows.  In Section~\ref{ObsSec}, we
summarize the available {\em Kepler} and {\em TESS} photometry and
discuss ground-based spectroscopic observations of the three {\em
Kepler} systems.  In Section~\ref{PhotodynamicalSec} we discuss our
photodynamical model and code that is used to analyze the
observational data.  In Section~\ref{SystemSec} we discuss the actual
modeling of the four systems of interest.  In
Section~\ref{DiscussionSec} we briefly discuss the stellar parameters
and age constraints from model isochrones, some basic long-term
dynamical properties, and~comparisons with previous work.  Finally,
Section~\ref{SummarySec} contains a brief~summary.

\section{Observational~Data}\label{ObsSec}

Table~\ref{targetdata} summarizes the observational data available for
our four targets.  We give below general summaries of the sources for
our observational data.  Additional details for each source are given
in Section~\ref{SystemSec}.

\begin{table}[H]
\tablesize{\footnotesize}
\caption{Observational Data Summary for Target~Systems.\label{targetdata}}
\begin{adjustwidth}{-\extralength}{0cm}
\centering 
\setlength{\tabcolsep}{4mm} \resizebox{\linewidth}{!}{\begin{tabular}{lllll}
\toprule
\textbf{Parameter} & 
\textbf{KIC 5255552} &  
\textbf{KIC 7668648} & 
\textbf{KIC 10319590} & 
\textbf{EPIC 220204960} \\
\midrule
TIC ID number  &  120316928   & 158215322 & 123358966 &  \ldots \\
Other designation & KOI-6545  & KOI-6899 &  SPOCS 2149 &  SDSS J004832.66+001015.2 \\
RA (J2000)        & 18 58 46.3378826448 & 19 05 06.4737051312 & 
    18 46 51.5603406384 &      00 48~32.6684739096  \\
DEC (J2000)       & +40 29 55.062482712 & +43 20 20.880322836 &
     +47 24 51.814823652 & 
    +00 10~15.224216232  \\
$G$ magnitude     & 15.21               & 15.25 & 13.70 &  17.00 \\
{\em Kepler} Quarters & 1--4, 6--8, 10--12, 14--16 & 1--17 & 1--17 &  \ldots\\ 
Time span used $^1$      & $-29.806$ to 1390.959 & $-23.717$ to 1424.012  &   $-35.3851$  to 520.2844 &  2393.744 to~2470.7501 \\
{\em TESS} Sectors  & 14, 26, 40, 41, 53, 54  & 14, 26, 40, 41, 53, 54, 55  &
     14, 26, 40, 41, 53, 54, 55 &   \ldots \\
Number of RV measurements & 8  & 7 & 6 & 6 \\
Time span of RV measurements $^1$  & 736.806 to 1460.808 & 
    1084.698 to 1459.866 & 740.907 to 1092.884 & 2650.743 to~2718.882 \\
\bottomrule
\end{tabular}}
\end{adjustwidth}
\noindent{\footnotesize{$^1$ BJD $-$2,455,000.}}
\end{table}

\subsection{{\em~Kepler}}

NASA's {\em Kepler} Mission was launched 7 March 2009 with the goal of
discovering and characterizing Earth-sized planets in the habitable
zone of their host stars
~\cite{Borucki_2010a,Borucki_2010b,Koch_2010}.  Science operations
started 13 May 2009 (Quarter 1, BJD 2,454,964.51) and continued
through the end of nominal spacecraft operation on May 11, 2013
(Quarter 17, BJD 2,456,424.00).  Nearly 200,000 targets were monitored
with high precision at a $\approx$$30$ minute cadence (hereafter long
cadence) with a duty cycle close to 92\% in some cases.  A~small
subset of the targets was observed with a cadence of 1 min (hereafter
short cadence).  For all the targets discussed here we downloaded the
latest version of the processed {\em Kepler} data in binary FITS
format (Data Release Version 25) from the Mikulski Archive for Space
Telescopes (MAST \endnote{\url{https://archive.stsci.edu}}).

The {\em Kepler} light curves sometimes have individual cadences with
bad or corrupted data.  Each cadence has a data quality flag, where a
flag value of zero indicates an observation with no known problem, and
a flag value of $>$0 indicates potential problems owing to certain
spacecraft events (such as a reaction wheel desaturation event) or
other events (such as cosmic rays near an aperture).  The MAST Kepler
Archive Manual
\endnote{\url{https://archive.stsci.edu/kepler/manuals/archive_manual.pdf}}
gives the definition of the qualified flags.  In~our experience,
individual cadences with quality flags with \mbox{values $>16$} are
generally unreliable and it has been our practice to remove those
cadences.  After the removal of individual cadences with quality flags
$>16$, the {\em Kepler} light curves generally need detrending to
remove instrumental drifts and occasionally modulations due to star
spots.  We used the iterative process described in
Orosz~et~al.\ \cite{Orosz_2019} arrive at the final normalized light
curves.  Briefly, the~light curves are initially detrended ``by hand''
using an interactive code to mask out eclipses and fit cubic splines
to the out-of-eclipse parts of the light curve, where the number of
nodes for the splines is typically between 10 and 50.  Once a good
solution is found, the~best-fitting model is used to precisely
determine the various times of ingress and egress and, hence, the
duration of each event.  We keep only portions of the light curve
centered on eclipse and transit events, where, as a rule of thumb, the
length of a ``chunk'' centered on an eclipse event is twice the
duration of the eclipse when possible.  A~fifth-order polynomial is
fit to each chunk where data that occur between ingress and egress are
given zero weight (other orders for the polynomials were tried,
but~generally, the fifth-order polynomials gave the best results).
The data in each chunk are normalized via division by the respective
best-fitting polynomial.  Finally, the~normalized chunks are
reassembled to obtain the final light~curve.

\subsection{K2}

The K2 mission was a follow-on to the {\em Kepler} mission
~\cite{Howell_2014} and was active between 4 February 2014 and 26
September 2018.  Over~that time span, there were 20 different
pointings along the ecliptic where the two remaining reaction wheels
and pressure from the solar wind were used to maintain the spacecraft
pointing.  The mission produced the same pixel-level and light-curve
data products as the original {\em Kepler} mission.  In addition, many
versions of the light curves with various levels of additional
processing were provided by various community-driven efforts and made
available as High-Level Science Products on the MAST
\endnote{\url{https://archive.stsci.edu/missions-and-data/k2}}.

\subsection{{\em~TESS}}

NASA's {\em Transiting Exoplanet Survey Satellite} (\emph{TESS}, see
Ricker et al.\ \cite{Ricker_2015}) has been conducting a photometric
survey of most of the sky since the start of nominal science
observations on 25 July 2018.  {\em TESS} observes a $24^{\circ}\times
96^{\circ}$ field (referred to as a ``Sector'') for $\approx$$27$ days
at a time.  During~the two-year primary mission, the~Southern
Hemisphere and the Northern Hemisphere were each divided into 13
Sectors.  Most of the {\em Kepler} field was observed in Sector 14 and
again in Sector 26.  A~small number of targets were observed with a
2-minute cadence.  Those that were not targeted for short cadence were
observed at a cadence of 30 min in the Full Frame Images (FFIs).  As
part of the various mission extensions, the~{\em Kepler} field was
observed in Sectors 40, 41, 53, 54, and~55 with a \mbox{10 min}
cadence.  For~the objects that we discuss here, the TESS light curves
were computed using the {\tt eleanor} package of
Feinstein~et~al.\ \cite{Feinstein_2019}, which makes use of the
TESScut package devised by Brasseur~et~al.\ \cite{Brasseur_2019}.
The~light curves were detrended in a similar manner as the {\em
Kepler} light curves described~above.

\subsection{Ground-Based Follow-Up~Observations}

Spectroscopic observations of KIC 5255552, 7668648, and~10319590 were
obtained as part of the NOAO Survey program 2011A-0022
(P.I. A. Pr\v{s}a) as described in Kirk~et~al.~\cite{Kirk_2016}.  The
echelle spectrograph attached to the Mayall 4 m telescope at the Kitt
Peak National Observatory was used, which provided a spectral
resolving power of $R\approx$ $20,000$ and a wavelength coverage
between 4600 and 9050~\AA, although~only the wavelength region between
4875 and 5858~\AA\ was used to extract the radial velocities.  The
typical observing sequence consisted of back-to-back exposures of
around 900 s, with~calibration exposures of a ThAr lamp at each
telescope position.  The IRAF (Image Reduction and Analysis Facility,
version V2.17) software package~\cite{Tody_1986,Tody_1993} was used to
perform the basic CCD calibration tasks and to extract, normalize,
and~merge the final spectra.  The~typical signal-to-noise ratios were
usually between about 15 and 50 per resolution~element.

The radial velocities were computed using the ``broadening function''
(BF) technique devised by Rucinski~\cite{Rucinski_1992,Rucinski_2002},
as this technique generally gives better results than a simple
cross-correlation analysis for double-lined spectra.  Prior to the
analysis, the merged spectra were normalized to their continua using
the ``continuum'' task in the IRAF package.  The wavelength range of
4875 to 5850~\AA\ was used to compute the radial velocities, as~this
region avoids strong Balmer lines and strong interstellar and telluric
lines, and~contains numerous metal lines.  High signal-to-noise
spectra of various bright radial velocity standard stars were used as
templates.  Gaussian functions were fitted to the BFs to determine
their centroids and~widths.

\subsection{Times of Eclipse, Transit, and~Occultation~Events}

After performing the final detrending step on the light curves as
described above, we are left with normalized chunks of data centered
on individual eclipse, transit, and~occultation events.  To~find the
mid-times of these events, we ideally want to have a smooth curve that
has the same shape as the observed eclipse profiles and ``slide'' it
across the data in time until the best match is found.  Gaussian
functions or simple polynomials will not exactly match the profile
shape, especially when the signal-to-noise ratio is large.  Instead,
we used the {\tt ELC} (Eclipsing Light Curve) code of Orosz \&
Hauschildt~\cite{Orosz_2000} to produce realistic model curves and its
Markov sampler based on the Differential Evolution Monte Carlo Markov
Chain (DE-MCMC) algorithm of Ter Braak~\cite{TerBraak_2006} to sample
parameter space and to find the uncertainties in the fitted times.  To
model an eclipse event, one needs to specify the time of conjunction,
which obviously determines the time of mid-eclipse.  This parameter
can be given a narrow prior.  The shape of the eclipse event is
determined by the orbital period (for fixed masses, this sets the
scale for the duration), the inclination, the~radius of the eclipsed
star, the~ratio of radii, and~limb-darkening parameters, and~if
necessary, the eccentricity parameters.  These parameters can be given
wide priors since for this purpose, we want a model with an eclipse
profile that matches the observed eclipse profile, and~we do not care
if the properties of the stars themselves are realistic.  Thirty-two
chains are typically evolved for 100,000 generations, and~posterior
samples are taken every 5000 generations.  The~final adopted time of
eclipse is taken to be the median of the posterior sample and its
uncertainty to be the larger of $|T_{84\%}-T_{median}|$ or
$|T_{16\%}-T_{median}|$, where $T_{84\%}$ and $T_{16\%}$ is the times
at which 84\% and 16\% of the samples are smaller, respectively.

\section{The Photodynamical~Model}\label{PhotodynamicalSec}

For this discussion, a~``photodynamical model'' is a model that
combines a numerical solution to the appropriate gravitational
equations of motion of a system with two or more bodies, along with a
detailed scheme to synthesize light curves, velocity curves, and other
observable properties of the system.  Our photodynamical model is
implemented in our {\tt ELC} code~\cite{Orosz_2000}.  Many details of
the numerical scheme used to solve the equations of motion are given
in Orosz~et~al.\ \cite{Orosz_2019}.  We summarize some of the key
features of the model and describe some recent~modifications.

The numerical integrator uses Cartesian coordinates relative to the
system's barycenter.  Given the masses of each body and the Keplerian
elements of each orbit (e.g.,\ the period $P$, the~time of barycenter
conjunction $T_{\rm conj}$, the~eccentricity $e$, the~argument of
periastron $\omega$, the~inclination $i$, and~the nodal angle
$\Omega$) at a specified reference time $T_{\rm ref}$, the~Cartesian
coordinates (where $x$ and $y$ are in the sky plane and $z$ is the
radial coordinate with the positive $z$-axis is pointed towards the
observer) are computed using algorithms given in Murray \& Dermott
~\cite{Murray_1999}.  Three basic configurations are currently
available.  The first configuration consists of primary central body
with up to 9 companions, each in successively larger orbits using
Jacobi coordinates.  There are no restrictions on the masses of the
bodies (e.g.,\ the central body does not have to be the most massive).
This configuration includes the \mbox{2 + 1 mode} where a close binary
has a more distant companion.  The second configuration is the
``double binary'' or 2 + 2 mode with 4 bodies consisting of two close
binaries in a much larger orbit about the system barycenter.  The
third configuration is the ``triple plus binary'' or ``3 + 2'' mode
with 5 bodies that has a close binary with a tertiary companion in a
wider orbit paired with another close binary.  The standard Newtonian
equations of motion are supplemented with extra force terms to account
for precession due to General Relativity (GR) and/or precession due to
tidal bulges on the stars ~\cite{Mardling_2002}.  There is also a
Cartesian-to-Keplerian converter so that one may track changes to the
orbital elements over~time.

The numerical solution of the equations of motion essentially yields a
function that gives either the sky coordinates of every body as a
function of time or the radial coordinates of every body as a function
of time (as discussed in Orosz~et~al.\ \cite{Orosz_2019} these
coordinates are corrected for light travel time).  The~time
derivatives of the coordinates as a function of time are also
returned.  The times of eclipses, transits, and occultations are taken
to be the times when the projected separation between the centers of
two given bodies on the sky plane is minimized.  In~a similar manner,
the~times of ingress and egress of the various eclipse, transit,
or~occultation events are taken to be the times when the projected
separation between the centers of two given bodies on the sky plane is
equal to the sum of the respective radii of the two~bodies.

The stars are assumed to be spherical for the photodynamical mode of
{\tt ELC}.  The~algorithm described by Short~et~al.\ \cite{Short_2018}
is used to compute the light curve.  This algorithm allows for several
of the common limb-darkening approximations to be used [e.g.,\ the
``quadratic'' law, the~``square-root'' law, the~``Power-2'' law (the
one used in this work), and the ``logarithmic'' law], and~can
account for more than two overlapping bodies (e.g.,\ syzygy events).

{\tt ELC} has several optimizers and Markov samplers to explore the
parameter space.  The~ones used in this work include (i) the genetic
algorithm based on the code given in
Charbonneau~\cite{Charbonneau_1995}, (ii) a simple Monte Carlo Markov
Chain (MCMC) algorithm outlined by
Tegmark~et~al.\ \cite{Tegmark_2004}, (iii) the Differential Evolution
MCMC (DE-MCMC) algorithm of Ter Braak~\cite{TerBraak_2006}, (iv) an
implementation of the Nested Sampling algorithm described by
Skilling~\cite{Skilling_2006}, and~(v) a recently developed ``hybrid''
code dubbed ``{\tt gennestELC}'' that combines elements of the genetic
and the Nested Sampling algorithms.  Given an $N$-dimensional
parameter space, scaled to the unit hypercube, one chooses $N_{\rm
  live}$ ``live points'' and computes their likelihood values.
The~genetic algorithm is used to evolve the live points for a
user-adjustable number of generations.  After~this evolution a
bounding hyper-ellipsoid is computed, and~a new set of live points is
picked by uniform sampling from the hyper-ellipsoid.  In the Nested
Sampling algorithm, the volume of the hyper-ellipsoid is scaled at a
rate proportional to $1/N_{\rm live}$ (see Skilling
~\cite{Skilling_2006}) so that the actual volume (its ``statistical
volume'') is much larger than the actual volume.  In the {\tt
  gennestELC} algorithm, the~volume of the hyper-ellipsoid is allowed
to shrink faster than the statistical rate but~slower than the actual
rate without any scaling.  This algorithm can find good solutions in
complex parameter spaces much faster than the standard Nested Sampling
algorithm can when the number of free parameters is more than
about~25.

All the algorithms discussed above are bounded in the sense that the
user must select a lower bound and an upper bound for each fitting
parameter.  Obviously, care must be taken so that the ``true'' value
of the parameter is between the lower and upper bound.  It is often
more efficient to fit various combinations of two parameters rather
than the parameters themselves.  For example, one can fit for
$\sqrt{e}\cos\omega$ and $\sqrt{e}\sin\omega$ rather than for $e$ and
$\omega$, or~the sum and difference of the radii of the stars in the
binary, and~so~on.

The likelihood is usually the standard $\chi^2$ (one can also use
the average deviation as a likelihood).  The~time-dependent observables
that one usually fits for are the light curves (where up to 8 different
bandpasses can be fit simultaneously), the~radial velocity curves
for up to 5 bodies and~the times of eclipses, transits,
and occultations.  Observable properties that are not  usually
time-dependent, such as the observed rotational velocity of the
stars, their spectroscopically measured surface gravities, etc.\
can also be used as extra~constraints.

Finally, the~ability to use the MIST [MESA (Modules for Experiments in
  Stellar Astrophysics) Isochrones and Stellar Tracks] models
~\cite{Choi_2016,Dotter_2016} in the likelihood has recently been
implemented.  For~a large grid of various metallicities and ages,
the~isochrones supply the stellar radius and temperature as a function
of the stellar mass.  For~each isochrone, the~distance in a
mass–radius–temperature cube from each star to the model curve is
computed and summed.  The~distance is computed for each model in the
grid, and the overall smallest summed distance $d_{\rm iso,min}$ is
found.  The age and metallicity of the best-fitting isochrone are also
saved.  Finally, $d_{\rm iso,min}$ is scaled by a weight $W_{\rm iso}$
and the product is added to the overall $\chi^2$.  When $W_{\rm iso}$
is small (say less than 0.05), the~isochrones do not really influence
the fit since only a small quantity is added to the overall $\chi^2$.
In~this case, the~isochrones can be used after the fitting is carried
out as an independent check on the overall consistency and
plausibility of the fit.  In cases where some stellar parameters may
not be particularly well constrained (for example, a three-body system
where the third star does not eclipse or transit), the~weighting
factor $W_{\rm iso}$ can be much larger (say 20) to ensure the stellar
masses, radii, and~temperatures are all simultaneously consistent with
evolutionary models.  At~this point, no attempt is made to interpolate
between models in the mass–radius–temperature cube.  Finally, the~user
can choose to use only the mass-radius plane and ignore
the~temperatures.

\section{The Systems of~Interest}\label{SystemSec}
\unskip

\subsection{KIC~10319590}

\subsubsection{Overview}

\begin{figure}[H]
\includegraphics[width=8.5cm, angle=-90]{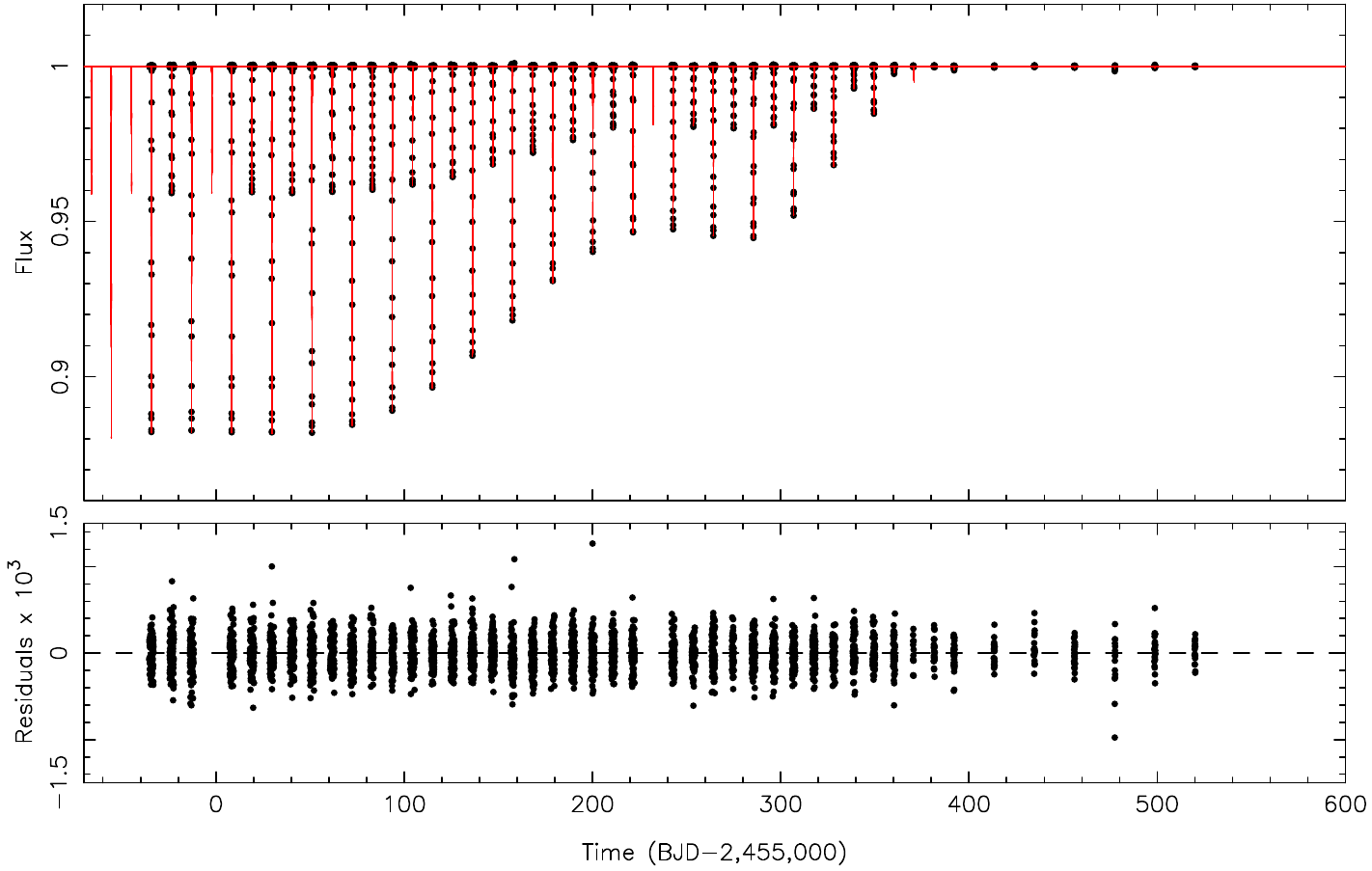}
\caption{\textbf{Top}: The normalized {\em Kepler} light curve of
KIC 10319590 (points) with the best-fitting model (red line).
\textbf{Bottom}: The residuals of the fit.
\label{1031fig01}}
\end{figure}

KIC 10319590 was first identified as an eclipsing binary with a period
of 21.34 days by Pr\v{s}a~et~al.\ \citep{Prsa_2011} using the first 44
days of {\em Kepler} data.  Slawson~et~al.\ \cite{Slawson_2011} found
significant eclipse time variations (ETVs) that suggested a dynamical
origin (as opposed to the light travel time effect or LTTE).  KIC
10319590 was among the more remarkable systems of the 39 triple-star
candidates found by Rappaport~et~al.\ \cite{Rappaport_2013} in that
the binary stopped eclipsing entirely after the first $\approx$$400$
days of {\em Kepler} observations (see Figure~\ref{1031fig01}).
Borkovits~et~al.\ \cite{Borkovits_2015} performed an analytic study of
the ETVs of KIC 10319590 and found an outer period of 451 days,
orbital eccentricities of 0.026(1) and 0.17(1) for the inner and outer
orbits, respectively, and~a retrograde outer orbit with a mutual
inclination of $135.4\pm 0.3^{\circ}$.  In~subsequent work,
Borkovits~et~al.\ \cite{Borkovits_2016} found a mutual inclination of
40.2(4) degrees for the outer~orbit.

\subsubsection{Photodynamical~Model}

KIC 10319590 was observed during all 17 Quarters of the nominal {\em
Kepler} mission, although~no eclipse events are seen in Quarter 6 or
any other Quarter thereafter.  Table~\ref{5255_times} in the Appendix
\ref{appendix_times} gives the eclipse times derived from the {\em
Kepler} data.  {\em TESS} observed KIC 10319590 in Sectors 14, 26,
40, 41, 53, 54, and~55.  No eclipse events were~seen.

\begin{figure}[H]
\includegraphics[width=10.0 cm, angle=-90]{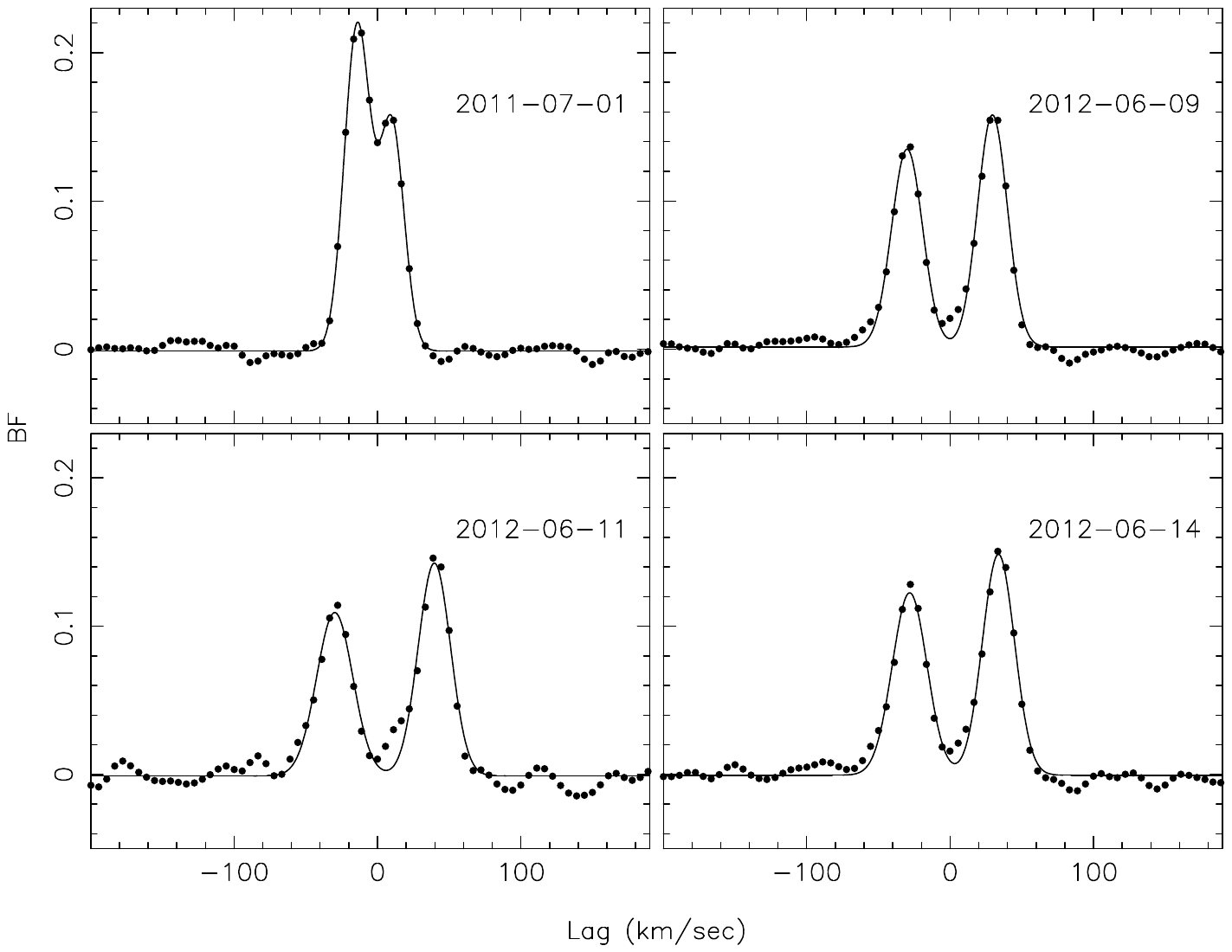}
\caption{Selected Broadening Functions (BFs) of KIC 10319590 (filled
circles) and their respective best-fitting double-Gaussian models
(solid lines).  The~stronger peak is due to the primary star in the
inner binary, and the other peak is due to the tertiary.  
\label{1031BF}}
\end{figure}

KIC 10319590 was observed nine times using the 4 m telescope at Kitt
Peak and the echelle spectrograph between 28 June 2011 and 18 June
2013.  A~spectrum of the F8V star HD 174912 was used as the template
for the BF analysis.  The spectra are double-lined (e.g.,\ two strong
peaks are detected), with the strongest set of lines attributable to
the primary star in the inner binary.  The~second set of lines does
not belong to the secondary in the inner binary but~rather to the
tertiary star.  The two peaks were blended in the spectra taken on 5
June 2012, 15 June 2013, and~18 June 2013.  A double-Gaussian model
was fitted to the remaining six BFs to determine the radial velocities
(see Figure~\ref{1031BF}), after~suitable heliocentric corrections
were applied (see Table~\ref{1031_RVtab}).  Using the areas under the
BF peaks as a proxy for the flux ratio
~\cite{Rucinski_2002,Tofflemire_2019}, we find a flux ratio (in this
case, tertiary flux divided by primary flux) in the spectral region
between 4875\AA\ and 5850\AA\ of $0.18\pm 0.05$.

\textls[15]{The photodynamical model has 30 free parameters.  There
are orbital parameters for the inner binary orbit, namely the period
$P_{\rm bin}$, a~time of primary conjunction $T_{\rm conj,bin}$, the
eccentricity parameters $e_{\rm bin}\cos\omega_{\rm bin}$, $e_{\rm
  bin}\sin\omega_{\rm bin}$, and the inclination $i_{\rm bin}$.
The~nodal angle of the binary is fixed at $\Omega_{\rm bin}=0$.
There are similar parameters for the outer orbit: $P_{3}$, $T_{\rm
  conj,3}$, $e_3\cos\omega_3$, $e_3\sin\omega_3$, $i_3$,
and~$\Omega_3$.  The~masses are parameterized by $M_1$,
\mbox{$Q_{\rm bin}\equiv M_2/M_1$}, and~$Q_3\equiv (M_1+M_2)/M_3$.
The~radii are parameterized by $R_1$, \mbox{$\rho_{\rm bin}\equiv
  R_1/R_2$} and \mbox{$\rho_3\equiv R_1/R_3$}.  The~stellar
effective temperatures are $T_1$, $\tau_{\rm bin}\equiv T_2/T_1$,
and~$T_3$.  The power-2 limb-darkening law with the Maxted
transformation for the coefficients was used
(see~\cite{Short_2019}), and~the limb-darkening coefficients $h_1$
and $h_2$ for the primary and for the secondary were allowed to
vary.  The~apsidal motion constants $k_{2,1}$ and $k_{2,2}$ for the
primary and secondary star, respectively, were free parameters.
Finally, there are four additional free parameters to account for
``seasonal contamination'': $s_0$ for Quarters 1, 5, 9, 13, and~17;
$s_1$ for Quarters 2, 6, 10, and~14; $s_2$ for Quarters 3, 7, 11,
and 15; and $s_3$ for Quarters 4, 8, 12, and~16.  The reference time
for the osculating parameters is BJD 2,454,800.  We fit the {\em
  Kepler} light curve and the radial velocity curve.  We also
include the measured times of all primary and secondary eclipses in
the likelihood~function.}

We used the genetic algorithm optimizer for the initial exploration of
parameter space.  Once a good solution was found, the~DE-MCMC code was
run 10 separate times for \mbox{60,000 generations} each, where each
run had a different seed for the random number generator.  After a
burn-in period of 2500 generations, posterior samples were generated
by sampling models every 3750 generations.  The~ten individual
posterior samples were combined, yielding sample sizes of 13,600 for
each fitting parameter and derived parameter.  Table~\ref{tab_1031fit}
gives the median values of the posterior distributions for the fitting
parameters and the corresponding $1\sigma$ uncertainty.  Except for
the apsidal constants $k_{1,2}$ and $k_{2,2}$ the fitted and derived
parameters are reasonably well constrained.  Table~\ref{tab_1031deriv}
gives the median values of the posterior distributions and the
corresponding uncertainties of several derived parameters.  The
initial Keplerian and Cartesian initial conditions for the
best-fitting model are given in Table~\ref{K1031_init} in the Appendix
\ref{appc}.

\begin{table}[H]
\caption{Fitting Parameters for KIC 10319590.\label{tab_1031fit}}
\begin{adjustwidth}{-\extralength}{0cm}
\centering 
\setlength{\tabcolsep}{8mm} \resizebox{\linewidth}{!}{\begin{tabular}{llllll}
\toprule
\textbf{Parameter $^1$} & \textbf{Median} & \textbf{$\boldsymbol{1\sigma}$} & \textbf{Parameter $^1$} & \textbf{Median} & \textbf{$\boldsymbol{1\sigma}$} \\
\midrule
$P_{\rm bin}$ (days) & 21.26551 & 0.000212     & $P_3$ (days) & 455.70 & 0.27 \\
$T_{\rm conj,bin}$  & $-204.86977~^2$ & 0.00037 & $T_{\rm conj,3}$ & $-189.33~^2$ & 0.32  \\
$e_{\rm bin}\cos\omega_{\rm bin}$ & 0.013740 &  0.000018 & $e_3\cos\omega_3$ & $-0.09768$ & 0.00046  \\
$e_{\rm bin}\sin\omega_{\rm bin}$ & 0.01841  & 0.00017 & $e_3\sin\omega_3$ & 0.09969 & 0.00041 \\
$i_{\rm bin}$ (deg) & 89.8536 & 0.0086 & $i_3$ (deg) & 104.78 & 0.10 \\
$\Omega_{\rm bin}$ (deg) & $0~^3$ & \ldots & $\Omega_3$ (deg) & 38.46  & 0.12 \\
$M_1$ ($M_{\odot}$) & 1.108 & 0.043 & $Q_{\rm bin}$ & 0.671 & 0.010  \\
$Q_3$          &  1.7651 & 0.0064 & & & \\
$R_1$ ($R_{\odot}$  & 1.590  & 0.019   & $\rho_{\rm bin}$ &  2.2151  & 0.0078  \\
$\rho_3$    &    1.140  & 0.099  & & & \\
$T_1$ (K)   & 5802 & 360 & $\tau_{\rm bin}$ 0.780 & 0.013 \\
$T_3$ (K) & 5836  & 462  & & & \\
$h_1$ primary & 0.7327 & 0.0058 & $h_2$ primary &  0.46 & 0.11 \\
$h_1$ secondary & 0.672 & 0.014 & $h_2$ secondary & 0.47 & 0.16 \\
$k_{2,1}$  & 0.049 & 0.032 & $k_{2,2}$ & 0.100 & 0.064 \\ 
$s_0$ & 0.024 & 0.020  & $s_1$ & 0.028 & 0.020 \\
$s_2$ & 0.029 & 0.020 & $s_3$ & 0.47 & 0.16 \\ 
\bottomrule
\end{tabular}}
\end{adjustwidth}
\noindent{\footnotesize{$^1$ Reference time for osculating Keplerian elements: 
BJD 2,454,800.  $^2$ BJD $-$2,455,000. $^3$ fixed.}}
\end{table}
\unskip

\begin{table}[H]
\caption{Derived Parameters for KIC 7668648 and KIC~10319590.\label{tab_1031deriv}}
\begin{adjustwidth}{-\extralength}{0cm}
\centering 
\setlength{\tabcolsep}{12mm} \resizebox{\linewidth}{!}{\begin{tabular}{lllll}
\toprule
 & \textbf{KIC 7668648} &  &  \textbf{KIC 10319590} & \\
\textbf{Parameter $^1$} & \textbf{Median} &$\boldsymbol{1\sigma}$  &  \textbf{Median} & $\boldsymbol{1\sigma}$ \\
\midrule
$M_1$ ($M_{\odot}$) & 0.8403  & 0.0090  & 1.108 & 0.043 \\
$M_2$ ($M_{\odot}$) & 0.8000  & 0.0085  & 0.743 & 0.023 \\
$M_3$ ($M_{\odot}$) & 0.2750  & 0.0029  & 1.049 & 0.038 \\
$R_1$ ($R_{\odot}$) & 1.0066  & 0.0036  & 1.590 & 0.019 \\
$R_2$ ($R_{\odot}$) & 0.8779  & 0.0032  &  0.7180 & 0.0086 \\
$R_3$ ($R_{\odot}$) & 0.2874  & 0.0010  &  1.39 & 0.11 \\
$a_{\rm bin}$ (AU) &  0.21179  & 0.00076  &  0.1844 & 0.0022 \\
$a_3$         (AU) &  0.8532   & 0.0030  &  1.653 & 0.020 \\
$e_{\rm bin}$      &  0.0489731  & 0.0000045  &  0.02297 & 0.00014 \\
$e_3$              &  0.2033187  & 0.0000072  &  0.13957 & 0.00041 \\
$\omega_{\rm bin}$ (deg)     &  258.7446  & 0.0021  &  53.27 & 0.22 \\
$\omega_3$         (deg)     &  166.4007  & 0.0025  &   134.42 & 0.19 \\
$i_{\rm bin}$ (deg)     &  91.49422  & 0.00022  &  89.8536 & 0.0086 \\
$i_3$         (deg)     &  89.13552  & 0.00051  &  104.779 & 0.099 \\
$\Omega_3$         (deg)     &  0.64016  & 0.00065  &  38.46 & 0.12 \\
$I$         (deg)     &   2.44402  & 0.00068  &   40.84 & 0.12 \\
\bottomrule
\end{tabular}}
\end{adjustwidth}
\noindent{\footnotesize{$^1$ Reference times for osculating Keplerian 
elements: BJD 2,454,750 (KIC 7668648); BJD 2,454,800 (KIC 10319590).}}
\end{table}

\begin{figure}[H]
\includegraphics[width=10.0 cm, angle=-90]{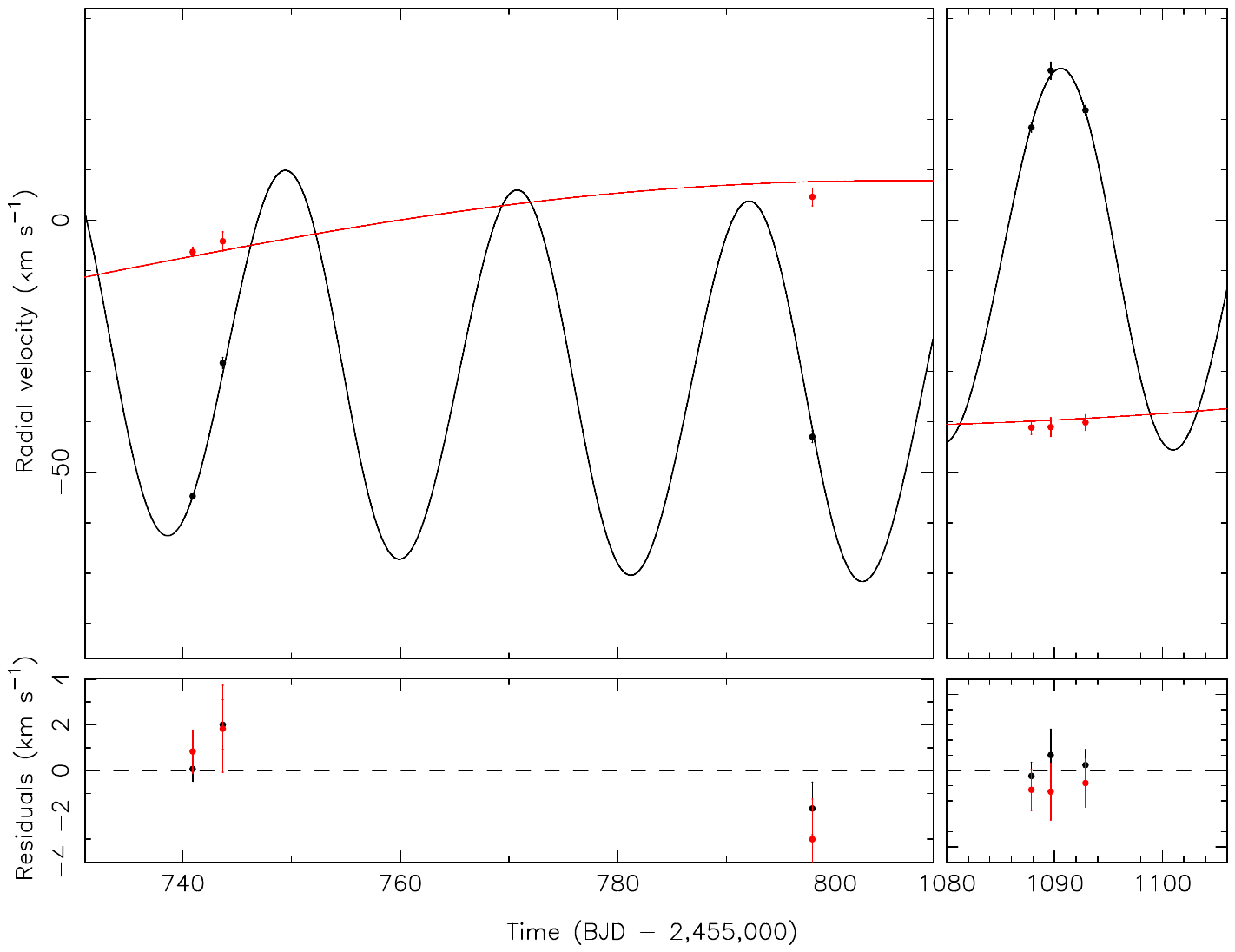}
\caption{\textbf{Top}: The radial velocity curve of the KIC 10319590
primary star (black points) with the best-fitting model (black line)
and the radial velocity curve of the KIC 10319590 tertiary star (red
points) with the best-fitting model (red line) \textbf{Bottom}: The
residuals of the fit.
\label{1031fig02}}
\end{figure}

\begin{figure}[H]
\includegraphics[width=10.0 cm, angle=-90]{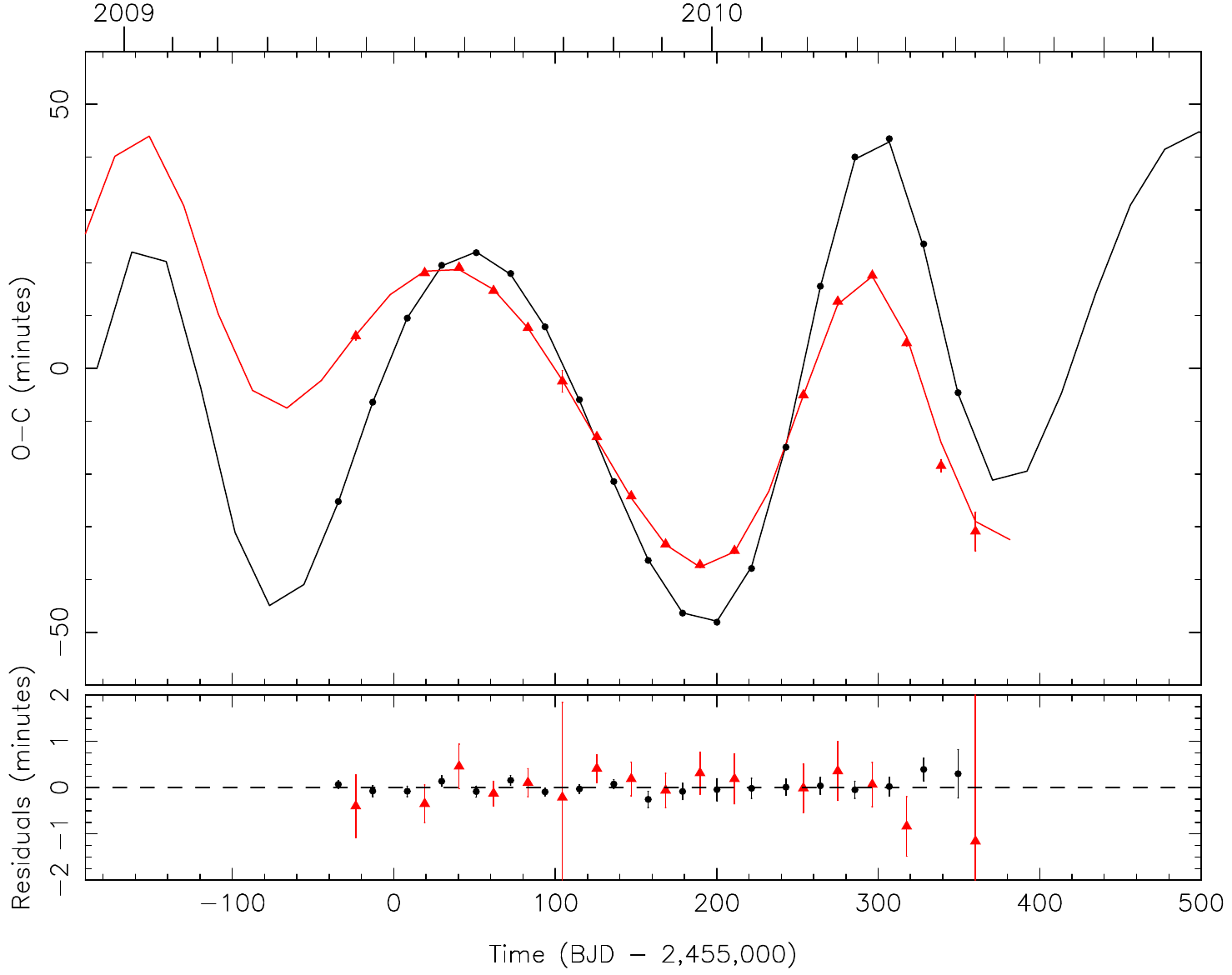}
\caption{\textbf{Top}: The Common-Period Observed Minus Computed
(CPOC) diagram for the KIC 10319590 primary eclipses (black points)
and secondary eclipses (red triangles), with~the respective
best-fitting models (black and red lines).  \textbf{Bottom}: The
residuals of the fit.
\label{1031fig03}}
\end{figure}

The best-fitting model light curve and the residuals of the fit are
shown in Figure~\ref{1031fig01} and the best-fitting radial velocity
curve and the residuals are shown in Figure~\ref{1031fig02}.
Figure~\ref{1031fig03} shows the Common-Period O-C (CPOC) diagram with
the measured eclipse times and the best-fitting model times.  Both
types of eclipses were not measurable in the {\em Kepler} data past
roughly day BJD 2,455,360.  In the best-fitting model, the last
primary eclipse is on day BJD 2,455,520.1, and the secondary eclipse
is on day BJD 2,455,381.6.

\subsection{KIC~7668648}
\unskip

\subsubsection{Overview}

KIC 7668648 was first noted in the catalog of
Pr\v{s}a~et~al.\ \cite{Prsa_2011}.  Using $\approx$$125$ days of {\em
  Kepler} data, Slawson~et~al.\ \cite{Slawson_2011} showed that there
were significant ETVs ($\approx$$\pm$$20$ min).  This object was one
of 39 triple-candidate systems identified by
Rappaport~et~al.\ \cite{Rappaport_2013}, who performed a basic
analytic analysis of the eclipse times that determined periods of
$P_1=27.8184$ days and $P_2=204.7$ days and masses of $0.02\le M_3\le
0.41\,M_{\odot}$ for the third star and $0.22\le M_{\rm bin}\le
3.65\,M_{\odot}$ for the binary, both at 90\% confidence.
Borkovits~et~al.\ 2015 and
Borkovits~et~al.\ 2016~\cite{Borkovits_2015, Borkovits_2016} noted the
presence of extra eclipse events (although they did not display them)
and did a much more comprehensive analytical analysis of the eclipse
times found from the full {\em Kepler} light curve.  The~latter
reported masses of $M_3=0.27(8)\,M_{\odot}$ for the third star and
$M_{\rm bin}=1.6(5)\,M_{\odot}$ for the binary, where the $1\sigma$
uncertainty in the last digit is given in the parentheses.  They also
found a mutual inclination (e.g.,\ the angle between the orbital plane
of the binary and the plane of the outer orbit) of 42(1) degrees.
Zhang~et~al.\ 2017~\cite{Zhang_2017} showed a total of 11 transit
events (their Figure~3) that are occultations and transits of the
third star.  They also show six additional possible transits (their
Figure~4), but~these appear to be spurious.  Finally,
Zhang~et~al.\ 2017~\cite{Zhang_2017} attempted to model the binary
eclipses without accounting for dynamical effects and,~as a result,
were unable to come up with a comprehensive and self-consistent model.
Indeed, Borkovits~et~al.\ 2016~\cite{Borkovits_2016} noted that for
systems like KIC 7668648 with large ETVs and extra eclipse events
``\ldots the accurate interpretation of such a system can be carried
out only by simultaneous modeling of its photometric and dynamical
properties, \ldots''.  We provide such a model~below.

\unskip   

\subsubsection{Photodynamical~Model}

\begin{figure}[H]
\includegraphics[width=9.0 cm, angle=-90]{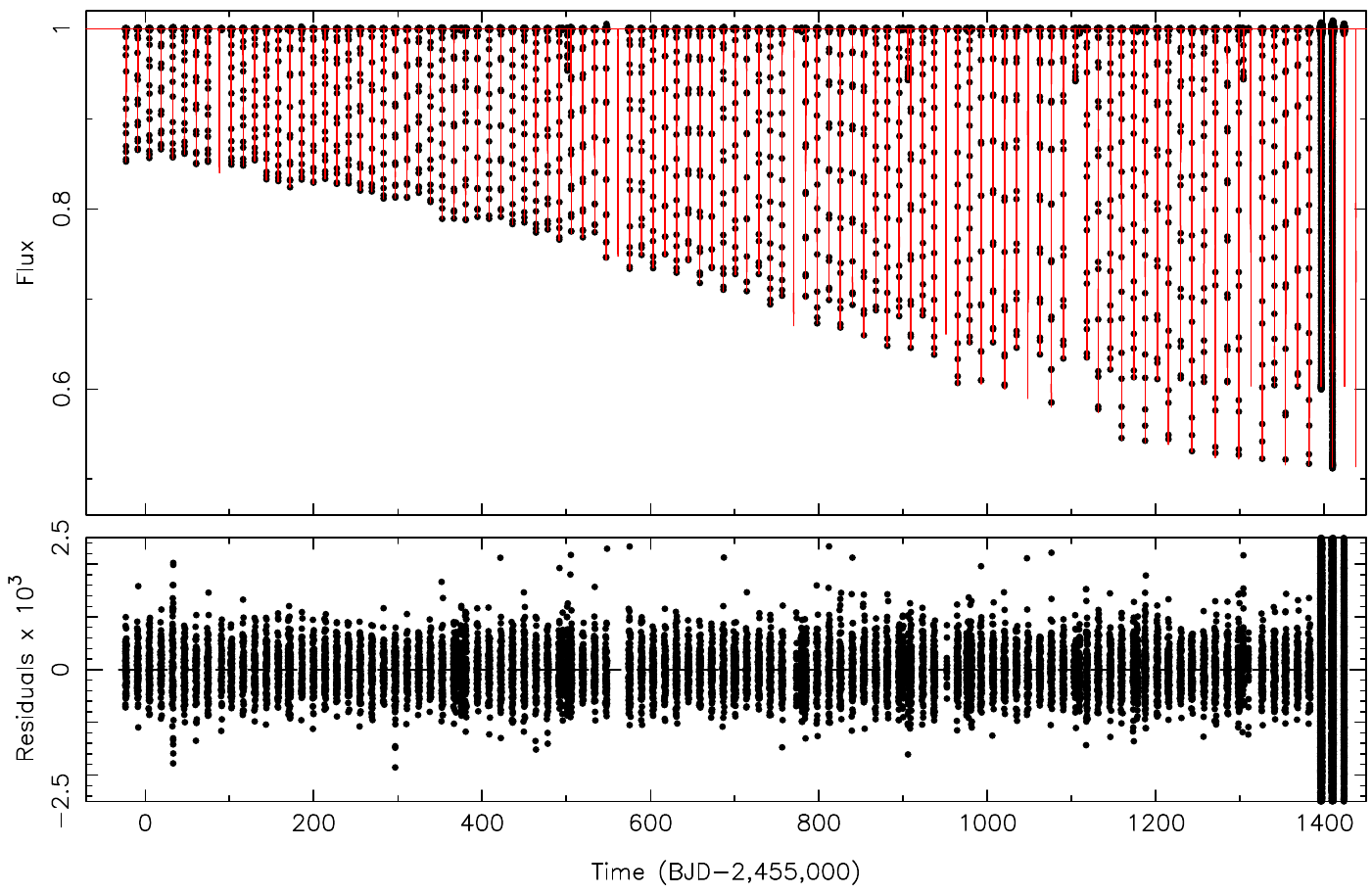}
\caption{\textbf{Top}: The normalized {\em Kepler} light curve of KIC
7668648 (points) with the best-fitting model (red line).
\textbf{Bottom}: The residuals of the fit.  Note that the last
portion of the light curve near day 1400 is in short cadence (1 min
``exposures'').
\label{7668fig01}}
\end{figure}

KIC 7668648 was observed in Long Cadence during Quarters 1 to 16 of
the nominal {\em Kepler} mission.  This target was placed on the list
of Short Cadence targets starting in Quarter 17 and~a little over one
month of data was taken before the nominal mission operations ceased
owing to a failed gyroscope.  The full {\em Kepler} light curve is
shown in Figure~\ref{7668fig01}.  There is a remarkable change in the
eclipse depths where the eclipses went from being $\approx$$15\%$ deep
at the start of the {\em Kepler} mission to being $\approx$$50\%$ deep
at the end.  {\em TESS} observed KIC 7668648 in Sectors 14 and 26 at
30 min cadence, and~Sectors 40, 41, 53, 54, and~55 at 10 min cadence.
The~source, also known as TIC 158215322, is relatively faint with a
TESS magnitude of 14.8, and~consequently, the light curves extracted
with {\tt eleanor} have relatively low signal-to-noise.  Three primary
eclipses and three secondary eclipses were identified.
Table~\ref{7668_times} in the Appendix \ref{appendix_times} gives the
eclipse times derived from the {\em Kepler} and {\em TESS} data.

\begin{figure}[H]
\includegraphics[width=10.0 cm, angle=-90]{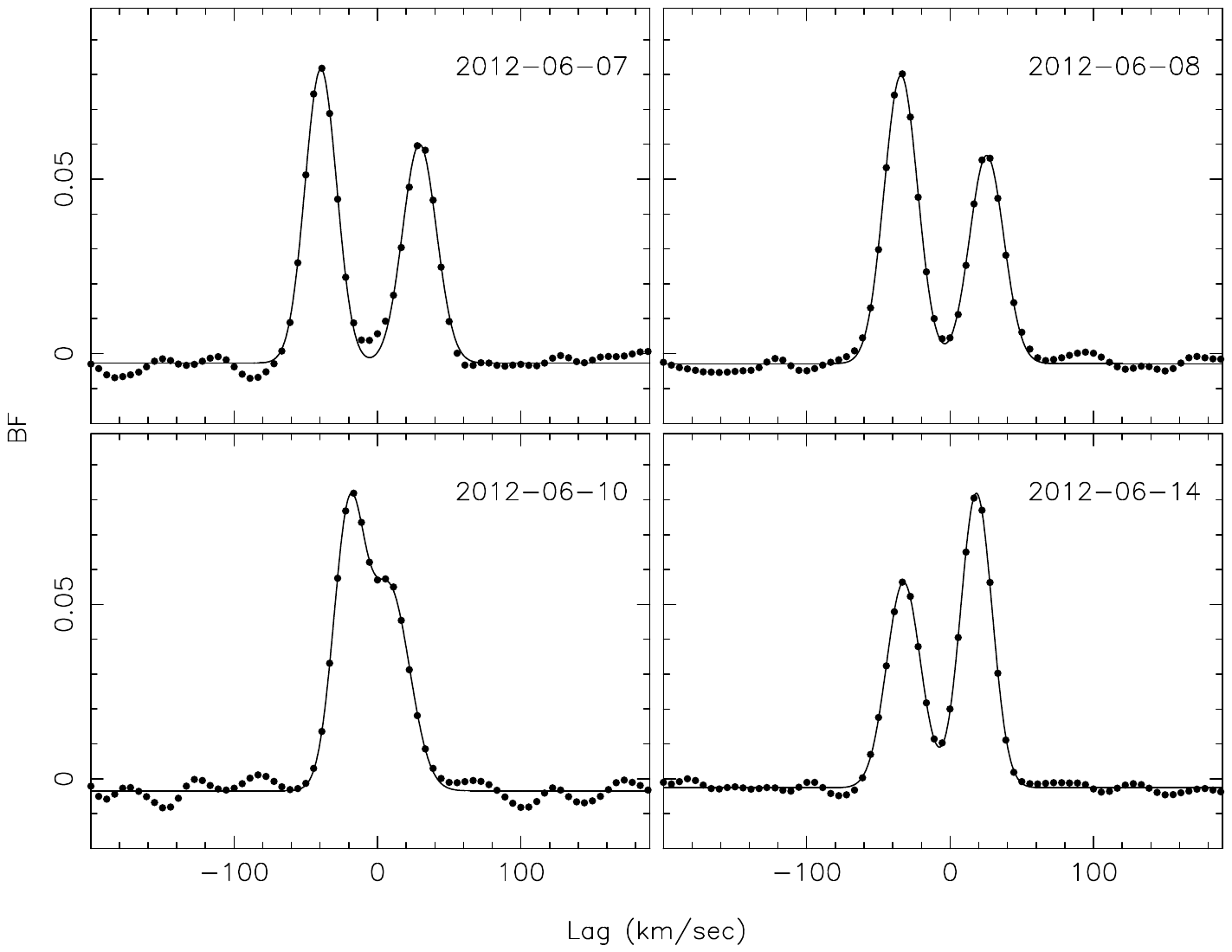}
\caption{Selected Broadening Functions (BFs) of KIC 7668648 (filled
circles) and their respective best-fitting double-Gaussian models
(solid lines).  The~stronger peak is due to the primary star in the
inner binary, and the other peak is due to the secondary star in the
inner binary.
\label{7668BF}}
\end{figure}

KIC 7668648 was observed seven times using the 4m telescope at Kitt
Peak and the echelle spectrograph between June 6, 2012 and June 16,
2013.  A~spectrum of the G8V star HD 181655 was used as the template
for the BF analysis.  The spectra are double-lined, with the strongest
set of lines attributable to the primary star in the inner binary,
while the other set of lines is attributable to the secondary star in
the inner binary.  A double-Gaussian model was fitted to the seven BFs
to determine the radial velocities (see Figure~\ref{7668BF}),
after~suitable heliocentric corrections were applied (see
Table~\ref{7668_RVtab}).  Using the areas under the BF peaks as a
proxy for the flux ratio (in this case, secondary flux divided by
primary flux), we find a flux ratio in the spectral region between
4875\AA\ and 5850\AA\ of $0.75\pm 0.05$.

The photodynamical model has 30 free parameters and~is very similar to
the one used for KIC 10319590, with~three changes: (i) tides have a
minimal effect on the binary, so only the extra force terms related to
GR were used; (ii) the radii of the primary and secondary stars were
parameterized by their fractional radii, where $f_1\equiv R_1/a$ and
$f_2\equiv R_2/a$; and (iii) the Power-2 limb-darkening coefficients
for the third star $h_1$ and $h_2$ were free parameters.  The
reference time for the osculating parameters is BJD 2,454,750.  We fit
the {\em Kepler} light curve and the radial velocity curve.  We also
include the measured times of all eclipse events in the likelihood
function.  There are some pretty significant instrumental changes in
the eclipse depths seen in the {\em TESS} light curve, so
consequently, only the measured eclipse times were used in
the~fitting.

The fitting procedure for KIC 7668648 was very similar to the fitting
procedure for \mbox{KIC 10319590} described above.  Once a good
initial solution was found, the~DE-MCMC code was run 10 separate times
for 30,500 generations, where each run of the DE-MCMC code had a
different seed for the random number generator.  After~a burn-in
period of \mbox{1250 generations}, posterior samples were generated by
sampling models every \mbox{750 generations}.  The~ten individual
posterior samples were combined to yield overall sample sizes of
32,000 for each fitting parameter and derived parameter.  The median
values of the posterior distributions and the corresponding $1\sigma$
uncertainties for the fitting parameters are given in
Table~\ref{tab_7668fit}.  Table~\ref{tab_1031deriv} gives the median
values of the posterior distributions and the corresponding
uncertainties of some key derived parameters.  The parameters are
generally well constrained, and~many have very small formal errors.
For~example, the~stellar masses have uncertainties of about 1\% and
the stellar radii have uncertainties of about 0.4\%.  The initial
Keplerian and Cartesian initial conditions for the best-fitting model
are given in Table~\ref{K7668_init} in the \mbox{Appendix \ref{appc}}.

\begin{table}[H]
\caption{Fitting Parameters for KIC~7668648.\label{tab_7668fit}}
\begin{adjustwidth}{-\extralength}{0cm}
\centering 
\setlength{\tabcolsep}{8mm} \resizebox{\linewidth}{!}{\begin{tabular}{llllll}
\toprule
\textbf{Parameter $^1$} & \textbf{Median} & $\boldsymbol{1\sigma}$  & \textbf{Parameter $^1$} & \textbf{Median} &$\boldsymbol{1\sigma}$ \\
\midrule
$P_{\rm bin}$ (days) & 27.796270 & 0.000019     & $P_3$ (days) & 208.00479 & 0.00028 \\
$T_{\rm conj,bin}$  & $-92.50434$ $^2$ & 0.00012 & $T_{\rm conj,3}$ & $-105.8123506$ $^2$ & 0.00057  \\
$e_{\rm bin}\cos\omega_{\rm bin}$ & $-0.0095588$ &  0.0000019 & $e_3\cos\omega_3$ & $-0.1976184$ & 0.0000083  \\
$e_{\rm bin}\sin\omega_{\rm bin}$ & $-0.0480312$  & 0.0000045 & $e_3\sin\omega_3$ & 0.0478065   & 0.0000079 \\
$i_{\rm bin}$ (deg) & 91.49422 & 0.00022 & $i_3$ (deg) & 89.13551 & 0.00051 \\
$\Omega_{\rm bin}$ (deg) & $0$ $^3$ & \ldots & $\Omega_3$ (deg) & 0.64016  & 0.00065 \\
$M_1$ ($M_{\odot}$) & 0.8403 & 0.0090 & $Q_{\rm bin}$ &  0.95198 & 0.00010  \\
$Q_3$          &  5.96517 & 0.00070 & & & \\
$f_1$ ($R_1/a$)  & 0.0221031  & 0.0000054   & $f_2$ ($R_2/a$)  &   0.019276  & 0.000012  \\
$\rho_3$    &    3.5019  & 0.0028  & & & \\
$T_1$ (K)   & 5191 & 82 & $\tau_{\rm bin}$ 0.96890 & 0.00057 \\
$T_3$ (K) & 2927  & 123  & & & \\
$h_1$ primary & 0.7530 & 0.0023 & $h_2$ primary &  0.529 & 0.038 \\
$h_1$ secondary & 0.7367 & 0.0026 & $h_2$ secondary & 0.447 & 0.053 \\
$h_1$ tertiary  & 0.58 & 0.24 &   $h_2$ tertiary & 0.26 & 0.19 \\ 
$s_0$ & 0.00549 & 0.00078  & $s_1$ & 0.00460 & 0.00078 \\
$s_2$ & 0.00834 & 0.00078 & $s_3$  & 0.00647 & 0.00078 \\ 
\bottomrule
\end{tabular}}
\end{adjustwidth}
\noindent{\footnotesize{$^1$ Reference time for  osculating Keplerian 
elements:  BJD 2,454,750.  $^2$  BJD $-$2,455,000.  $^3$  fixed.}}
\end{table}

\begin{figure}[h!]
\includegraphics[width=8.0 cm, angle=-90]{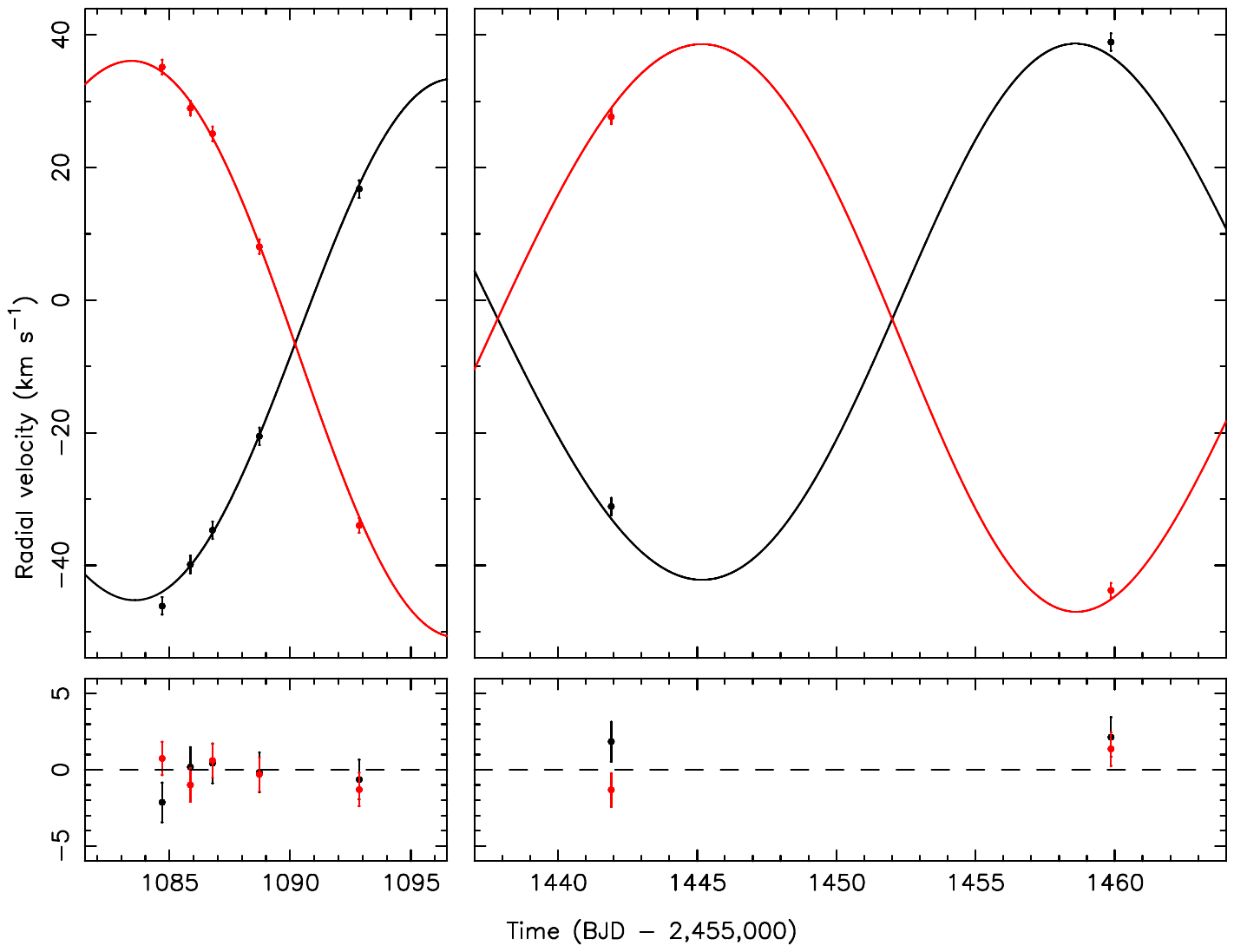}
\caption{\textbf{Top}: The radial velocity curve of the KIC 7668648
primary star (black points) with the best-fitting model (black line)
and the radial velocity curve of the secondary star star (red
points) with the best-fitting model (red line) \textbf{Bottom}: The
residuals of the fit.
\label{7668fig02}}
\end{figure}

\begin{figure}[H]
\includegraphics[width=9.0 cm, angle=-0]{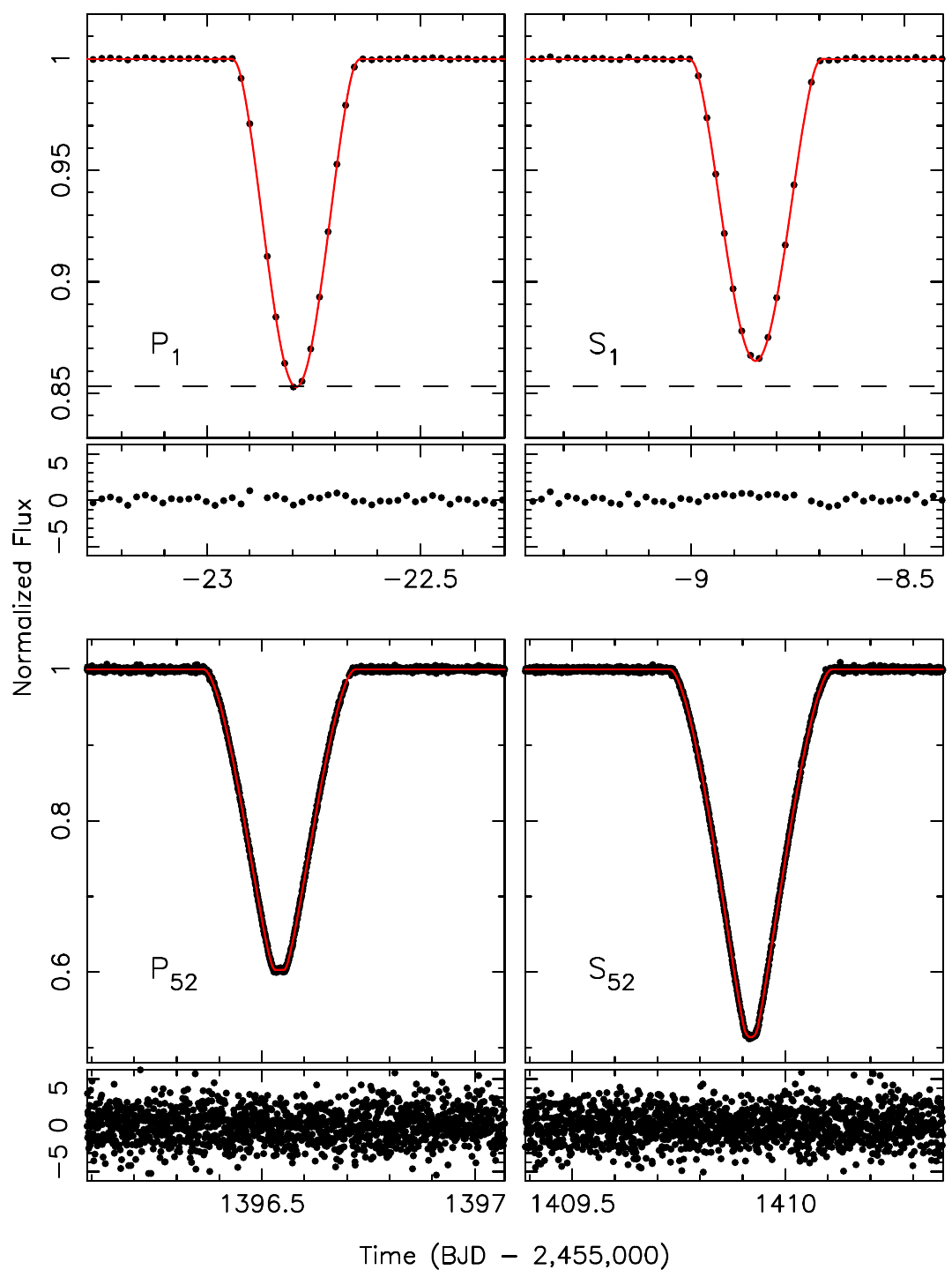}
\caption{\textbf{Top}: Close-up views of the first pair of
primary (\textbf{top left}) and secondary (\textbf{top right})
eclipses of KIC 7668648 (these were observed on Long Cadence).
\textbf{Bottom}: Close-up views of a pair of eclipses 51 cycles
later (these were observed in Short Cadence).  Please note that both
events are overall much deeper than the first pair of events.
In~addition, the~``secondary'' eclipse at cycle 52 is deeper than
the corresponding ``primary'' eclipse.  In each case, the
observations are shown with the black points and the best-fitting
model is the red line.  The residuals of the fit are shown in the
thin panels.
\label{7668fig05}}
\end{figure}

The best-fitting model radial velocity curve models are shown in
Figure~\ref{7668fig02}.  The best-fitting light-curve model is
displayed in Figure~\ref{7668fig01}.  As~noted before, there is a
striking change in the depths of the eclipses from the start of the
{\em Kepler} mission to the end of its nominal operation.
In~addition, there is another feature that we have no easy explanation
for, namely, the roles of the primary and secondary changed between
the beginning of the light curve and the end.  This is shown in
Figure~\ref{7668fig05}.  The~first eclipse event seen in the light
curve (near day $-23$, labeled $P_1$ in the figure) is deeper than the
next eclipse event (near day $-9$, labeled $S_1$ in the figure), so
the former was called a ``primary'' eclipse and the latter was called
a ``secondary'' eclipse.  The~last pair of eclipse events seen in the
{\em Kepler} light curve occur near days 1396.5 (labeled $P_{52}$ in
the figure) and 1410 (labeled $S_{52}$ in the figure).  There are
\mbox{51 orbital} cycles between events $P_1$ and $P_{52}$ and between
events $S_1$ and $S_{52}$.  Thus, one would have to label event
$P_{52}$ as a ``primary'' eclipse and event $S_{52}$ as a secondary
eclipse, even though the latter is clearly deeper than the~former.

\begin{figure}[H]
\includegraphics[width=8.5 cm, angle=-90]{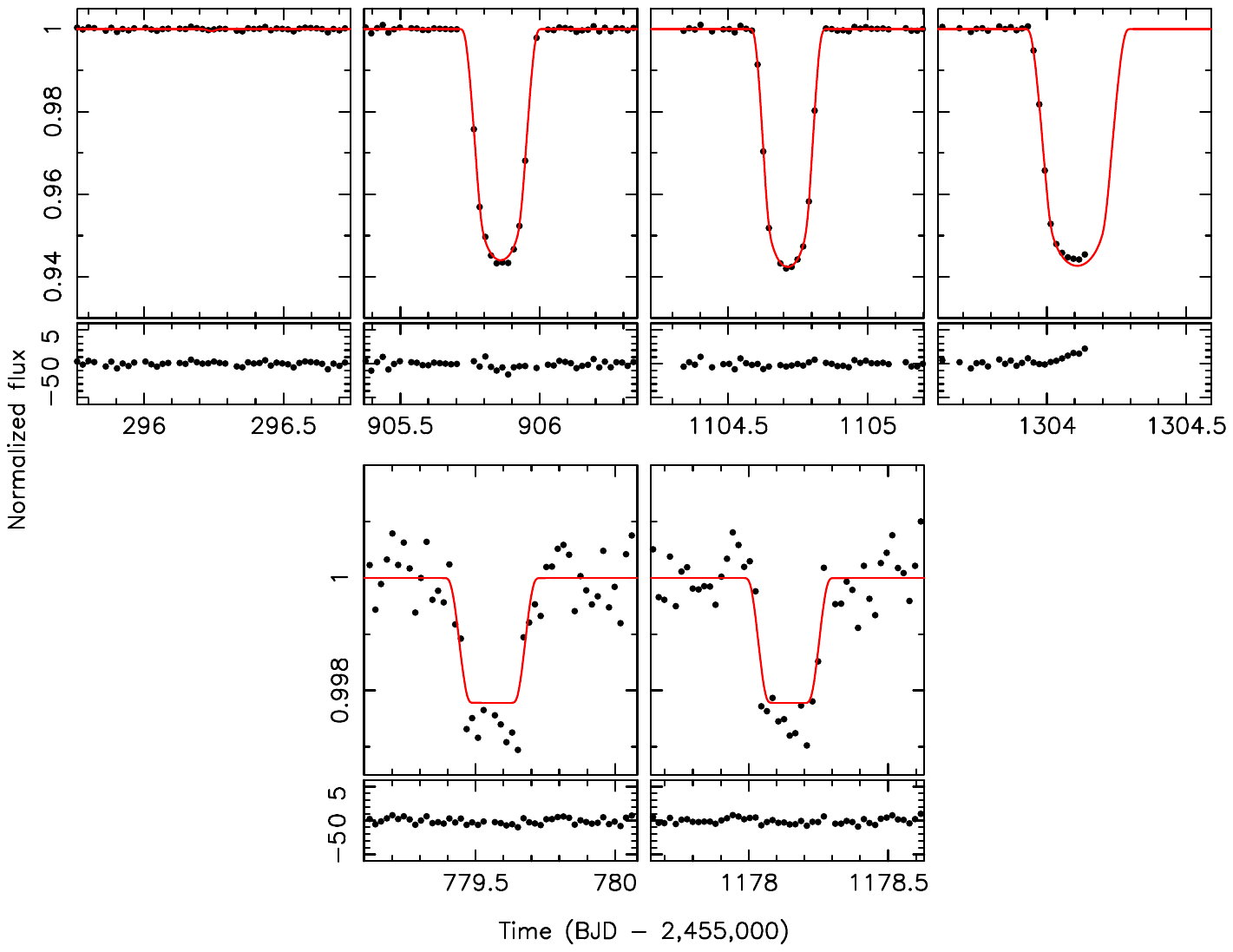}
\caption{\textbf{Top}: Transits of the primary star by the tertiary
star in KIC 7668648.  The~transit near day 296.3 is grazing.
\textbf{Bottom}: Occultations of the third star by the primary star
in KIC 7668648.  In each case, the observations are shown with the
black points and the best-fitting model is the red line.  The
residuals of the fit are shown in the thin panels.
\label{7668fig03}}
\end{figure}

\begin{figure}[h!]
\includegraphics[width=8.5 cm, angle=-90]{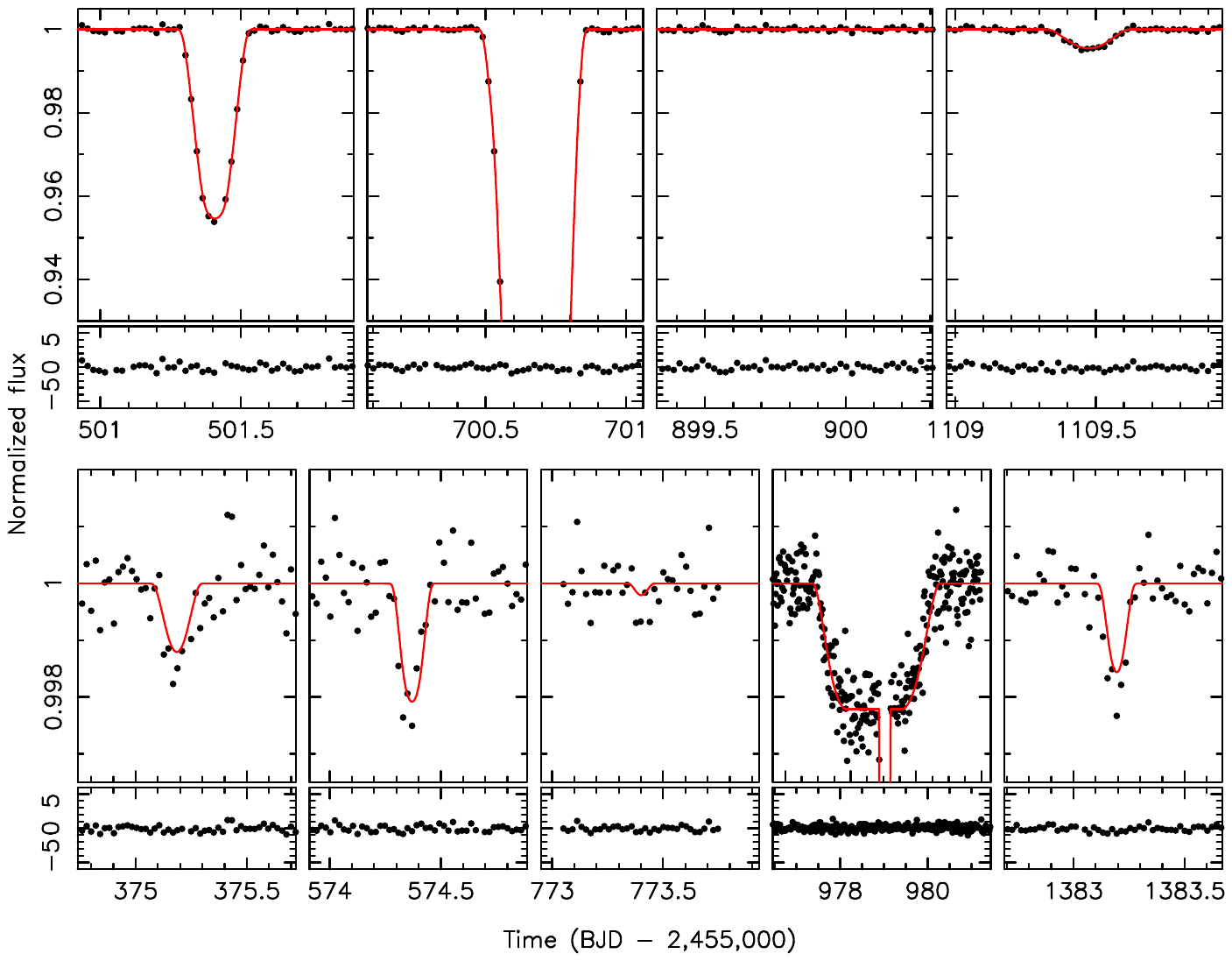}
\caption{\textbf{Top}: Transits of the secondary star by the tertiary
star in KIC 7668648.  The~transit near day 700.5 is blended with a
secondary eclipse and the event near day 900 is grazing.
\textbf{Bottom}: Occultations of the third star by the secondary
star in KIC 7668648.  The occultation near day 979 occurred near a
secondary eclipse, and~consequently, the occultation event is almost
4 days long.  In each case, the observations are shown with the
black points and the best-fitting model is the red line.  The
residuals of the fit are shown in the thin panels.
\label{7668fig04}}
\end{figure}

As noted earlier, there are numerous tertiary eclipse events seen in
the light curve.  Figure~\ref{7668fig03} shows the transits of the
primary (top of the figure) and occultations of the tertiary star by
the primary (bottom of the figure).  Figure~\ref{7668fig04} shows the
corresponding figure for the secondary+tertiary events.  In~general,
the~model fits these observed events well.  Owing to the complex
geometry brought about by the strong dynamical interactions between
the tertiary star and the binary, the~tertiary events can have a wide
variety of impact parameters and, hence, different
depths.\vspace{-3pt}

\begin{figure}[H]
\includegraphics[width=10.5 cm, angle=-90]{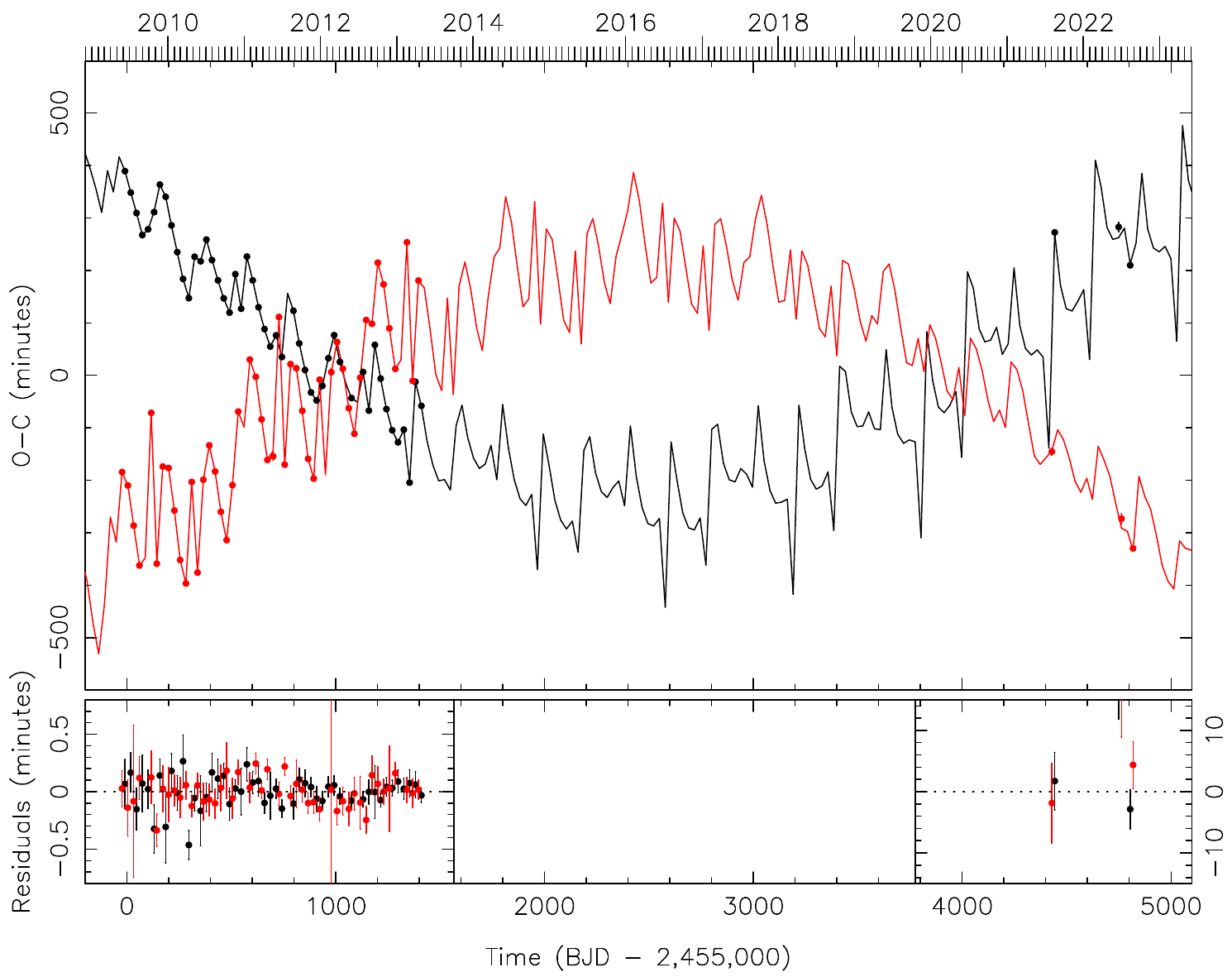}
\caption{\textbf{Top}: The Common-Period Observed Minus Computed
(CPOC) diagram for the \mbox{KIC 7668648} primary eclipses (black
points) and secondary eclipses (red points), with~the respective
best-fitting models (black and red lines).  \textbf{Bottom}: The
residuals of the fit.  Note the different scales on the $y$-axis for
the {\em TESS} measurements in the lower panel on the right.
\label{7668fig06}}
\end{figure}

Finally, KIC 7668648 has one of the more extreme ETV signals of any
known binary, where the instantaneous period (e.g.,\ the interval
between consecutive primary eclipses or the interval between
consecutive secondary eclipses) can vary by amounts as large as
$\approx$$5$ h.  For perspective, the~average orbital period is around
27.8 days, so 5 h is $\approx$$0.75$\% of the orbital period.
Figure~\ref{7668fig06} shows the model and observed CPOC signals.
As~discussed below, the~apsidal period is around 27 years, so about
half of the apsidal cycle is covered between the beginning of the {\em
  Kepler} observations and the end of the {\em TESS} observations.
The amplitudes of the primary and secondary CPOC signals are
$\approx$$13$ h.  In~spite of the large ETVs, the residuals of the
fits are less than 30 s for the {\em Kepler} observations.

\unskip   

\subsection{KIC~5255552}
\unskip

\subsubsection{Overview}

This object was first identified as a binary by
Pr\v{s}a~et~al.\ \citep{Prsa_2011} using the first 44 days of {\em
Kepler} data.  Using $\approx$$125$ days of {\em Kepler} data,
Slawson~et~al.\ \cite{Slawson_2011} were able to determine an orbital
period of $P=32.46$ days.  Using the entire {\em Kepler} data set
spanning $\approx$$1450$ days, Borkovits~et~al.\ (2013)
\cite{Borkovits_2015} performed an analytic study of the eclipse time
variations (ETVs).  They found an outer orbital period of $P_2\approx
863$ days and orbital eccentricies of $e_1\approx 0.3$ and $e_2\approx
0.4$.  Finally, they noted the presence of extra eclipses but~did not
display them or discuss them in detail.
Zhang~et~al.\ \cite{Zhang_2018} noted that the three sets of extra
eclipses were broadly consistent with being caused by a third body on
the $\approx$$863$ day period identified by Borkovits, but~were
otherwise unable to find a comprehensive and self-consistent solution.
Likewise, Getley~et~al.\ \cite{Getley_2020} performed a stability
analysis of KIC 5255552 and attempted to model the light curves.
There is a set of four extra eclipses that span $\approx$$25$ days,
and these events proved to be especially challenging to model since a
single star on an \mbox{$\approx$$863$ day} orbit would not linger
near conjunction for that long.  Getley~et~al.\ \cite{Getley_2020}
speculated that \mbox{KIC 5255552} might contain two binaries with
only the first binary eclipsing.  As~we show below, that is indeed
the~case.

\subsubsection{Photodynamical~Model}

\begin{figure}[H]
\includegraphics[width=8.7cm, angle=-90]{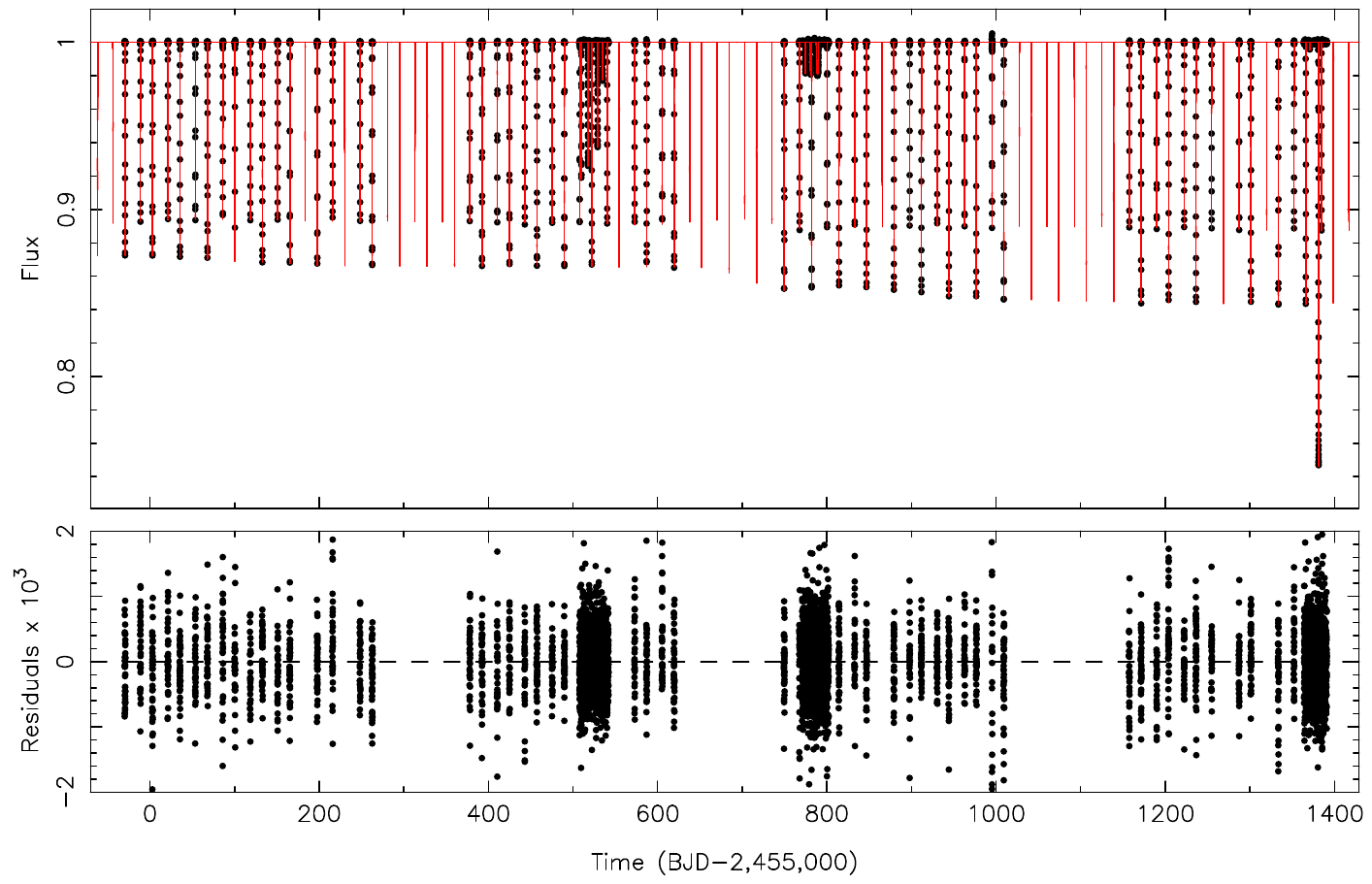}
\caption{\textbf{Top}: The normalized {\em Kepler} light curve of KIC
5255552 (points) with the best-fitting model (red line).  Module 3
failed on 9 January 2010, and~consequently, no data were obtained
for KIC 5255552 during Quarters 5, 9, 13, and~17.  \textbf{Bottom}:
The residuals of the fit.
\label{fig1}}
\end{figure}

KIC 5255552 was observed by the {\em Kepler} mission from the start of
{\em Kepler} Quarter 1 through the end of {\em Kepler} Quarter 16.
Module 3 of the {\em Kepler} detector failed on 9 January 2010,
and~unfortunately KIC 5255552 was not observed during Quarters 5, 9,
13, and~17 (see Figure~\ref{fig1}).  The~source, also known as TIC
120316928, was observed by {\em TESS} in Sectors 14 and 26 at 30 min
cadence and Sectors 40, 41, 53, and~54 at 10 min cadence.  A~total of
four primary eclipses and five secondary eclipses appear in the {\em
TESS} light curve.  Table~\ref{5255_times} in the Appendix
\ref{appendix_times} gives the eclipse times derived from the {\em
Kepler} and {\em TESS} data.

\begin{figure}[H]
\includegraphics[width=10.0 cm, angle=-90]{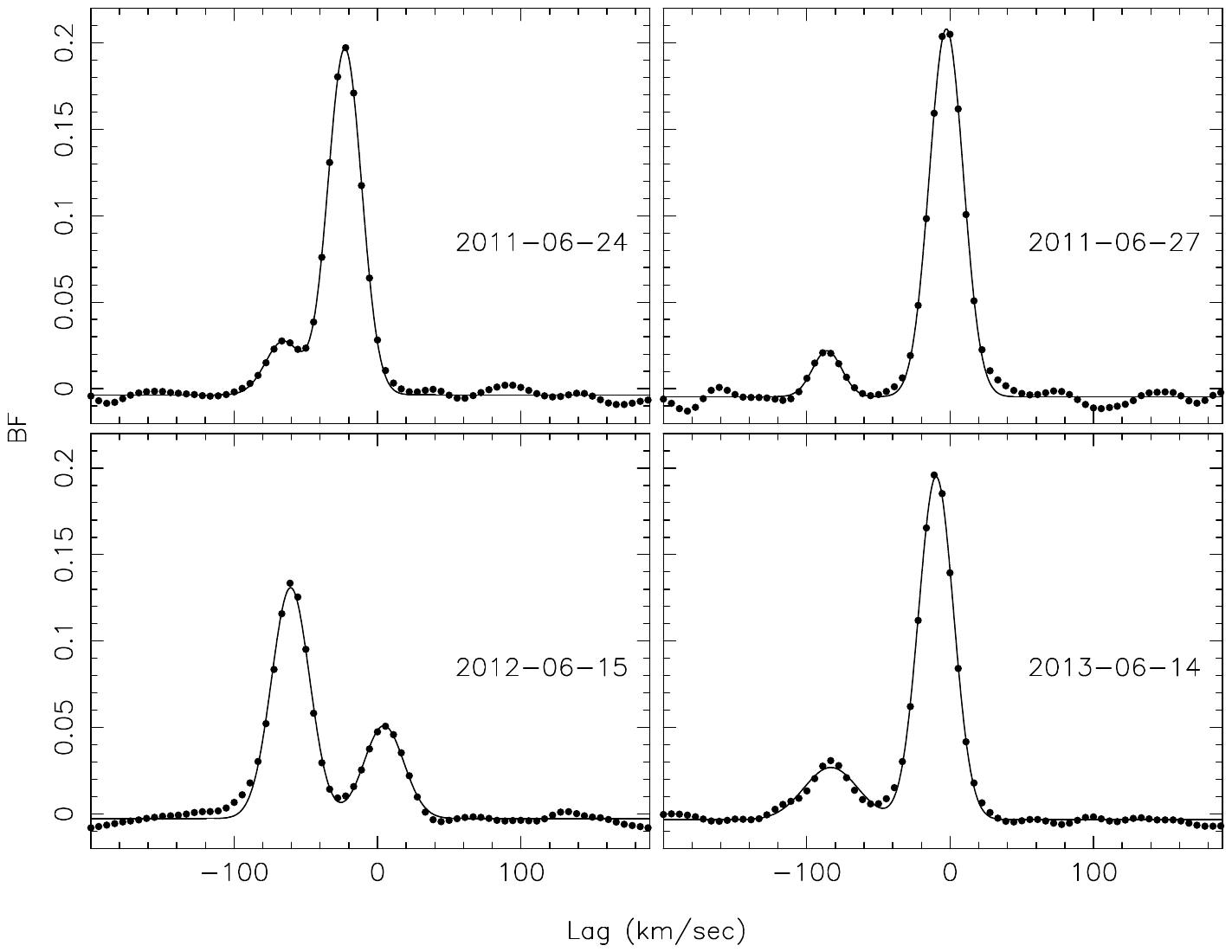}
\caption{Selected Broadening Functions (BFs) of KIC 5255552 (filled
circles) and their respective best-fitting double-Gaussian models
(solid lines).  The~stronger peak is due to the primary star in the
first binary and the weaker peak is due to the secondary star in the
first binary.
\label{5255BF}}
\end{figure}

KIC 5255552 was observed 11 times using the 4 m telescope at Kitt Peak
and the echelle spectrograph between 24 June 2011 and 14 June 2013.
Two spectra from 15 June 2013 were unusable owing to extremely poor
signal-to-noise ratios.  A spectrum of the G8V star HD 181655 was used
as the template for the BF analysis of the remaining nine spectra.
Except for the spectrum from 13 June 2013, the~BFs have one strong
peak and a second peak that is considerably weaker.  These two peaks
can be attributed to the primary and secondary of the eclipsing
binary, respectively.  A double-Gaussian model was fitted to the eight
BFs with double peaks to determine the radial velocities (see
Figure~\ref{5255BF}), after~suitable heliocentric corrections were
applied (see Table~\ref{5255_RVtab}).  Using the areas under the BF
peaks as a proxy for the flux ratio (in this case, secondary flux
divided by primary flux), we find a flux ratio in the spectral region
between 4875\AA\ and 5850\AA\ of $0.18\pm 0.04$.

After a considerable effort, we were unable to find a suitable fit to
the light curve using the 2 + 1 model that was successfully used for
KIC 10319590 and KIC 7668648.  Thus, we instead used the ``double
binary'' model, which has 49 free parameters.  We have three orbits:
(i) the orbit of the eclipsing binary (hereafter ``binary \#1''),
with~a period $P_{\rm bin,1}$, a time of primary conjunction $T_{\rm
  bin,1}$, eccentricity parameters $e_{\rm bin,1}\cos\omega_{\rm
  bin,1}$ and $e_{\rm bin,1}\sin\omega_{\rm bin,1}$, inclination
$i_{\rm bin,1}$, and~nodal angle $\Omega_{\rm bin,1}\equiv 0$; (ii)
the orbit of the non-eclipsing binary (hereafter ``binary \#2'') with
similar Keplerian elements (in this case, the nodal angle is a free
parameter); and (iii) the outer orbit with similar Keplerian elements
(where again the nodal angle is a free parameter).  The stellar masses
for binary \#1 are parameterized by the primary mass $M_1$ and the
binary mass ratio $Q_{\rm bin,1}\equiv M_2/M_1$.  The~stellar masses
for binary \#2 are parameterized by $Q_3\equiv M_1/M_3$ and $Q_4\equiv
M_1/M_4$ \endnote{Note this definition of the mass ratio differs from
  what was used for KIC 7668648 and KIC 10314940}.  The~stellar radii
for binary \#1 are parameterized by $R_1$ and $\rho_{\rm bin,1}\equiv
R_1/R_2$, and~the stellar radii for binary \#2 are parameterized by
$\rho_3\equiv R_1/R_3$ and $\rho_4\equiv R_1/R_4$.  The stellar
effective temperatures are $T_1$ and $\tau_{\rm bin, 1}\equiv T_2/T_1$
for binary \#1 and $T_3$ and $T_4$ for binary \#2.  We have
light-curve data in two bandpasses.  Since all four stars are involved
in eclipse events in the {\em Kepler} light curve, each star has a
pair of limb-darkening coefficients $h_1$ and $h_2$ for the Power-2
limb-darkening law with the Maxted transformations.  The {\em TESS}
light curve only has eclipse events involving the stars in binary \#1,
and~this gives a pair of limb-darkening coefficients $h_1$ and $h_2$
for each star in that binary.  We also have the seasonal contamination
parameters for the {\em Kepler} as described above and~similar
parameters for the {\em TESS} light curve: $c_0$ for the time interval
between days 3680 and 3705 (in units of BJD $-$2,455,000), $c_1$ for
the time interval between days 4010 and 4030, $c_2$ for the time
interval between days 4415 and 4440, and~$c_3$ for the time interval
between days 4760 and 4795.  As~was the case for TIC 7668648, it was
found that tides have a minimal effect, so only the extra force terms
related to GR were used.  We fit the {\em Kepler} light curve, the
{\em TESS} light curve, and~the radial velocity curves for the stars
in binary \#1, the~measured eclipse times for the primary and
secondary eclipses of binary \#1, and~for the measured times for the
extra eclipse events (whose origins were not known at first).

Finding a solution for KIC 5255552 proved to be extremely
challenging. The~basic properties of binary \#1 are {\rm roughly}
known since it is eclipsing.  Likewise, {\em some} of the basic
properties of the outer orbit are {\em roughly} known from a basic
analysis of the eclipse times of binary \#1.  On~the other hand, the
geometrical properties of binary \#2 are almost completely unknown,
except for two observed constraints: (i) the binary apparently does
not eclipse, which eliminates some small and complex regions of
parameter space, and~(ii) the two stars in binary \#2 are less
luminous than the secondary in binary \#1 since only two peaks are
seen in the BFs.  We discovered that one cannot really treat the
second binary as a point mass for detailed work.  That is, two nearly
identical models that have everything the same apart from the time of
primary conjunction for the second binary would have differences in
the eclipse times for the first binary that was much larger than the
uncertainties in the measured times.  This explains in part why fits
with three-body models, in an effort to get ``close'' to the solution
as a starting point for a four-body model, do not really work as well
as one might have expected.  Finally, as~others have
noted~\cite{Zhang_2018,Getley_2020}, the timing of the extra eclipse
events depends sensitively on precise positions of the bodies
involved, and~small deviations in various orbital elements can result
in a large change in the timing of the extra~events.

We initially tried two strategies.  First, if~one ignored the extra
eclipse events and tried to get a very good model for the eclipses in
binary \#1, then at some point, the model would have the extra
eclipses at the correct times.  How one weights the measured times of
the extra events can also be altered, so by setting the uncertainties
to 1 day instead of \mbox{$\approx$$0.004$ days}, the extra events are
almost ``ignored''.  Several runs with the genetic algorithm and later
the hybrid genetic algorithm/nested sampling code were done,
but~without success.  The~other approach was to make the uncertainties
on a few of the extra events very small, so that those events were
always matched after {\tt gennestELC} converged.  This approach also
did not really work since subsequent attempts to refine these
solutions with the DE-MCMC code almost always resulted in little
overall~improvement.
  
In the end, the successful strategy for finding a solution was
relatively straightforward but required an incredible brute-force
computing effort.  Several different searches were conducted with {\tt
  gennestELC} where the prior ranges for the period of binary \#2 and
the time of conjunction of binary \#2 were severely restricted.
The~uncertainties on the extra eclipse events were set to 1 day.
Then, after many runs were completed over a wide range of possible
orbital periods for binary \#2, the solutions were ranked by the
overall $\chi^2$, and the DE-MCMC code was deployed using the
top-ranked models as starting places.  A~few of these starting models
were ``close'' enough so that a solution that matched the light curve
with all the extra eclipse events was~found.

After a good solution was found, the~DE-MCMC code was used in a manner
similar to what was used for KIC 7668648 and KIC 10319590 as described
above.  The~code was run 10 times with different initial random number
seeds for a total of 30,000 generations.  After a burn-in period of
4000 generations, posterior samples were taken every 2000 generations.
The~samples were combined, yielding sample sizes of 22,400 for each
fitting parameter and each derived parameter.  Table~\ref{tab_5255fit}
gives the median values of the posterior distributions for the fitting
parameters and their corresponding $1\sigma$ uncertainty.
Table~\ref{tab_5255deriv} gives the same quantities for selected
derived parameters.  The initial Keplerian and Cartesian initial
conditions for the best-fitting model are given in
Table~\ref{K5255_init} in the Appendix \ref{appc}.

Figures~\ref{fig2}--\ref{fig4} show close-up views of the {\em Kepler}
light curve near the times when the extra eclipse events occur.
Figure~\ref{fig2} shows the sequence of four extra events that
\mbox{span $\approx$ $25$ days} that Getley~et~al.\ \cite{Getley_2020}
discussed.  The~first two events in this sequence are caused by the
primary star of binary \#2 transiting the secondary of binary \#1.
The~last two events in the sequence are the secondary star in binary
\#2 transiting the secondary star in binary \#1.  The~three extra
eclipse events near day 780 shown in Figure~\ref{fig3} are
occultations of the stars in binary \#2 by stars in binary \#1.
Finally, the two extra eclipse events near day 1375 shown in
Figure~\ref{fig4} are transits of the primary star in binary \#1 by
the secondary star in binary \#2.  Figure~\ref{plotTESS} shows the
{\em TESS} data (which features four primary eclipses and five
secondary eclipses of binary \#1) and the best-fitting model.
Figure~\ref{fig5} shows the radial velocities and best-fitting radial
velocity curve for binary \#1, and~Figure~\ref{fig6} shows the CPOC
diagram for binary \#1.  The~individual signals have modulations of up
to \mbox{$\approx$$5$ h}.  In~addition, the~diverging signals indicate
significant apsidal motion, where the rate is $\approx$$0.218$ degrees
per~cycle.

\begin{table}[H]
\caption{Fitting Parameters for KIC~5255552.\label{tab_5255fit}}
\begin{adjustwidth}{-\extralength}{0cm}
\centering 
\setlength{\tabcolsep}{7mm} \resizebox{\linewidth}{!}{\begin{tabular}{llllll}
\toprule
\textbf{Parameter $^1$} & \textbf{Median} & $\boldsymbol{1\sigma}$  & \textbf{Parameter $^1$} & \textbf{Median} &$\boldsymbol{1\sigma}$ \\
\midrule
$P_{\rm bin,1}$ (days) & 32.454276 & 0.000045     & $T_{\rm bin,1}$ & $-191.803437~^2$ & 0.000063 \\
$e_{\rm bin,1}\cos\omega_{\rm bin,1}$ & $0.0843826$ &  0.0000047 & 
                  $e_{\rm bin,1}\sin\omega_{\rm bin,1}$ & $-0.201273$ & 0.000065  \\
$i_{\rm bin,1}$ (deg) & 88.9812 & 0.0034 & $\Omega_{\rm bin,1}$ (deg) & $0.0~^3$ & \ldots \\
$M_1$ ($M_{\odot}$) & 0.950 & 0.018 & $Q_{\rm bin,1}$ &  0.78393 & 0.00034  \\
$R_1$ ($R_{\odot}$) & 0.9284  & 0.0063 & $\rho_{\rm bin,1}$ & 1.3472 & 0.0059 \\
$P_{\rm bin,2}$ (days) & 33.74293 & 0.00046     & $T_{\rm bin,2}$ & $-199.4782~^2$ & 0.0056 \\
$e_{\rm bin,2}\cos\omega_{\rm bin,2}$ & $-0.15815$ &  0.00012 & 
                  $e_{\rm bin,2}\sin\omega_{\rm bin,2}$ & $0.03343$ & 0.00015  \\
$i_{\rm bin,2}$ (deg) & 86.147 & 0.013 & $\Omega_{\rm bin,2}$ (deg) & $-$1.054 & 0.010 \\
$Q_3$  & 1.96665 & 0.00099 & $Q_4$ &  1.8728 & 0.0010  \\
$\rho_3$  & 2.0008  & 0.0057 & $\rho_4$ & 1.9551 & 0.0046 \\
$P_{\rm out}$ (days) & 878.4725 & 0.0057     & $T_{\rm out}$ & $-344.4563~^2$ & 0.0070 \\
$e_{\rm out}\cos\omega_{\rm out}$ & $-0.3125385$ &  0.0000039 & 
                  $e_{\rm out}\sin\omega_{\rm out}$ & $-0.259190$ & 0.000082  \\
$i_{\rm out}$ (deg) & 89.89368 & 0.00032 & $\Omega_{\rm out}$ (deg) & $-0.3743$ & 0.0027 \\
$h_1$ star 1, Kepler & 0.6862 & 0.0035 & $h_2$ star 1, Kepler &  0.652 & 0.046 \\
$h_1$ star 1  TESS & 0.547 & 0.035 & $h_2$ star 1, TESS &  0.30  & 0.18 \\
$h_1$ star 2, Kepler & 0.6183 & 0.0096 & $h_2$ star 2, Kepler &  0.149 & 0.036 \\
$h_1$ star 2  TESS & 0.954 & 0.045 & $h_2$ star 2, TESS &  0.50 & 0.32 \\
$h_1$ star 3, Kepler & 0.28 & 0.47 & $h_2$ star 3, Kepler &  0.10 & 0.33 \\
$h_1$ star 4, Kepler & 0.588 & 0.058 & $h_2$ star 4, Kepler &  0.35 & 0.20 \\
$s_0$ & 0.0079 & 0.0042  & $s_1$ & 0.0110 & 0.0039 \\
$s_2$ & 0.0079 & 0.0039 & $s_3$  & 0.0051 & 0.0038 \\ 
$c_0$ & 0.295 & 0.021  & $c_1$ & 0.008 & 0.010 \\
$c_2$ & 0.321 & 0.011 & $c_3$  & 0.236 & 0.015 \\ 
\bottomrule
\end{tabular}}
\end{adjustwidth}
\noindent{\footnotesize{$^1$ Reference time for  osculating Keplerian 
elements:  BJD 2,454,800.  $^2$ BJD $-$2,455,000.  $^3$ fixed.}}
\end{table}
\unskip

\begin{figure}[H]
\includegraphics[width=8.5 cm, angle=-90]{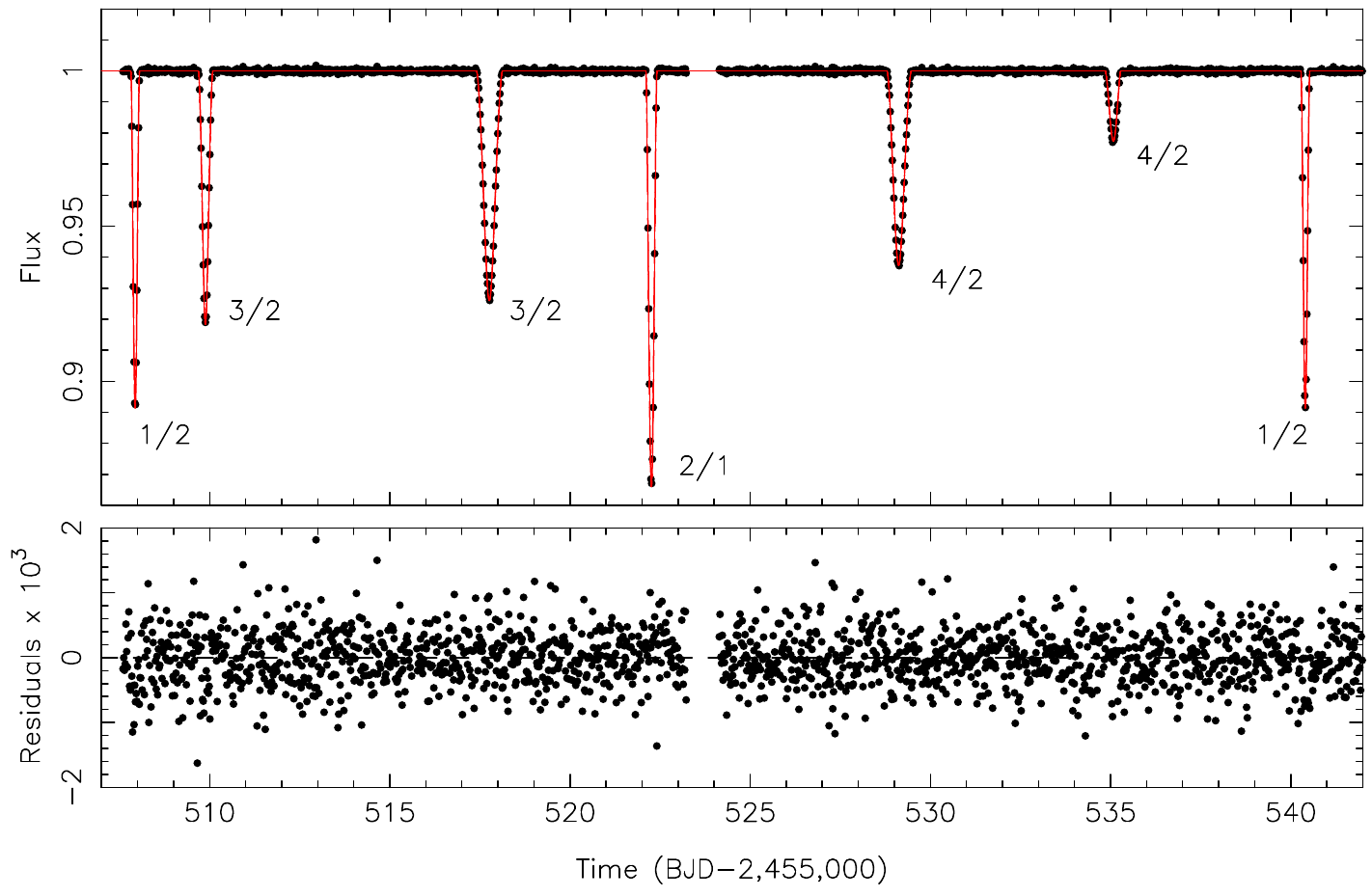}
\caption{\textbf{Top}: A close-up of the normalized {\em Kepler} light
curve of KIC 5255552 (points) with the best-fitting model (red
line).  Each eclipse is marked with a fraction of the form $M/N$,
where $M$ is the
\label{fig2}}
\end{figure}
\vspace{-12pt}
{\captionof*{figure}{label of the body in front
and $N$ is the label of the body in the back during the event.
For example, ''2/1'' means ``body \#2 is in front of body \#1'',
which, in this case, is a primary eclipse of the first binary.
The first two of the four  extra eclipses are caused by the primary in the
second (non-eclipsing)
binary passing in front of the secondary in the first (eclipsing) binary.
The other two extra eclipses are caused secondary in the second binary
passing in front of the secondary in the first binary.
\textbf{Bottom}: The residuals of the fit.}}

\begin{figure}[H]
\includegraphics[width=8.8 cm, angle=-90]{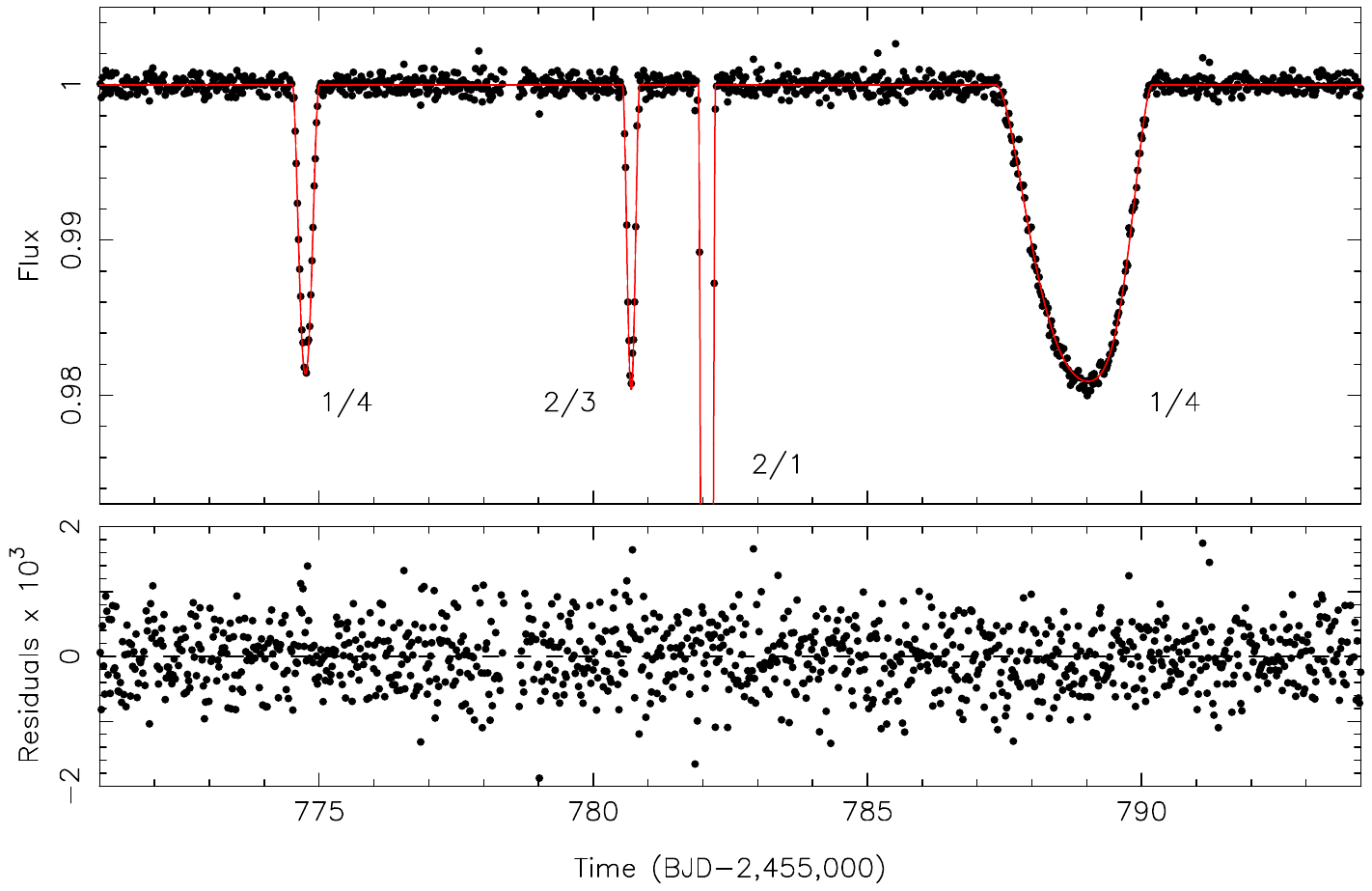}
\caption{Similar to Figure \protect{\ref{fig2}}, but~for the second
set of extra eclipses seen in KIC 5255552.  Here, the~first
(eclipsing) binary is passing in front of the second (non-eclipsing)
binary.
\label{fig3}}
\end{figure}
\unskip   

\begin{figure}[H]
\includegraphics[width=9.0 cm, angle=-90]{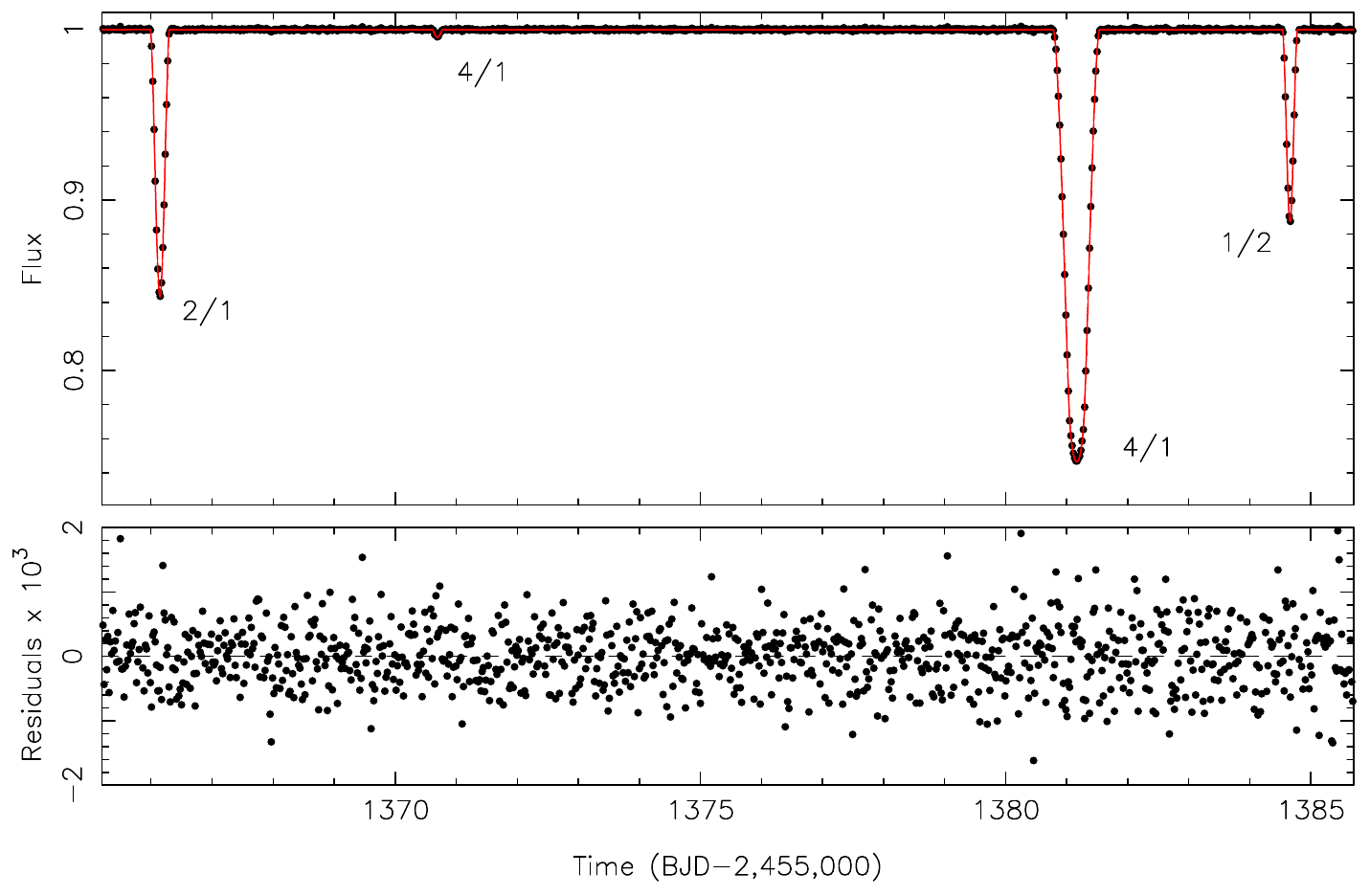}
\caption{Similar to Figure \protect{\ref{fig2}}, but~for the third set
of extra eclipses seen in KIC 5255552.  As was the case for Figure
\protect{\ref{fig2}}, the~second (non-eclipsing) binary is passing
in front of the first (eclipsing) binary.
\label{fig4}}
\end{figure}
\unskip   

\begin{table}[H]
\caption{Derived Parameters for KIC 5255552 and EPIC~2202.
\label{tab_5255deriv}}
\begin{adjustwidth}{-\extralength}{0cm}
\centering 
\setlength{\tabcolsep}{11mm} \resizebox{\linewidth}{!}{\begin{tabular}{lllll}
\toprule
 & \textbf{KIC 5255552} &  &  \textbf{EPIC 2202} & \\
\textbf{Parameter $^1$} & \textbf{Median} & $\boldsymbol{1\sigma}$  &  \textbf{Median} & \textbf{$\boldsymbol{1\sigma}$ Range} \\
\midrule
$M_1$ ($M_{\odot}$) & 0.950   & 0.018  & 0.542 & 0.502, 0.583 \\
$M_2$ ($M_{\odot}$) & 0.745   & 0.014  & 0.461 & 0.425, 0.497 \\
$M_3$ ($M_{\odot}$) & 0.483   & 0.010  & 0.380 & 0.345, 0.417 \\
$M_4$ ($M_{\odot}$) & 0.507   & 0.010  & 0.380 & 0.346, 0.417 \\
$R_1$ ($R_{\odot}$) & 0.9284  & 0.0063 & 0.462 & 0.448, 0.478 \\
$R_2$ ($R_{\odot}$) & 0.6891  & 0.0051 & 0.365 & 0.343, 0.408 \\
$R_3$ ($R_{\odot}$) & 0.4640  & 0.0036 & 0.400 & 0.382, 0.416 \\
$R_4$ ($R_{\odot}$) & 0.4749  & 0.0031 & 0.365 & 0.349, 0.382 \\
$a_{\rm bin,1}$ (AU) &  0.2374  & 0.0015  &  0.1098 & 0.1072, 0.1124 \\
$a_{\rm bin,2}$ (AU) &  0.2810  & 0.0018  &  0.1309 & 0.1278, 0.1340 \\
$a_3$         (AU) &  2.454   & 0.016  &  2.15 & 1.56, 3.13 \\
$e_{\rm bin,1}$      &  0.218246  & 0.000058  &  0.119 & 0.107, 0.133 \\
$e_{\rm bin,2}$      &  0.161644  & 0.000098  &  0.0700 & 0.0629, 0.0787 \\
$e_3$              &    0.406029  & 0.000051  &  0.45 & 0.25, 0.65 \\
$\omega_{\rm bin,1}$ (deg)     &  292.7457  & 0.0074  &  240 & 236, 243 \\
$\omega_{\rm bin,2}$ (deg)     &  168.066  & 0.058  &  240 & 235, 243 \\
$\omega_3$         (deg)     &  219.6691  & 0.0092  &   284 & 245, 310 \\
$i_{\rm bin,1}$ (deg)     &  88.9812  & 0.0034  &  89.85 & 89.64, 89.99 \\
$i_{\rm bin,2}$ (deg)     &  86.147  & 0.013  &  89.95 & 89.77, 90.22 \\
$i_3$         (deg)     &  89.89368  & 0.00032  &  85 & 80, 99 \\
$\Omega_{\rm bin,2}$         (deg)     &  $-1.054$  & 0.010  &  $ -22$ & $-69$, 7 \\
$\Omega_3$         (deg)     &  $-0.3743$  & 0.0027  &  137 & 18, 151 \\
$I_{\rm bin_1-bin_2}$         (deg)     &   3.024  & 0.016  &   27 & 9, 69 \\
$I_{\rm bin_1-3}$         (deg)     &   0.9862  & 0.0028  &   137 & 26, 150 \\
\bottomrule
\end{tabular}}
\end{adjustwidth}
\noindent{\footnotesize{$^1$ Reference times for osculating
Keplerian elements: BJD 2,454,800 (KIC 5255552); BJD 2,457,390 (EPIC 2202).}}
\end{table}
\unskip

\begin{figure}[H]
\includegraphics[width=10.5 cm, angle=-90]{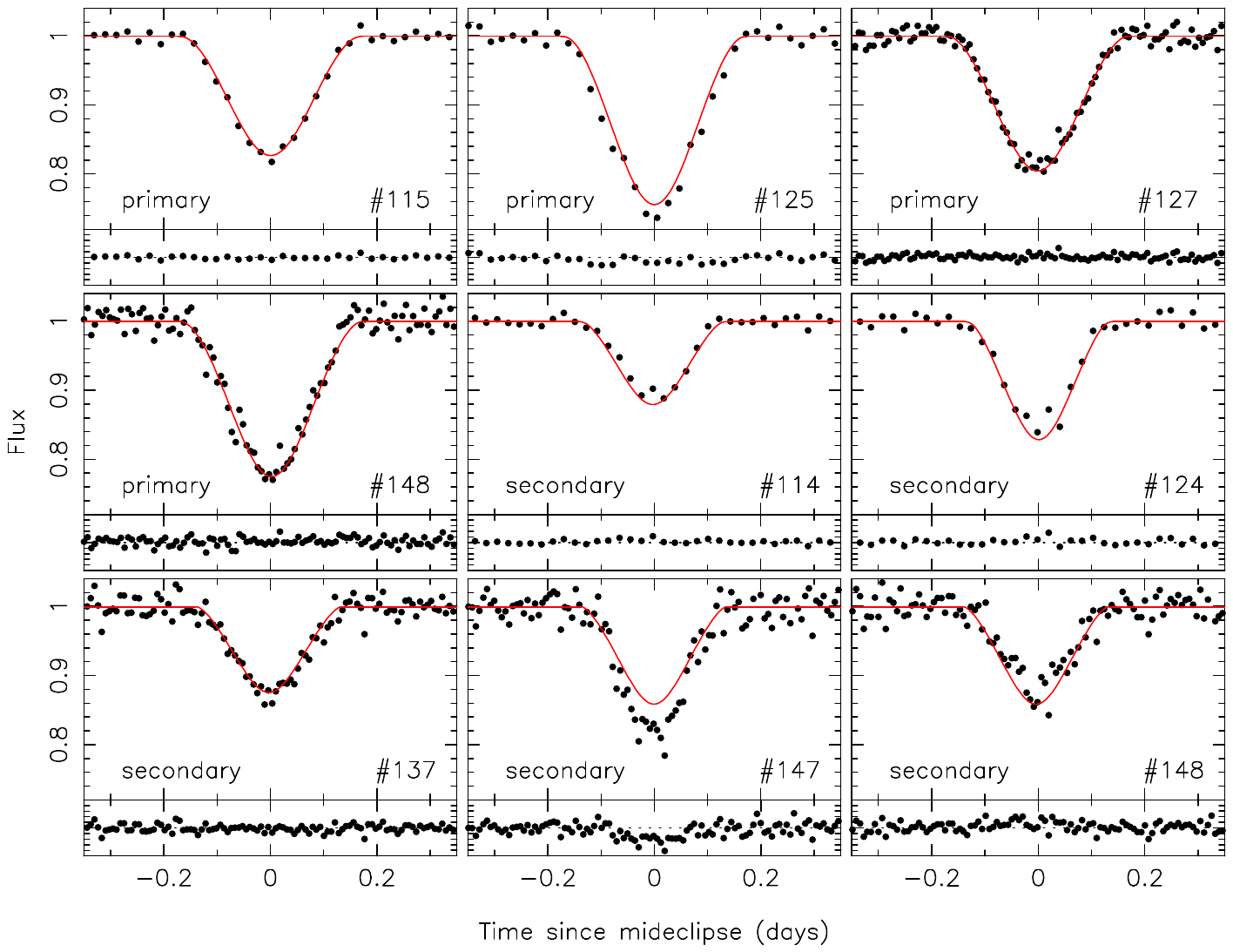}
\caption{{\em TESS} data for KID 5255552 with the best-fitting models.
See Table \protect\ref{5255_times} in the Appendix
\ref{appendix_times} for the eclipse times.  The~vertical scale on
the panels with the residuals is $\pm 0.1$.
\label{plotTESS}}
\end{figure}   
\unskip

\begin{figure}[H]
\includegraphics[width=9.0 cm, angle=-90]{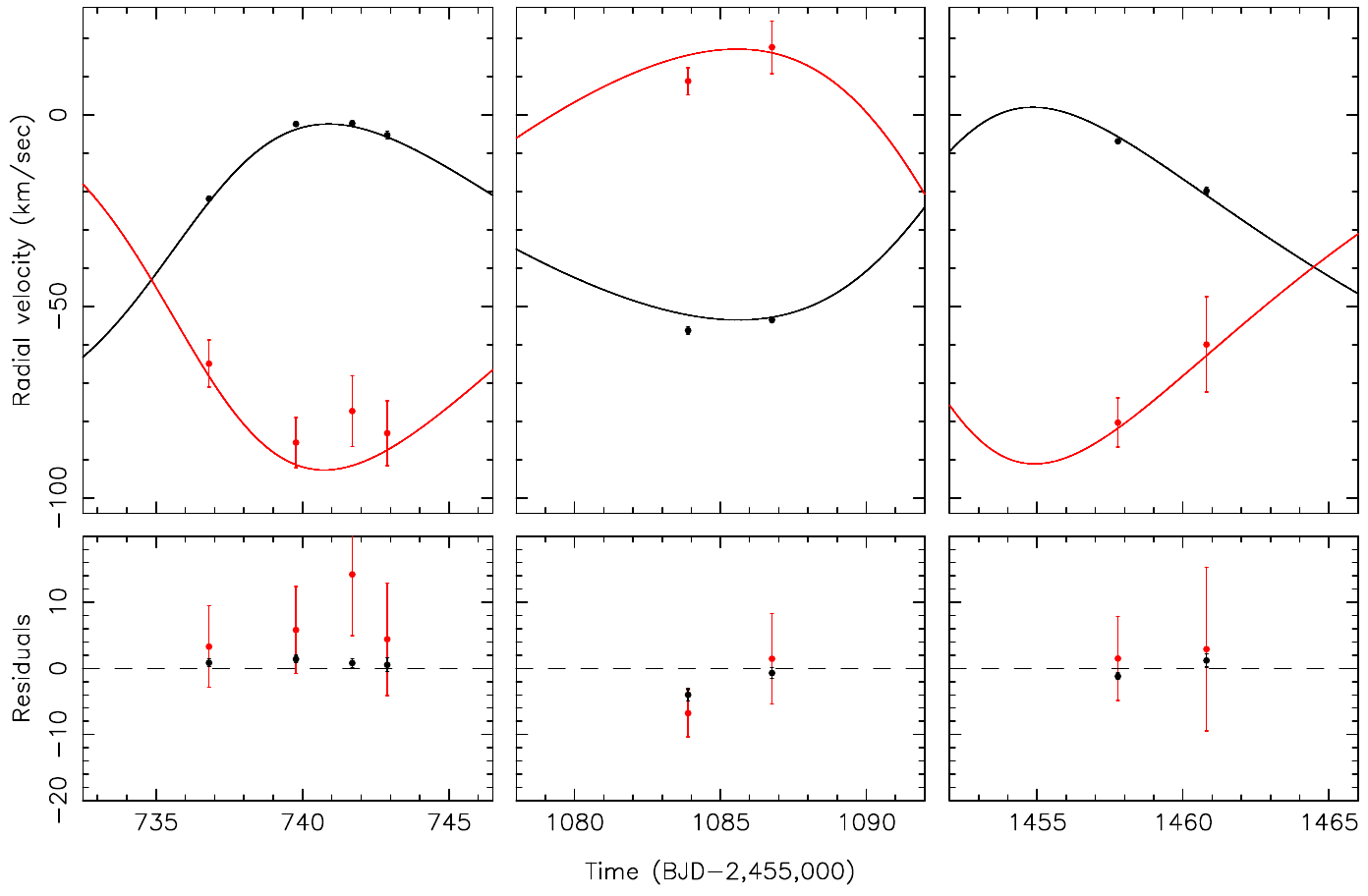}
\caption{\textbf{Top}: The radial velocity curve of the KIC 5255552
primary star (black points) with the best-fitting model (black line)
and the radial velocity curve of the secondary star star (red
points) with the best-fitting model (red line) \textbf{Bottom}: The
residuals of the fit.
\label{fig5}}
\end{figure}
\unskip   

\begin{figure}[H]
\includegraphics[width=9.5 cm, angle=-90]{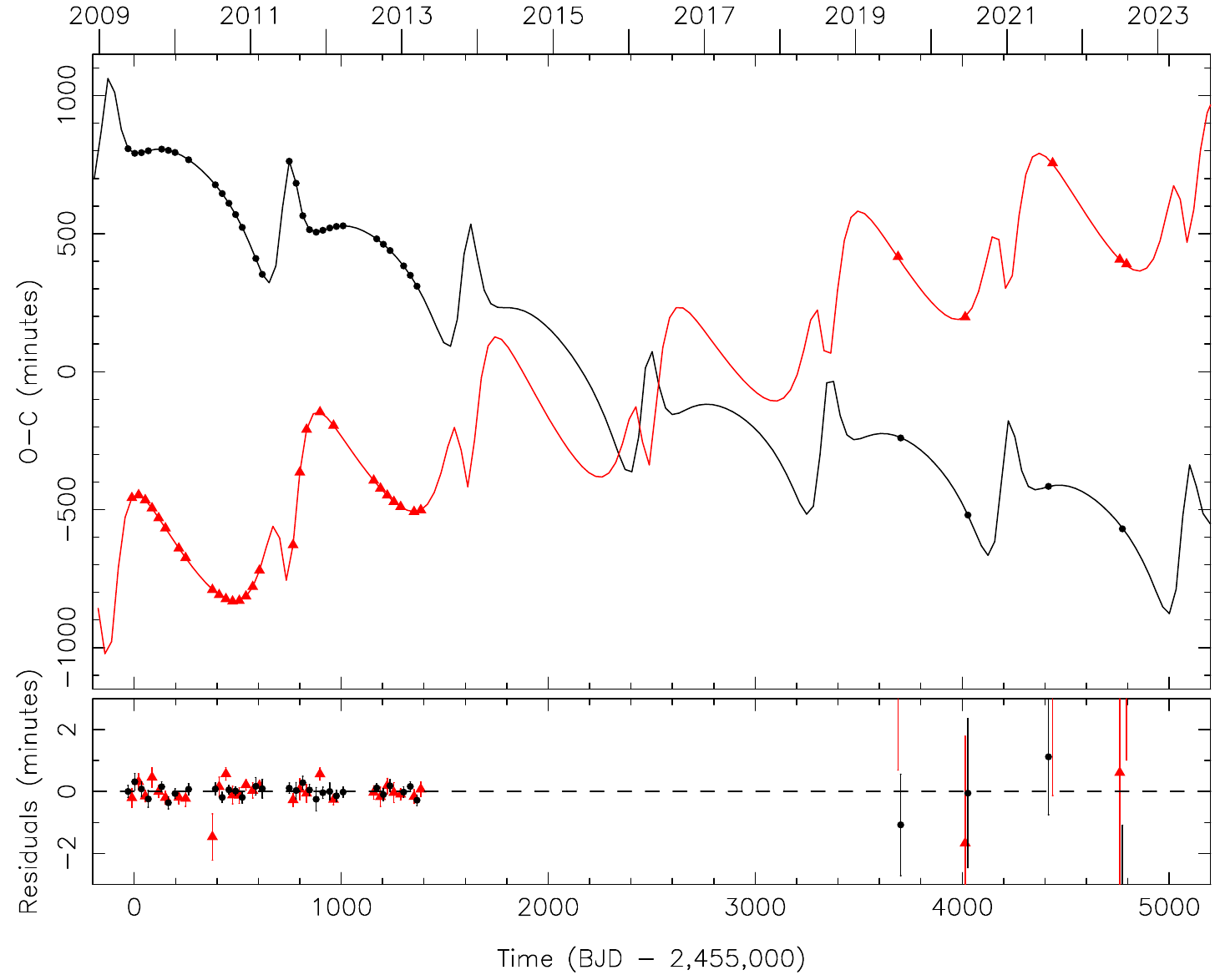}
\caption{\textbf{Top}: The Common-Period Observed Minus Computed
(CPOC) diagram for the KIC 5255552 primary eclipses (black points)
and secondary eclipses (red triangles), with~the respective
best-fitting models (black and red lines).  \textbf{Bottom}: The
residuals of the fit.
\label{fig6}}
\end{figure}
\unskip   

\subsection{EPIC~220204960}
\unskip

\subsubsection{Overview}

EPIC 220204960 (hereafter EPIC 2202) was observed for $\approx$$80$
days in Campaign 8 of the K2 mission~\cite{Howell_2014}.
Rappaport~et~al.\ \cite{Rappaport_2017} found that EPIC 2202 consists
of two EBs with periods of 13.27 days and 14.41 days.  They obtained
follow-up spectroscopic observations and measured radial velocities
for all four components.  The~two EBs individually show strong ETVs,
and~given all of this Rappaport~et~al.\ \cite{Rappaport_2017}
concluded that EPIC 2202 is a compact quadruple system in a
binary-binary configuration (a 2 + 2 similar to that of \mbox{KIC
  5255552}) with all 4 stars having masses near $0.4\,M_{\odot}$.
They performed a series of numerical simulations and an analytic
analysis of the outer orbit (which assumed roughly coplanar orbits)
and concluded the outer orbital period is very likely between 300 and
\mbox{500 days} with a possibility of being as long as two to four
years (e.g.,\ $\approx$$730$ to 1460 days).  Using the photometric and
radial velocity data provided by
Rappaport~et~al.\ \cite{Rappaport_2017}, we perform our own
independent photodynamical analysis with the goal of seeing what
constraints, if~any, we could place on the properties of the
outer~orbit.

\subsubsection{Photodynamical~Model}

Saul Rappaport kindly sent us the processed K2 light curve they used
in their paper.  We normalized the flux to 1.0 using a simple spline
fitting routine where data in the various eclipses were given zero
weight.  The radial velocity data were taken from
Rappaport~et~al.\ \cite{Rappaport_2017} (their Table~3).  In our
notation, the primary of ``binary \#1'' has the radial velocity
measurements denoted by ``Star A-1'' in their Table~3 and the
secondary of binary \#1 has the radial velocity measurements denoted
by ``Star A-2'' in their Table~3.  The primary and secondary of
``binary \#2'', respectively, have measurements from the columns
labeled ``Star B-1'' and ``Star B-2'' from that same table.  The
uncertainties on all the radial velocity measurements were taken to be
3 km s$^{-1}$, with~two exceptions.  First, the radial velocities of
all four stars are relatively close together from the observation on
day 2713.9 (in units of BJD $-$2,455,000), which indicates the
cross-correlation peaks were even more blended than usual.  Initial
fits showed evidence of ``peak pulling'' for both binaries
(see~\cite{Rucinski_1992}).  That is, the~star with the larger
redshift at that time had a negative residual, and the star with the
small redshift had a positive residual.  Thus, for~that day, the
uncertainties on the measurements for all four stars were set to 6 km
s$^{-1}$.  Likewise, the~radial velocities for binary \#2 from day
7698.0 showed similar evidence of peak pulling, so the uncertainties
on those measurements were set to 6 km s$^{-1}$.  Binary \#1 was near
the quadrature phase on that day, so the radial velocity measurements
for that binary appear to be~unaffected.
  
The photodynamical model has 42 free parameters.  The Keplerian
elements for binary \#1 and binary \#2 have the same parameterization
as they did for KIC 5255552.  For~the outer orbit, parameterization of
the period, time of conjunction, and~the eccentricity elements were
the the same as they were for KIC 5255552.  However, instead of
fitting for the outer orbit's inclination $i_{\rm out}$ and nodal
angle $\Omega_{\rm out}$, we fit for the sum and difference of those
quantities: $\Sigma_{{\rm out},i,\Omega}\equiv i_{\rm out}+\Omega_{\rm
  out}$, $\Delta_{{\rm out},i,\Omega}\equiv i_{\rm out}-\Omega_{\rm
  out}$.  For each binary, the~masses were parameterized by the sum of
the masses of each component and the mass ratio (e.g.,\ $\Sigma_{\rm
  bin_1,mass} \equiv M_1+M_2, Q_{\rm bin,1}\equiv M_2/M_1$,
with~similar quantities for binary \#2).  Likewise, the~stellar radii
for each binary were parameterized by their sum and ratio, as~in
\mbox{$\Sigma_{\rm bin_1,rad} \equiv R_1+R_2, \rho_{\rm bin,1}\equiv
  R_1/R_2$} for binary \#1 and $\Sigma_{\rm bin_2,rad} \equiv R_3+R_4,
\rho_{\rm bin,2}\equiv R_3/R_4$ for binary \#2.  The stellar
temperatures are parameterized by $T_1$, $T_2$, $T_3$, and~$T_4$.
Each star has a pair of limb-darkening coefficients $h_1$ and $h_2$
for the Power-2 limb-darkening law with the Maxted transformations.
The source is heavily blended with a nearby brighter star and~a single
contamination parameter $c_0$ was used for the entire light curve.
Finally, extra force terms for GR and tidal effects were used, and
consequently, each star had an apsidal constant as a free parameter:
$k_{2,1}$, $k_{2,2}$, $k_{2,3}$, and~$k_{2,4}$.

\textls[25]{We again used a brute-force method to explore parameter
space.  Both binaries eclipse, so consequently, we have reasonably
good constraints on the orbital periods, the~times of conjunction,
the eccentricity parameters, and~the inclinations.  The~nodal angle
of binary \#1 was fixed at zero, and~the nodal angle of binary \#2
was given a prior range of $-90^{\circ}\le\Omega_{\rm out}\le
90^{\circ}$.  On the other hand, there are very few constraints on
the outer orbit that are immediately obvious.  The~outer period
probably cannot be less than \mbox{$\approx$$100$ days} or so or
otherwise the system would be unstable.  Since there are strong ETVs
seen in both binaries, the~outer period cannot be excessively long.
Given all of this we adopt fairly large prior ranges for the
parameters of the outer orbit: \mbox{$1700 \le T_{\rm out} \le
4400$} in units of BJD $-$2,455,000, $100\le P_{\rm out}\le 3000$
in units of days, \mbox{$-0.61 \le e_{\rm out}\cos\omega_{\rm out}
\le 0.61$}, \mbox{$-0.61 \le e_{\rm out}\sin\omega_{\rm out} \le
0.61$}, $30\le\Sigma_{{\rm out},i,\Omega}\le 290$ in units of
degrees, and \mbox{$-180\le\Delta_{{\rm out},i,\Omega}\le 180$} in
units of degrees.  The {\tt gennestELC} code was run 22 separate
times, each with different initial seeds for a random number
generator.  After~convergence, the~DE-MCMC code was run for 100,000
generations, using the best model from {\tt gennestELC} as the basis
for generating initial models.  After a burn-in period of 25,000
generations, posterior samples were taken every 12,500th generation.
The~separate posterior samples were combined, yielding sample sizes
of 20,608 for the fitting and derived parameters. }

\begin{figure}[H]
\includegraphics[width=8.8 cm, angle=-90]{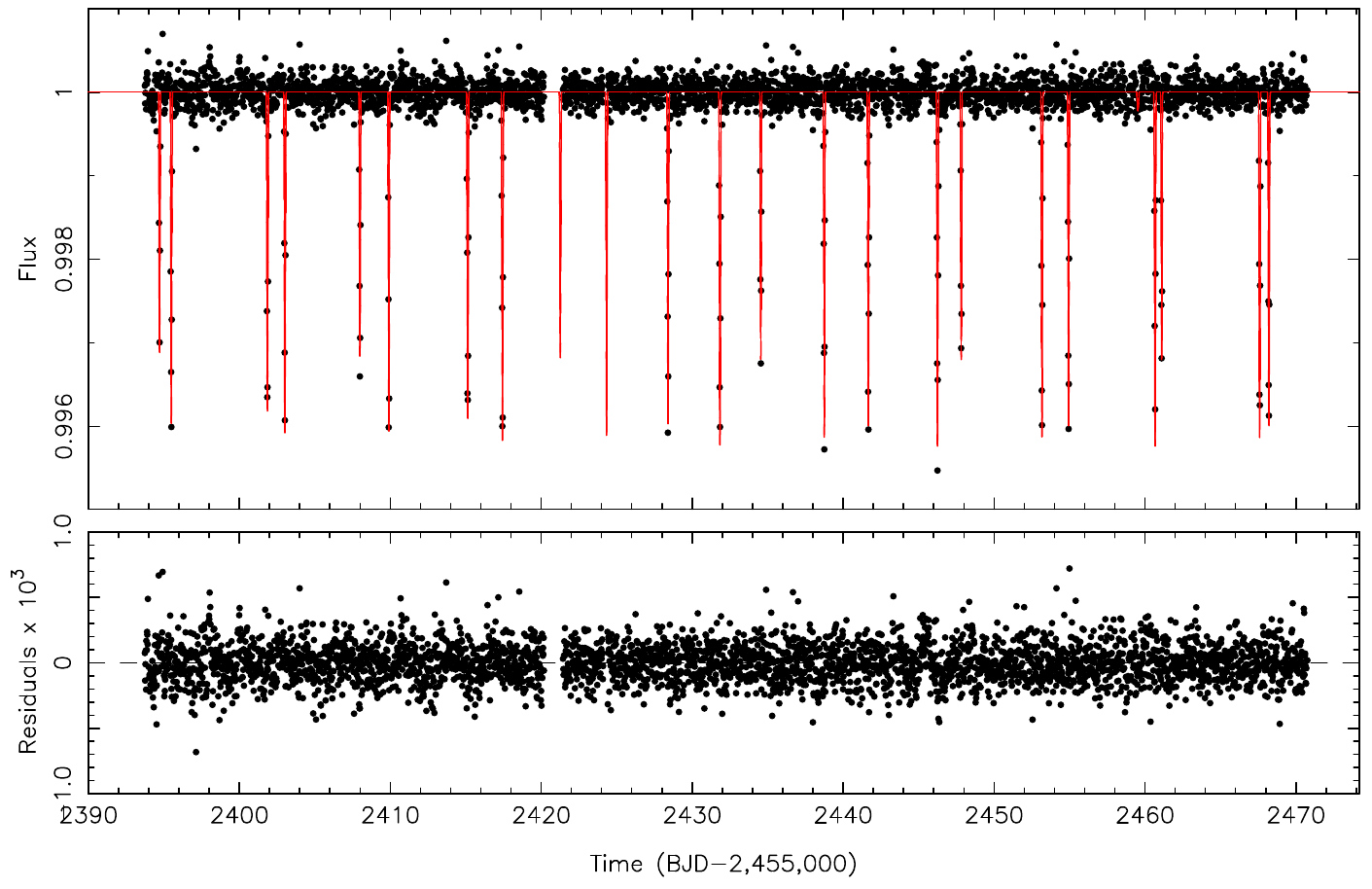}
\caption{\textbf{Top}: The normalized K2 light curve of EPIC 2202
(points) with the best-fitting model (red line).  
\label{epicfig01}}
\end{figure}

\begin{figure}[H]
\includegraphics[width=9.0 cm, angle=-0]{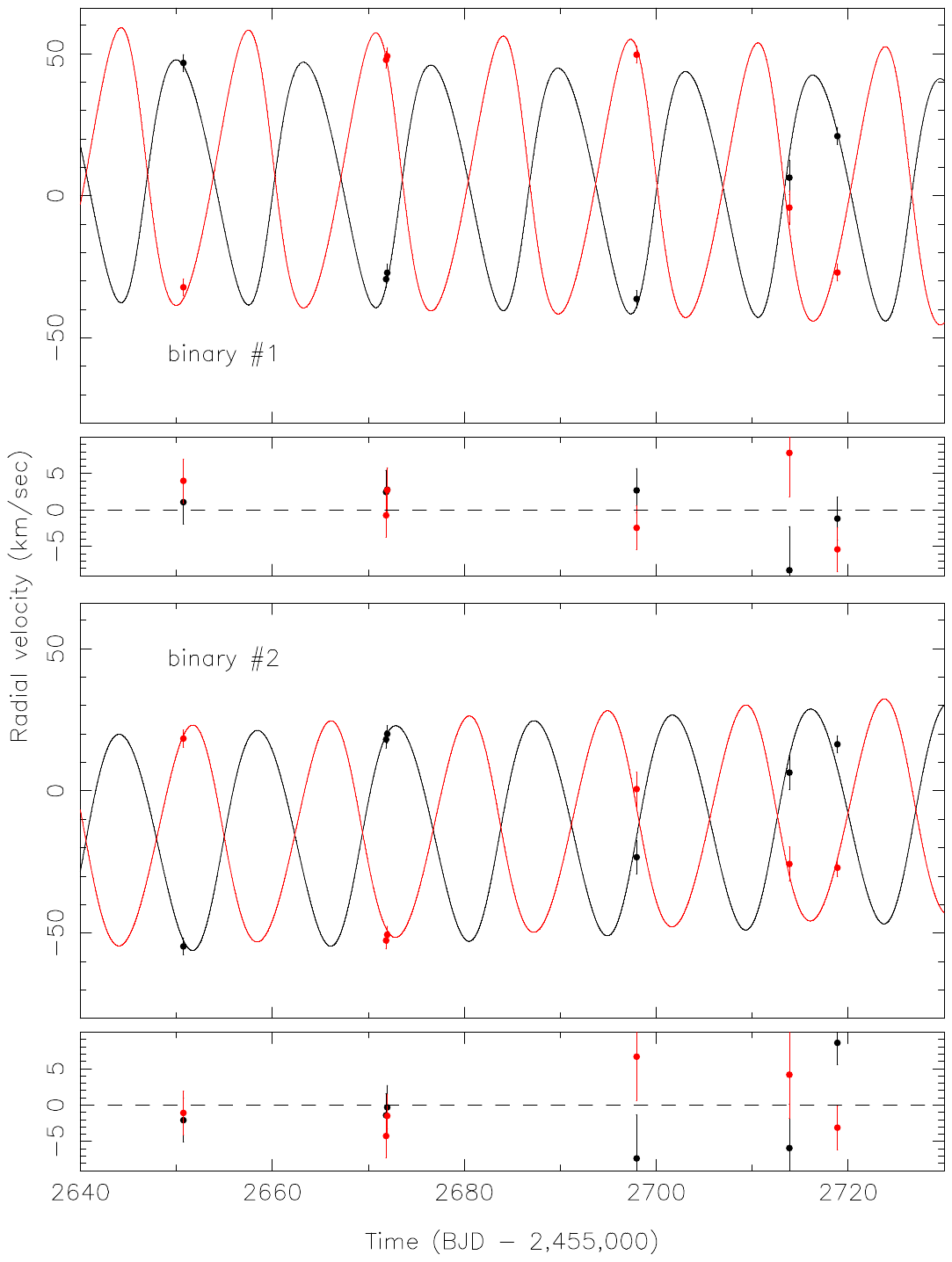}
\caption{\textbf{Top}: The radial velocity curve of the EPIC 2202
binary \#1 primary star (black points) with the best-fitting model
(black line) and the radial velocity curve of the binary \#1
secondary star (red points) with the best-fitting model (red line).
Second: The residuals of the fit.  Third: Similar to the Top above,
but~for the EPIC 2202 binary \#2.  \textbf{Bottom}: The residuals of
the fit.
\label{epicfig02}}
\end{figure}

\begin{figure}[H]
\includegraphics[width=9.0 cm, angle=-90]{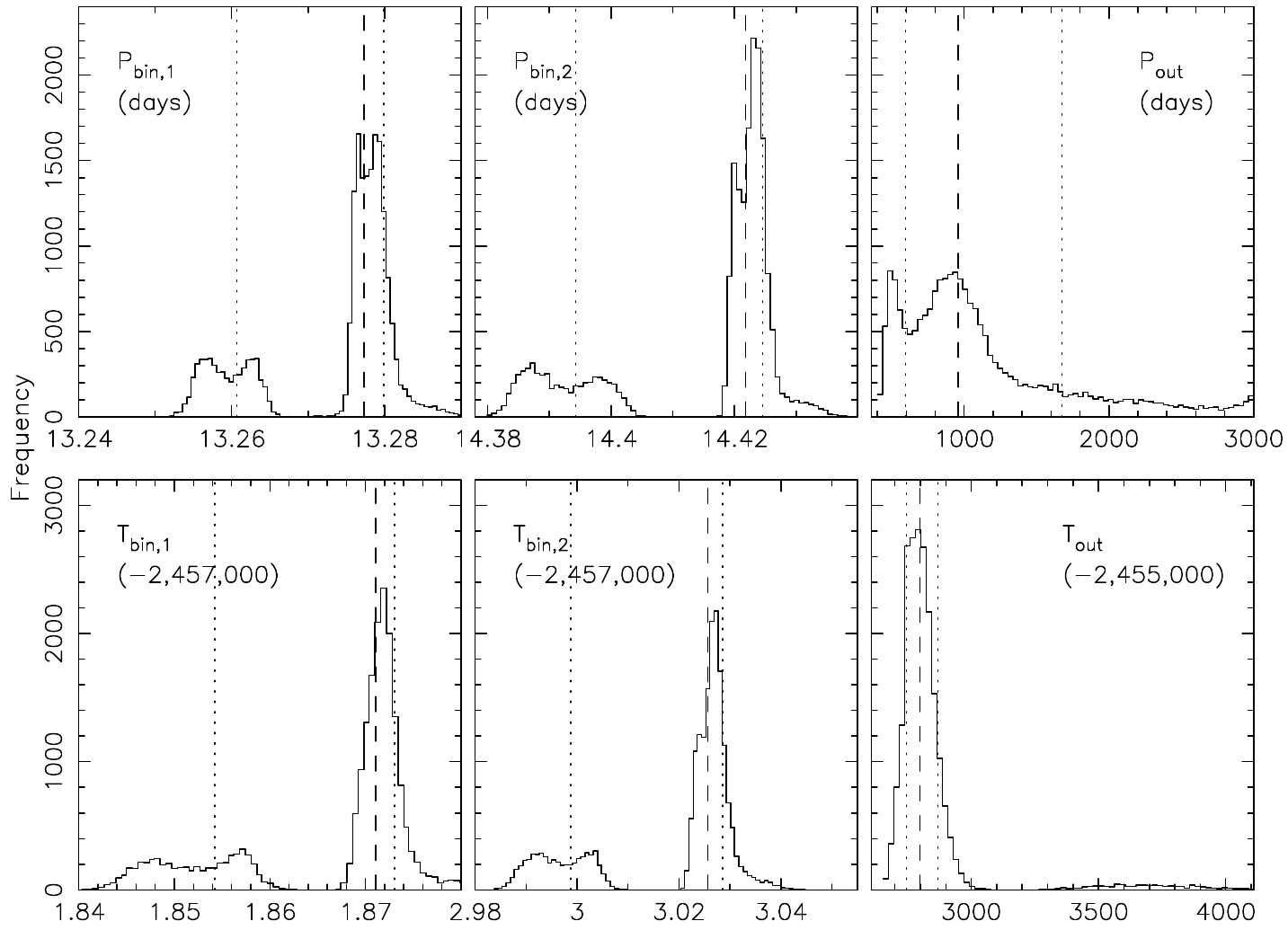}
\caption{\textbf{Top}: Posterior samples for EPIC 2202 showing,
from~left to right, the~period of binary \#1, the period of binary
\#2, and~the period of the outer orbit.  \textbf{Bottom}: Posterior
samples for EPIC 2202 showing, from~left to right, the~time of
primary conjunction of binary \#1, the time of conjunction of binary
\#2, and~the time of conjunction of the outer orbit.  The units of
the times are BJD minus the offset given in each panel.  In each
panel, the vertical dashed lines indicate the sample median, the two
vertical dotted lines delineate the central region that contains
68\% of the sample.
\label{plotpost_epic02}}
\end{figure}

\begin{figure}[H]
\includegraphics[width=9.5 cm, angle=-90]{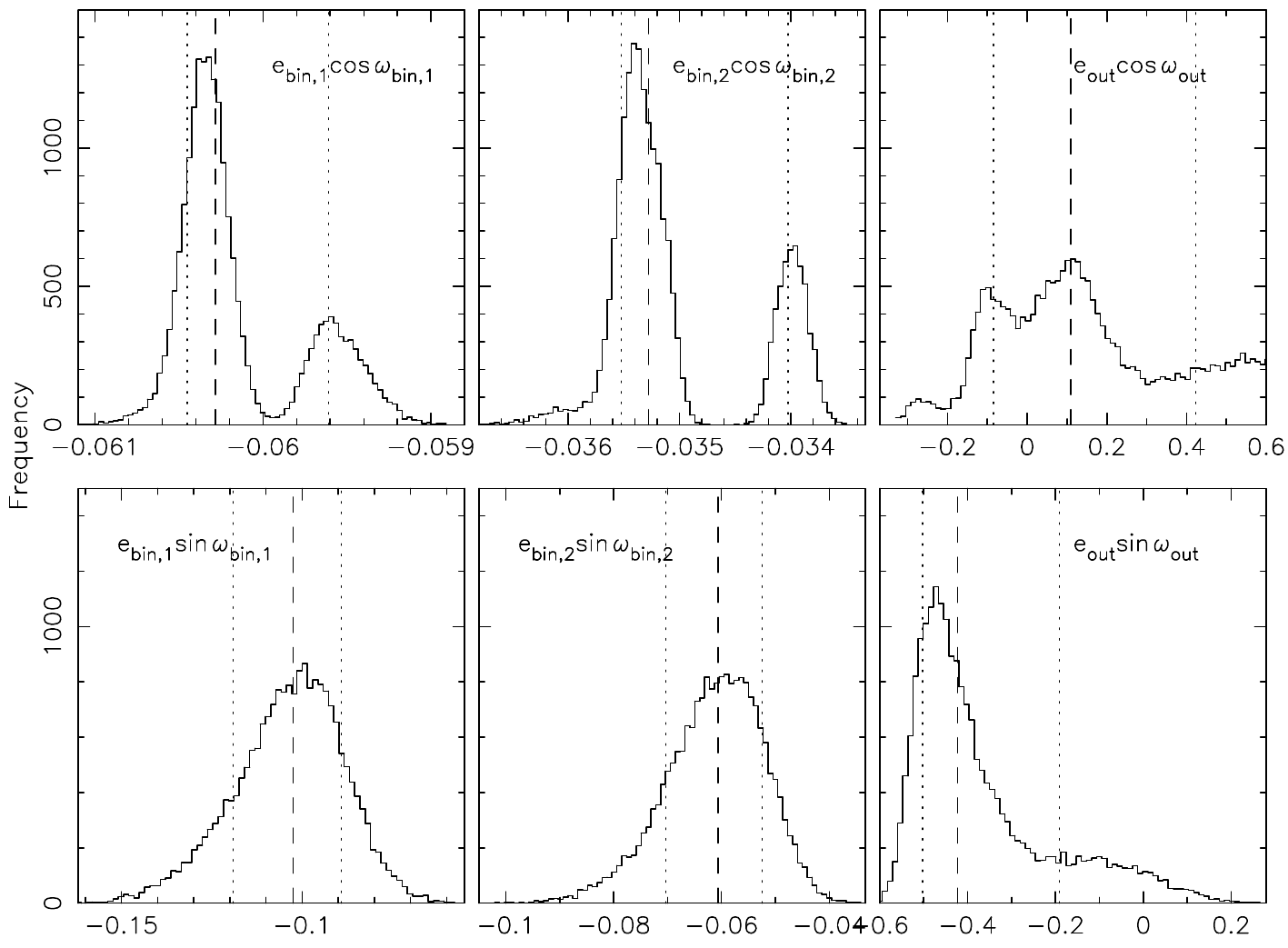}
\caption{Similar to Figure \protect{\ref{plotpost_epic02}},
but~$e_{\rm bin,1}\cos\omega_{\rm bin,1}$, $e_{\rm
  bin,2}\cos\omega_{\rm bin,2}$, and~$e_{\rm out}\cos\omega_{\rm
  out}$ (\textbf{top}) and $e_{\rm bin,1}\sin\omega_{\rm bin,1}$,
$e_{\rm bin,2}\sin\omega_{\rm bin,2}$, and~$e_{\rm
  out}\sin\omega_{\rm out}$ (\textbf{bottom}).
\label{plotpost_epic03}}
\end{figure}

\begin{figure}[H]
\includegraphics[width=9.5 cm, angle=-90]{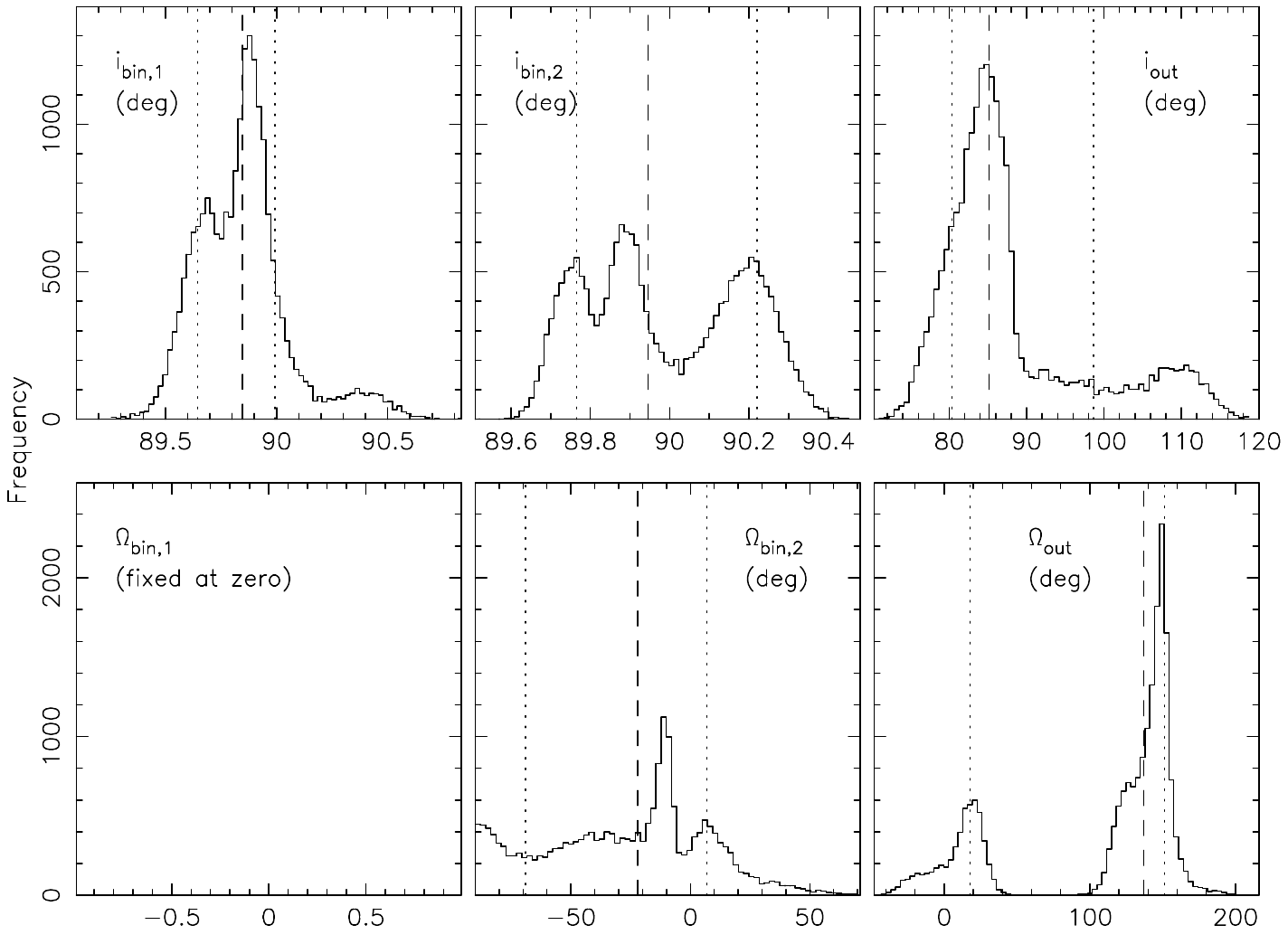}
\caption{Similar to Figure \protect{\ref{plotpost_epic02}},
but~$i_{\rm bin,1}$, $i_{\rm bin,2}$, and~$i_{\rm out}$
(\textbf{top}) and $\Omega_{\rm bin,1}$ (fixed at zero),
$\Omega_{\rm bin,2}$, and~$\Omega_{\rm out}$ (\textbf{bottom}).
\label{plotpost_epic04}}
\end{figure}

\begin{figure}[H]
\includegraphics[width=9.5 cm, angle=-90]{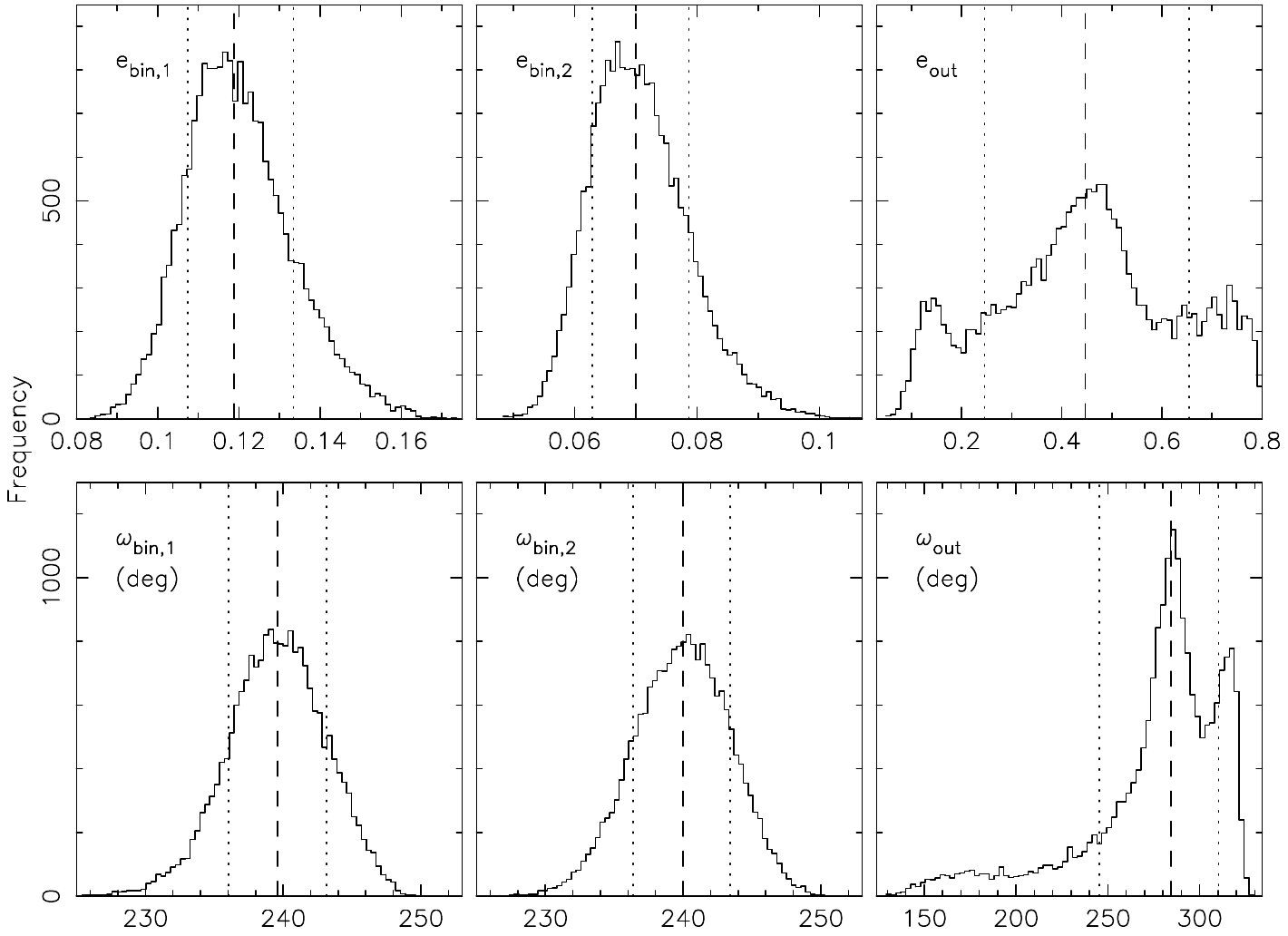}
\caption{Similar to Figure \protect{\ref{plotpost_epic02}},
but~$e_{\rm bin,1}$, $e_{\rm bin,2}$, and~$e_{\rm out}$
(\textbf{top}) and $\omega_{\rm bin,1}$, $\omega_{\rm bin,2}$,
and~$\omega_{\rm out}$ (\textbf{bottom}).
\label{plotpost_epic05}}
\end{figure}

\begin{figure}[H]
\includegraphics[width=9.5 cm, angle=-90]{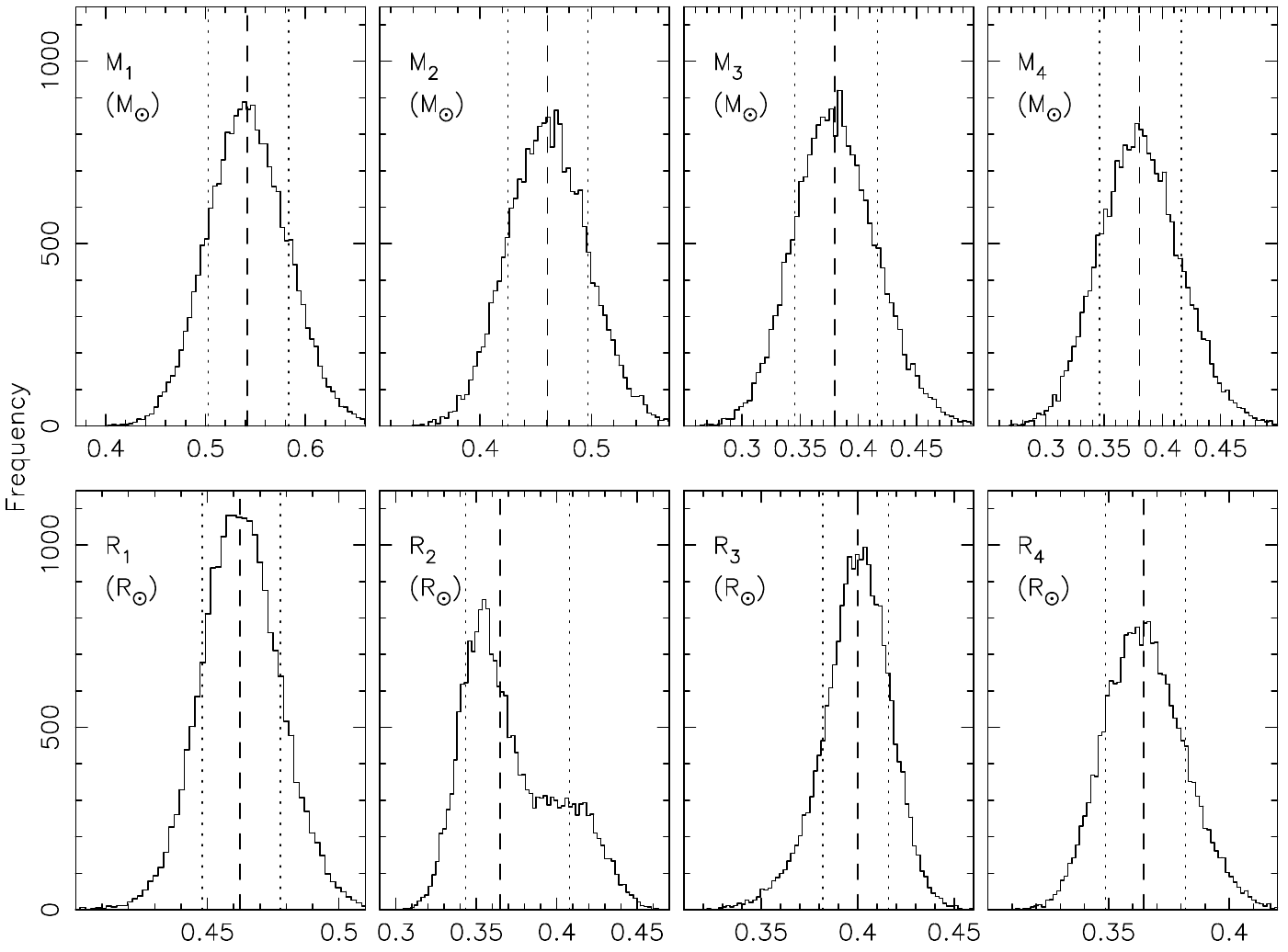}
\caption{Similar to Figure \protect{\ref{plotpost_epic02}}, but~for
the stellar masses (\textbf{top}) and stellar radii
(\textbf{bottom}).
\label{plotpost_epic01}}
\end{figure}

\begin{figure}[H]
\includegraphics[width=7.0 cm, angle=-90]{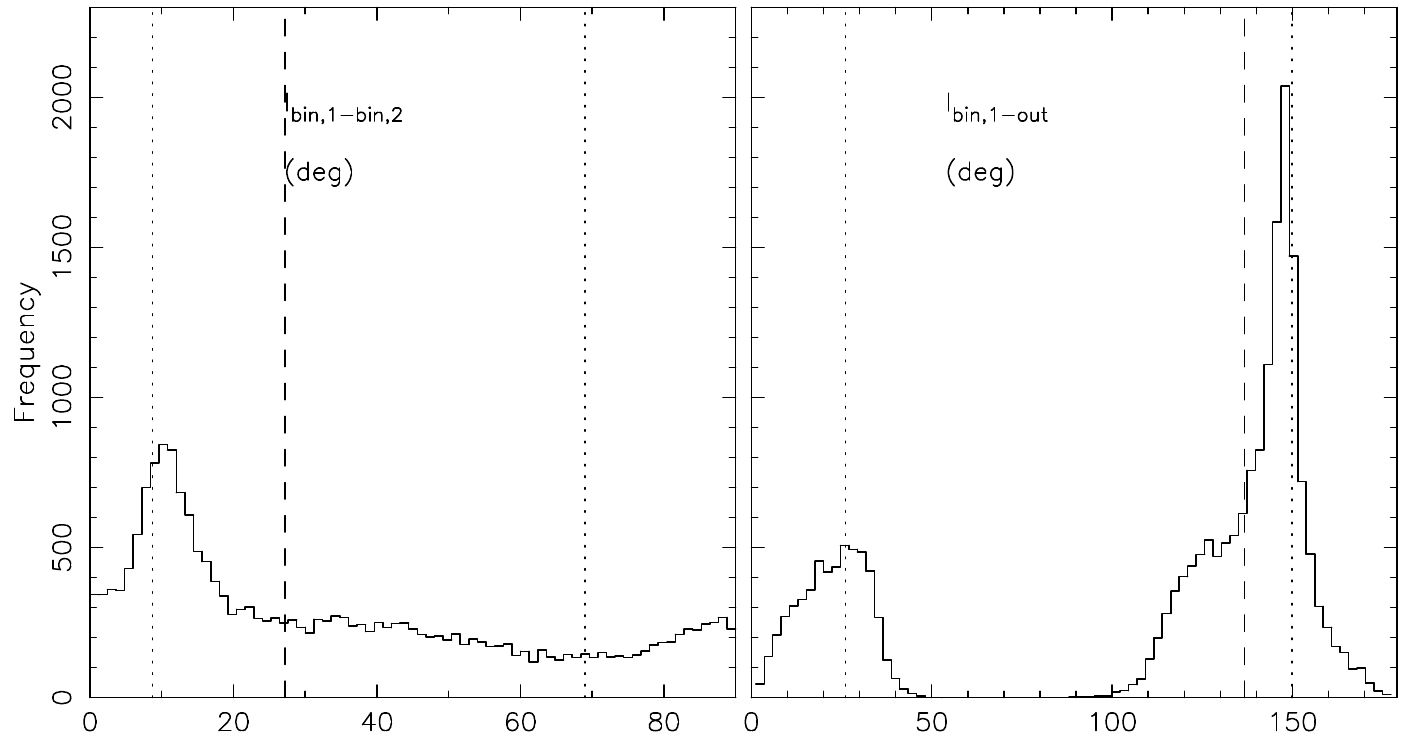}
\caption{Similar to Figure \protect{\ref{plotpost_epic02}}, but~for
the mutual inclination between the two binary orbital planes
(\textbf{left}) and for the mutual inclination between the orbital
plane of binary \#1 and the orbital plane of the outer orbit
(\textbf{right}).
\label{plotpost_epic06}}
\end{figure}

The best-fitting model light curve is shown in Figure~\ref{epicfig01}
and the best-fitting model radial velocity curves are shown in
Figure~\ref{epicfig02}.  The initial Keplerian and Cartesian initial
conditions for the best-fitting model are given in
Table~\ref{EPIC01_init} in the Appendix \ref{appc}.  Many of the
posterior distributions are complex, often with two or more modes.
Consequently, when reporting values of the fitted and derived
parameters (Tables \ref{tab_5255deriv} and \ref{tab_epicfit},
respectively) we quote the sample median ${\cal P}_{\rm med}$ (where
${\cal P}$ is a generic parameter) and a range ${\cal P}_{16\%}$ to
${\cal P}_{84\%}$, where 16\% of the sample is smaller than ${\cal
  P}_{16\%}$ and 16\% of the sample is larger than ${\cal P}_{84\%}$.
The posterior distributions for some of the key orbital elements are
shown in Figures~\ref{plotpost_epic02}--\ref{plotpost_epic04}.  The
posterior distributions for some derived quantities are shown in
Figure~\ref{plotpost_epic05} (the eccentricity $e$ and the argument of
periastron $\omega$), Figure~\ref{plotpost_epic01} (the stellar masses
and radii), and~Figure~\ref{plotpost_epic06} (the mutual inclination
between the orbital planes of the two binaries and the mutual
inclination between binary \#1 and the outer orbit).

\begin{table}[H]
\caption{Fitting Parameters for EPIC~2202.\label{tab_epicfit}}
\begin{adjustwidth}{-\extralength}{0cm}
\centering 
\setlength{\tabcolsep}{6mm} \resizebox{\linewidth}{!}{\begin{tabular}{llllll}
\toprule
\textbf{Parameter $^1$} & \textbf{Median} & \textbf{$\boldsymbol{1\sigma}$ Range}  & \textbf{Parameter $^1$} & \textbf{Median} & \textbf{$\boldsymbol{1\sigma}$ Range} \\
\midrule
$P_{\rm bin,1}$ (days) & 13.277 & 13.261, 13.280     & $T_{\rm bin,1}$ & $2401.871~^2$ & 2401.854, 2401.873 \\
$e_{\rm bin,1}\cos\omega_{\rm bin,1}$ & $-0.06028$ &  $-0.06045$, $-0.05961$ & 
                  $e_{\rm bin,1}\sin\omega_{\rm bin,1}$ & $-0.102$ & $-0.119$,  $-0.089$  \\
$i_{\rm bin,1}$ (deg) & 89.85 & 89.64, 89.99 & $\Omega_{\rm bin,1}$ (deg) & $0.0~^3$ & \ldots \\
$\Sigma_{\rm bin_1,mass}$ ($M_{\odot}$) & 1.002 & 0.932, 1.076 & $Q_{\rm bin,1}$ &   0.850 & 0.805, 0.897  \\
$\Sigma_{\rm bin_1,rad}$  ($R_{\odot}$) & 0.832  &  0.801, 0.870 & $\rho_{\rm bin,1}$ & 1.272 & 1.124, 1.351 \\
$P_{\rm bin,2}$ (days) & 14.4218 & 14.3943, 14.4246   & $T_{\rm bin,2}$ & $2403.0257~^2$ & 2402.9988, 2403.0286 \\
$e_{\rm bin,2}\cos\omega_{\rm bin,2}$ & $-0.0353$ &  $-0.0355$, $-0.0340$ & 
                  $e_{\rm bin,2}\sin\omega_{\rm bin,2}$ & $-0.0606$ & $-0.0703$, $-0.0524$  \\
$i_{\rm bin,2}$ (deg) & 89.95 &  89.77,  90.22 & $\Omega_{\rm bin,2}$ (deg) & $-22.0$ & $-69.0$, 6.8 \\
$\Sigma_{\rm bin_2,mass}$  & 0.761 & 0.697, 0.828 & $Q_{\rm bin,2}$ &  1.002 & 0.931, 1.076  \\
$\Sigma_{\rm bin_2,rad}$  & 0.764  & 0.739, 0.789 & $\rho_{\rm bin_2}$ & 1.105 & 1.033, 1.157 \\
$P_{\rm out}$ (days) & 957 & 595, 1674     & $T_{\rm out}$ & $2799~^2$ & 2745, 2868 \\
$e_{\rm out}\cos\omega_{\rm out}$ & $0.10$ & $-0.08$,  0.42 & 
                  $e_{\rm out}\sin\omega_{\rm out}$ & $-0.42$ & $-0.50$,  $-0.19$  \\
$\Sigma_{{\rm out},i,\Omega}$ (deg) & 223 & 123, 234 & $\Delta_{{\rm out},i,\Omega}$ (deg) & $-51$ & $-70$, 88 \\
$T_1$ (K)    & 3602 & 3509, 3667 & $T_2$ (K) & 3428  & 3340, 3545    \\
$T_3$ (K)    & 3592 & 3496, 3662 & $T_4$ (K) & 3532  & 3435, 3618  \\ 
$h_1$ star 1 & 0.52 & 0.32, 0.74 & $h_2$ star 1  & 0.24 & 0.09, 0.50 \\
$h_1$ star 2 & 0.73 & 0.37, 0.92 & $h_2$ star 2  & 0.33 & 0.07, 0.70 \\
$h_1$ star 3 & 0.76 & 0.61, 0.90 & $h_2$ star 3  & 0.49 & 0.25, 0.70 \\
$h_1$ star 4 & 0.91 & 0.81, 0.97 & $h_2$ star 4  & 0.50 & 0.17, 0.79 \\
$k_{2,1}$  & 0.10 & 0.03, 0.16   & $k_{2,2}$    & 0.10 & 0.04, 0.17 \\
$k_{2,3}$  & 0.10 & 0.03, 0.16   & $k_{2,3}$    & 0.10 & 0.04, 0.17 \\
$c_0$ & 0.9815 & 0.9807, 0.9826  &   &   &   \\
\bottomrule
\end{tabular}}
\end{adjustwidth}
\noindent{\footnotesize{$^1$ Reference time for  osculating Keplerian 
elements:  BJD 2,457,390 $^2$ BJD $-$2,455,000 $^3$ fixed.}}
\end{table}
\unskip

\unskip   

\unskip
\unskip   

\unskip   

\unskip   

\unskip   

\unskip   

\unskip   

\section{Discussion}\label{DiscussionSec}
\unskip

\subsection{Stellar Parameters and Age~Constraints}

\begin{figure}[H]
\includegraphics[width=10 cm, angle=-90]{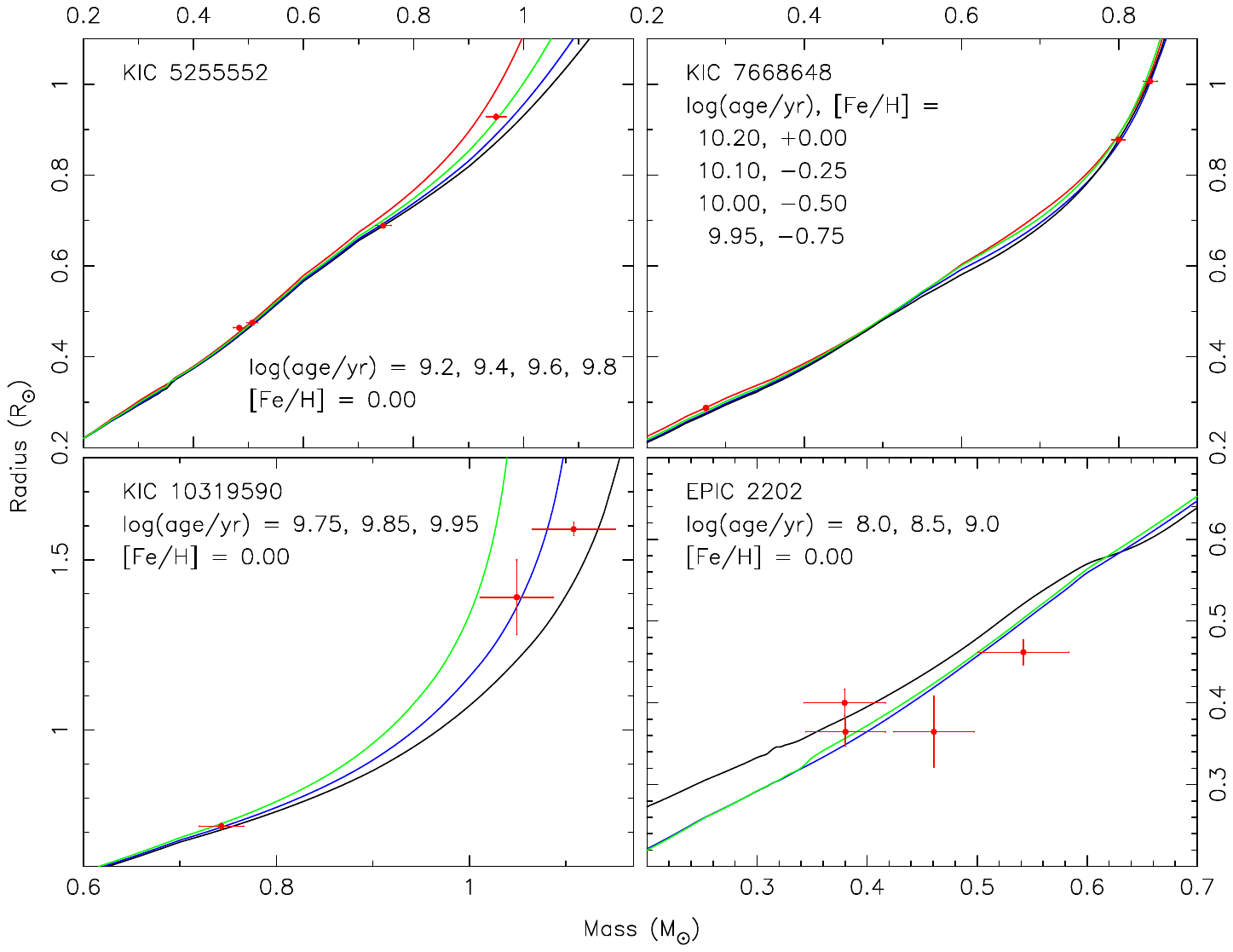}
\caption{MIST isochrone models in the mass-radius plane for KIC
5255552 (\textbf{upper left}), KIC 7668648 (\textbf{upper right}),
KIC 10319590 (\textbf{lower left}) and EPIC 2202 (\textbf{lower
right}).  For each panel, the observed masses and radii and their
uncertainties for each star are indicated with the red crosses, and
the~isochrone ages are indicated by the various colored lines,
with~black being the youngest age, then blue, then green, and~then
finally red being the oldest age.  The solar metallicity isochrone
($[{\rm Fe}/{\rm H}]=0$) with an age of $\log(\tau/{\rm yr})=9.6$
($\approx$$4$ Gyr) provides a close match for all four stars in KIC
5255552.  KIC 7668648 appears to be old ($\tau\approx 10$ Gyr) and
somewhat metal poor, where the metallicities greater than $[{\rm
Fe}/{\rm H}] - 0.5$ give ages older than the galaxy.  Solar
metallicity isochrones with ages of $\log(\tau/{\rm yr})=9.75$
($\approx$$5.6$ Gyr) and $\log(\tau/{\rm yr})=9.85$ ($\approx$$7.1$
Gyr) provide plausible matches in the mass-radius plane for KIC
10319590.  The uncertainties on the stellar masses and radii for
EPIC 2202 are too large to give meaningful constraints.
\label{plotiso01}}
\end{figure}

We can place the stars in the four systems studied herein in the
mass-radius plane and see what, if~any, constraints can be placed on
their ages (hereafter denoted as $\tau$).  For~this analysis, we use
the MIST models ~\cite{Choi_2016,Dotter_2016}.  For~a given age (where
$\log(\tau/{\rm yr})$ ranges from 5.0 to 10.3 in 0.05 dex steps) and
metallicity (parameterized by $[{\rm Fe}/{\rm H}]\equiv
\log_{10}(N_{\rm Fe}/N_H)_* -\log_{10}(N_{\rm Fe}/N_H)_{\odot}$, which
ranges from $-4.00$ to $-2.00$ in 0.50 dex steps and from $-2.00$ to
$+0.50$ in 0.25 dex steps), various stellar parameters are computed as
a function of the stellar mass.  In this way, we can plot various
model curves in the mass-radius plane, and assuming the stars in a
given system have the same age and that no mass exchange has happened,
the~measured masses and radii of all the stars should be consistent
with a single isochrone model in the ideal case.  In KIC 5255552, all
four stars have mass measurements with $\approx$$2\%$ uncertainties
and radius measurements with uncertainties $<1\%$.  All three stars in
KIC 7668648 have measured masses with uncertainties of $\approx$$1\%$
and measured radii with formal uncertainties of $\approx$$0.3\%$.
Thus, taking these measurements and their uncertainties at face value,
the constraints on the ages could potentially be relatively tight.
Figure~\ref{plotiso01} shows the mass-radius plane for these two
systems (top panels).  A~solar metallicity isochrone (e.g.,\ $[{\rm
    Fe}/{\rm H}]=0$) with $\log(\tau/{\rm yr})=9.6$ ($\approx$$4$ Gyr)
provides a decent match to all four stars in KIC 5255552.  For KIC
7668648, a~solar metallicity isochrone with $\log(\tau/{\rm yr})=10.2$
($\approx$$16$ Gyr) provides a good fit.  However, for~obvious
reasons, this isochrone is not physically realistic.  Likewise,
an~isochrone with a metallicity of $[{\rm Fe}/{\rm H}]-0.25$ and an
age of $\log(\tau/{\rm yr})=10.1$ ($\approx$$12.6$ Gyr) provides a
good match in the mass-radius plane, but~as before, the age is
unrealistic.  Isochrones with $\tau=10$ Gyr and a metallicity of
$[{\rm Fe}/{\rm H}]-0.5$ and with $\log(\tau/{\rm yr})=9.95$
($\approx$$8.9$ Gyr) and a metallicity of $[{\rm Fe}/{\rm H}]=-0.75$
give good matches in the mass-radius~plane.

The masses of the three stars in KIC 10319590 have uncertainties of
$\approx$$4$\%. The~two stars in the eclipsing binary have radii
determined to $\approx$$1$\% accuracy.  Since there are no extra
eclipse events involving the third star, its radius is poorly
determined.  The masses and radii of the four stars in EPIC 2202 are
determined with accuracies of $\approx$$10$\% and $\approx$$5$\%,
respectively.  The bottom panels of Figure~\ref{plotiso01} show these
two systems in the mass-radius plane.  Solar metallicity isochrones
with ages of $\log(\tau/{\rm yr})=9.75$) ($\approx$$5.6$ Gyr) and
$\log(\tau/{\rm yr})=9.85$ ($\approx$$7.1$ Gyr) provide plausible
matches in the mass-radius plane for KIC 10319590.  The uncertainties
on the stellar masses and radii for EPIC 2202 are too large to give
any meaningful constraints on the age of the~system.
\subsection{Dynamical~Properties}

\begin{figure}[H]
\includegraphics[width=8.4 cm, angle=-0]{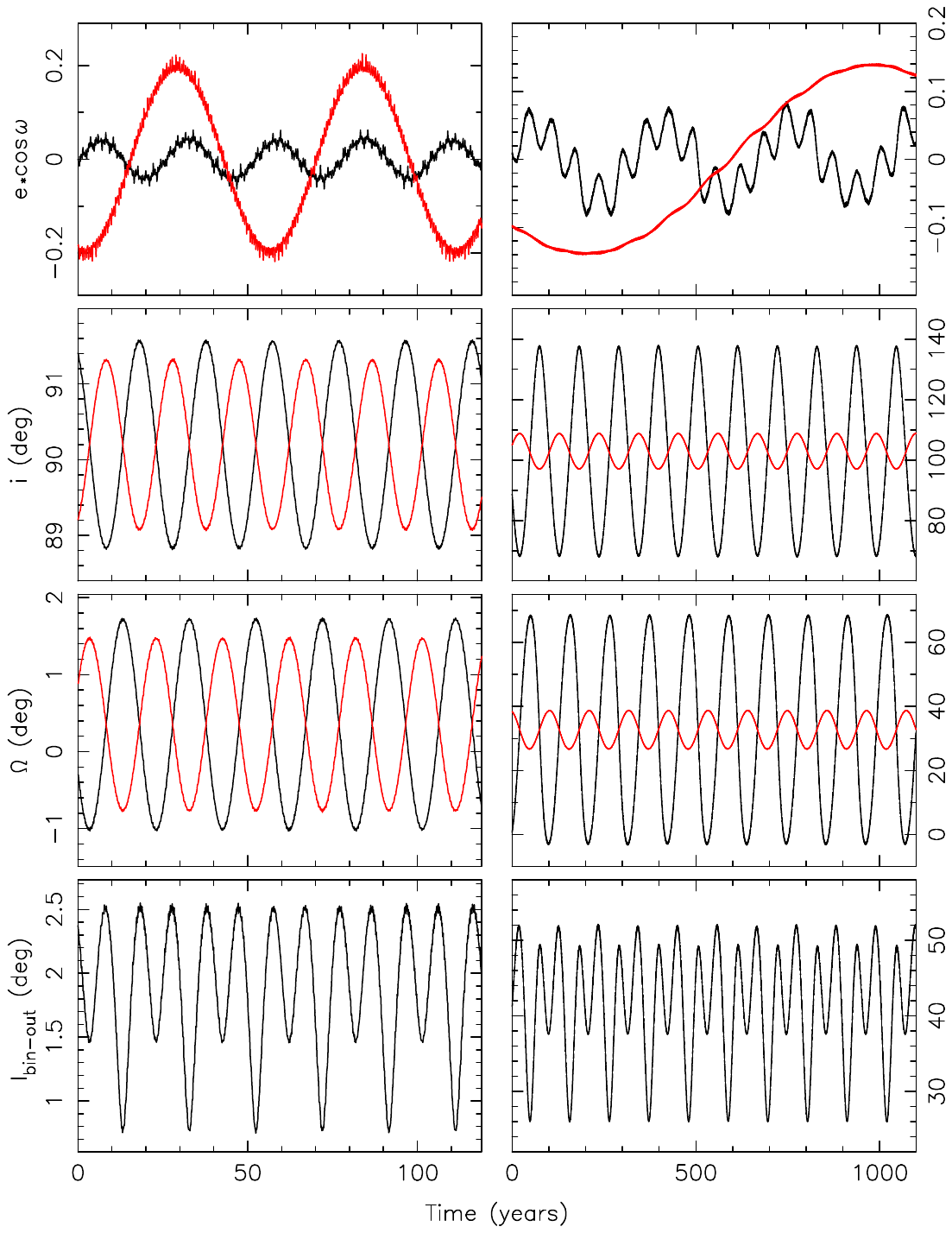}
\caption{\textls[15]{The long-term evolution of key orbital parameters
for KIC 7668648 (\textbf{left} panels) and KIC 10319590
(\textbf{right} panels). From~\textbf{top} to \textbf{bottom}, the
quantities shown are $e\cos\omega$, the inclination $i$ in }
\label{plotdyn01}}
\end{figure}

\begin{figure}[H]
\includegraphics[width=8.8 cm, angle=-0]{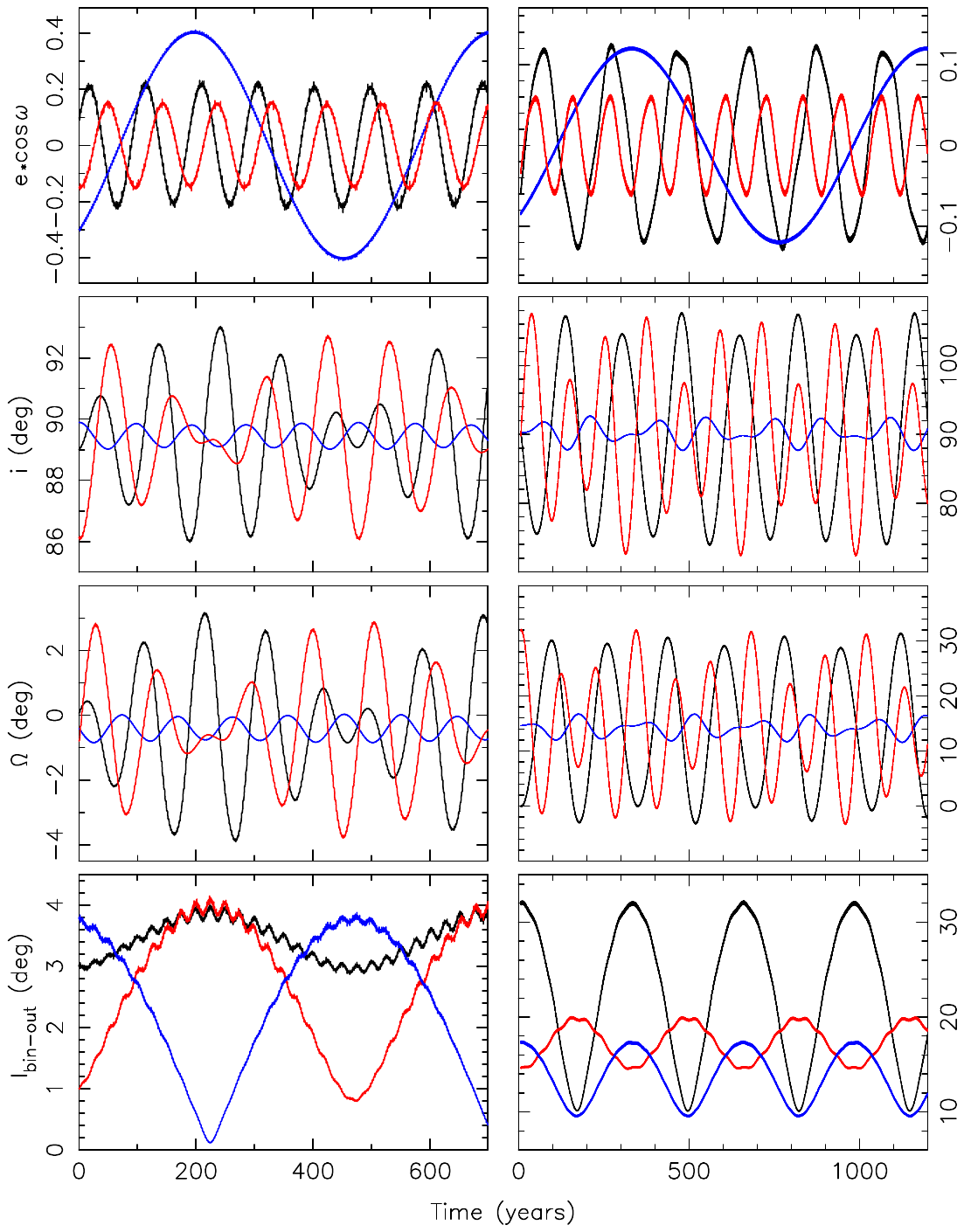}
\caption{The long-term evolution of key orbital parameters for KIC
5255552 (\textbf{left} panels) and EPIC 2202 (\textbf{right}
panels). From~\textbf{top} to \textbf{bottom}, the quantities shown
are $e\cos\omega$, the inclination $i$ in degrees, the~nodal angle
$\Omega$ in degrees, and~the mutual inclination between the two
orbital planes $I_{\rm bin-out}$ in degrees.  The~black lines
indicate the quantities for binary \#1, and the red lines indicate
the quantities for binary \#2, and~the blue lines indicate the
quantities for the outer orbit.
\label{plotdyn02}}
\end{figure}

Although a thorough investigation of the dynamical properties of the
four systems is beyond the scope of this paper, we can compute some
basic quantities that illustrate how the orbits can change over time.
Owing to gravitational interactions between the various stars in these
systems, the~orbits are not perfect Keplerian orbits, and~the orbital
``elements'' (e.g.,\ the eccentricity $e$, the~argument of periastron
$\omega$, the~inclination $i$, etc.) will change.  We can use the
dynamical integrator in {\tt ELC} and its Cartesian-Keplerian
converter to integrate a model over a long time scale and record the
values of $e\cos\omega$, $i$, $\Omega$, and~the mutual inclinations
the various orbital planes over time.  A~simple Fast Fourier Transform
routine can be used to find periodicities in these curves.  Hereafter,
we will call the time it takes the argument of periastron $\omega$ to
change by $360^{\circ}$ as the ``apsidal period'' (in practice, this
is the main periodicity of the $e\cos\omega$ or $e\sin\omega$ curves).
Likewise, we will call the most significant periodicity of the $i$
vs.\ time curve or the $\Omega$ vs.\ time curve the ``nodal period''.
After the equations of motion are solved over the requested time
interval, the~eclipse-finding routine can find the number of eclipse
events (e.g.,\ primary eclipses, secondary eclipses, transits of the
tertiary star in front of the primary star, etc.) in that interval.
The~``eclipse fraction'' is then defined as the number of actual
eclipse events divided by the ``expected'' number, which is simply the
time interval divided by the average orbital period over that
interval.  For the three {\em Kepler} systems that have
well-constrained parameters, we used the initial conditions from all
the posterior models and integrated them over long intervals to
determine the periodicities of the various curves.
Table~\ref{tab_dyn1} gives the results for the 2 + 1 systems KIC
7668648 and KIC 10319590, and~Figure~\ref{plotdyn01} shows these
curves for the respective best-fitting models.  Table~\ref{tab_dyn2}
gives the results for the 2 + 2 systems KIC 5255552 and EPIC 2202,
where for EPIC 2202, we only give rough results for the best-fitting
model since the code that computes the various periodicities of models
in the posterior sample does not reliably work when the parameters of
the outer orbit are poorly constrained.  Figure~\ref{plotdyn02} shows
the long-term behavior of the key dynamical elements for the
best-fitting models for the 2 + 2~systems.

\begin{table}[H]
\caption{Long-Term Dynamical Parameters. 2 + 1~Systems.\label{tab_dyn1}}
\begin{adjustwidth}{-\extralength}{0cm}
\centering 
\setlength{\tabcolsep}{10mm} \resizebox{\linewidth}{!}{\begin{tabular}{lllll}
\toprule
 & \textbf{KIC 7668648 $^1$} &  &  \textbf{KIC 10319590 $^2$} & \\
\textbf{Parameter $^1$} & \textbf{Median} & $\boldsymbol{1\sigma}$  &  \textbf{Median} & $\boldsymbol{1\sigma}$ \\
\textbf{ } & \textbf{(yr)} & \textbf{(yr)}  &  \textbf{(yr)} & \textbf{(yr)} \\
\midrule
$e_{\rm bin}\cos\omega_{\rm bin}$  & 26.0619   & 0.0021  & 352 & 22 \\
$e_{\rm bin}\sin\omega_{\rm bin}$  & 26.0619   & 0.0021  & 352 & 22 \\
$i_{\rm bin}                    $  & 19.61793   & 0.00045  & 107.67 & 0.66 \\
$\Omega_{\rm bin}               $  & 19.61793   & 0.00045  & 107.67 & 0.66 \\
$e_{\rm out}\cos\omega_{\rm out}$  & 55.3400   & 0.0011  & 1692 & 40 \\
$e_{\rm out}\sin\omega_{\rm out}$  & 55.3400   & 0.0011  & 1692 & 40 \\
$i_{\rm out}                    $  & 19.61793   & 0.00045  & 107.67 & 0.66 \\
$\Omega_{\rm out}               $  & 19.61793   & 0.00045  & 107.67 & 0.66 \\
primary eclipse fraction           & 1.0      & \ldots    &  0.7571       &  0.0015     \\
secondary eclipse fraction         & 1.0      & \ldots    &  0.7571   & 0.0015      \\
transit$_{3-1}$ fraction             & 0.2888   & 0.0021 &   0.0     &  \ldots     \\
transit$_{3-2}$ fraction             & 0.2659   & 0.0021 &   0.0     &  \ldots     \\
occultation$_{1-3}$ fraction       &  0.2929    & 0.0021 &   0.0     &  \ldots    \\
occultation$_{2-3}$ fraction       &  0.2618    & 0.0010 &   0.0     &  \ldots    \\
\bottomrule
\end{tabular}}
\end{adjustwidth}
\noindent{\footnotesize{$^1$ Time span = 200,250 days; 
$^2$ Time span = 2,500,200 days.}}
\end{table}
\unskip

\begin{table}[H]
\caption{Long-Term Dynamical Parameters. 2 + 2~Systems.\label{tab_dyn2}}
\begin{adjustwidth}{-\extralength}{0cm}
\centering 
\setlength{\tabcolsep}{10mm} \resizebox{\linewidth}{!}{\begin{tabular}{lllll}
\toprule
 & \textbf{KIC 5255552 $^1$} &  &  \textbf{EPIC 2202 $^2$} & \\
\textbf{Parameter $^1$} & \textbf{Median} & $\boldsymbol{1\sigma}$  &  \textbf{Median} & $\boldsymbol{1\sigma}$ \\
\textbf{ } & \textbf{(yr)} & \textbf{(yr)}  &  \textbf{(yr)} & \textbf{(yr)} \\
\midrule
$e_{\rm bin,1}\cos\omega_{\rm bin,1}$  & 96.716   & 0.016  & 202 & \ldots \\
$e_{\rm bin,1}\sin\omega_{\rm bin,1}$  & 96.716   & 0.016  & 201 & \ldots \\
$i_{\rm bin,1}                    $  & 96.3524   & 0.0052  & 171 & \ldots \\
$\Omega_{\rm bin,1}               $  & 96.3524   & 0.0052  & 171 & \ldots \\
$e_{\rm bin,2}\cos\omega_{\rm bin,2}$  & 93.281   & 0.062  & 113 & \ldots \\
$e_{\rm bin,2}\sin\omega_{\rm bin,2}$  & 93.281   & 0.062  & 113 & \ldots \\
$i_{\rm bin,2}                    $  & 96.3540   & 0.0055  & 112 & \ldots \\
$\Omega_{\rm bin,2}               $  & 96.3538   & 0.0055  & 112 & \ldots  \\
$e_{\rm out}\cos\omega_{\rm out}$  & 591.520   & 0.065  & 858 & \ldots \\
$e_{\rm out}\sin\omega_{\rm out}$  & 591.364   & 0.066  & 873 & \ldots \\
$i_{\rm out}                    $  & 96.3524   & 0.0053  & 170 & \ldots \\
$\Omega_{\rm out}               $  & 96.3524   & 0.0053  & 343 & \ldots \\
EB \#1 primary eclipse fraction           & 1.0      & \ldots    &  \ldots       &  \ldots     \\
EB \#1 secondary eclipse fraction         & 1.0      & \ldots    &  \ldots   & \ldots     \\
EB \#2 primary eclipse fraction           & 0.1557      & 0.0014    &  \ldots       &  \ldots     \\
EB \#2 secondary eclipse fraction         & 0.1560      & 0.0011    &  \ldots   & \ldots      \\
transit$_{3-1}$ fraction             & 0.0864   & 0.0087 &   \ldots     &  \ldots     \\
transit$_{3-2}$ fraction             & 0.1025   & 0.0058 &   \ldots     &  \ldots     \\
transit$_{4-1}$ fraction             & 0.00214   & 0.00017 &   \ldots     &  \ldots     \\
transit$_{4-2}$ fraction             & 0.00236   & 0.00017 &   \ldots     &  \ldots     \\
occultation$_{1-3}$ fraction       &  0.1068    & 0.0073 &   \ldots     &  \ldots    \\
occultation$_{2-3}$ fraction       &  0.0790    & 0.0073 &   \ldots     &  \ldots    \\
occultation$_{1-4}$ fraction       &  0.00264    & 0.00017 &   \ldots     &  \ldots    \\
occultation$_{2-4}$ fraction       &  0.00186   & 0.00022 &   \ldots     &  \ldots    \\
\bottomrule
\end{tabular}}
\end{adjustwidth}
\noindent{\footnotesize{$^1$ Time span = 600,200 days; $^2$ Time span = 
2,500,200 days.}}
\end{table}
\vspace{-15pt}

\vspace{-12pt}
{\captionof*{figure}{degrees, the~nodal
angle $\Omega$ in degrees, and~the
mutual inclination between the two orbital
planes $I_{\rm bin-out}$ in degrees.  The~black
lines indicate the quantities for the binary,
and the red lines indicate the quantities
for the outer orbit.}}

For KIC 7668648, we see that the inclination of the two orbits and the
nodal angles of the two orbits do not change by more than $\approx$$
3^{\circ}$.  Consequently, the~mutual inclination between the two
orbital planes is never more than $2.5^{\circ}$, and~the primary and
second eclipse fractions are each 1.0.  The various eclipse events
involving the third star occur between $\approx$$26\%$ and
$\approx$$29$ of the time.  The~nodal precession period of the inner
binary (and of the outer orbit) is only $19.61793\pm 0.00045$ years.
The~apsidal period of the inner binary is somewhat longer at
$26.0619\pm 0.0021$ years, while the apsidal period of the outer orbit
is $55.3400\pm 0.0011$ years.

For KIC 10319590, the~change in the inclination of the inner binary
over time is much more dramatic where the inclination is in the range
of $70^{\circ}$ to $149^{\circ}$.  The nodal angle of the binary has a
similarly large range of between $0^{\circ}$ and $70^{\circ}$.  On the
other hand, those two quantities for the outer orbit have a much
smaller range of around $\pm 10^{\circ}$.  Since the nodal angle of
the outer orbit hovers around $30^{\circ}$, the~mutual inclination
between the two orbitals is never less than about $25^{\circ}$.
The~nodal period is about 108 years, and~the eclipse fraction is about
0.76.  Given these two quantities, the~current non-eclipsing state of
the binary will last $\approx$$0.24\times 108\approx 26$ years.  Owing
to the relatively large mutual inclination, there were no eclipse
events involving the third star in the posterior sample.  Finally,
the~way the argument of periastron of KIC 10319590 changes over time
is more complicated than it is for KIC 7668648.  For~the inner binary
of KIC 10319590, the~apsidal period (e.g.,\ the time for $\omega$ to
change by $360^{\circ}$) is $349\pm 21$ years.  However, a~smaller
amplitude modulation with a period of around 55 years is also evident
in the $e\cos\omega$ curve.  The~apsidal period is
\mbox{$\approx$$1600$ years} for the outer~orbits.

For KIC 5255552, the~various inclinations $i$ and the nodal angles
$\Omega$ vary by less than $\approx$$6^{\circ}$ over the time frame
investigated.  As a result, the~mutual inclination between the two
binary orbital planes and the mutual inclinations between the outer
orbit and the binary planes are always less than $\approx$$4^{\circ}$.
As seen in Table~\ref{tab_dyn2}, binary \#1 in KIC 5255552 always
eclipses, whereas the eclipse fractions of binary \#2 are
$\approx$$0.16$.  Figures~\ref{fig2}--\ref{fig4} show the various
extra eclipse events seen in KIC 5255552.  The~secondary star in
binary \#2 transited both the primary star in binary \#1 and the
secondary star in binary \#1.  Surprisingly, these types of events are
rare with a transit fraction of $\approx$$0.2\%$ (see
Table~\ref{tab_dyn2}).  The secondary star in binary \#1 transited the
primary star in binary \#2, and~vice versa.  Table~\ref{tab_dyn2}
shows that these events are somewhat more common with transit
fractions of $\approx$$8\%$ and $\approx$$10\%$.  Finally,
the~secondary star in binary \#2 transited the secondary star in
binary \#1, and~this type of event has a transit fraction of only
$\approx$$0.2\%$.  Given the low transit fractions of the various
observed extra eclipse and occultation events, it seems we were
exceedingly lucky that these events occurred during the relatively
short {\em Kepler} mission, and~that KIC 5255552 was not in the field
of view of the failed CCD module when the events~occurred.

As noted above, the~code that determines the periodicities of the
various dynamical quantities using models in a posterior sample does
not reliably work for cases like EPIC 2202, where the properties of
the outer model are poorly constrained.  Consequently, we give rough
periodicities of the model whose initial conditions are given in
Table~\ref{EPIC01_init} in the Appendix \ref{appc}.

KIC 5255552 and KIC 7668648 have nearly coplanar orbital planes.
On~the other hand, the mutual inclination between the two orbital
planes in KIC 10319590 is much larger and is always within
$\approx$$12^{\circ}$ of $40^{\circ}$.  There is a mechanism known as
Kozai cycles with tidal friction, whereby the orbital separation of a
binary can shrink due to the combined effects of secular perturbations
from a distant companion and tidal friction that become more and more
important as the fractional radii (e.g.,\ $R/a$) of the stars
increases with time ~\cite{Kiseleva_1998,Mazeh_1979,Fabrycky_2007}.
An expected statistical outcome of this process is that short-period
binaries will have distant companions with a distributions of mutual
inclinations between the two orbital planes strongly peaked near
$40^{\circ}$ and $140^{\circ}$.  The~present-day configuration of KIC
10319590 seems to be consistent with this expected outcome.
Although~the uncertainties are much larger for EPIC 2202,
the~distribution of the mutual inclinations between the orbital plane
of the first binary and the orbital plane of the outer orbit has peaks
near $40^{\circ}$ and $140^{\circ}$. Even though EPIC 2202 is a 2 + 2
system and not a 2 + 1 system, could a similar Kozai cycles with tidal
friction process be shrinking the binary orbits while widening the
outer orbit?  A thorough analytical and numerical study of this
process in 2 + 2 systems would be interesting, and is certainly beyond
the scope of the present work.

\subsection{Comparison to Previous~Studies}

When comparing the results of a comprehensive photodynamical model to
previous work, one must use care since many of the parameters change
with time. Thus, a quantity like the orbital period at a specific
reference time that can have a very small formal error could be many
$\sigma$ off when compared to a value found for a different reference
time or a value with a different meaning (for example, a period that
is some kind of an average over a long interval).  Also, the~previous
studies of the three {\em Kepler} systems did not have access to
radial velocity measurements.  Nevertheless, it is still interesting
and useful to make a few cursory~comparisons.

KIC 10319590 is a relatively uncomplicated 2 + 1 system with a fairly
large period ratio ($P_{\rm out}/P_{\rm bin}=21.4$ at the reference
time of BJD 2,454,800) and no extra eclipse events.  On~the other
hand, the~number of useful {\em Kepler} data is limited since the
binary stopped eclipsing early in the {\em Kepler} mission.  Borkovits
et al.\ \cite{Borkovits_2016} found masses of $1.5\pm 0.2\,M_{\odot}$
for the binary and $0.8\pm 0.1\,M_{\odot}$ for the third star,
respectively.  We find a total mass for the binary of $1.86\pm
0.04\,M_{\odot}$ and a tertiary mass of $1.05\pm 0.04\,M_{|odot}$.
Both quantities are $\approx$$2\sigma$ larger than the corresponding
quantities found by Borkovits et al.~\cite{Borkovits_2016}.  They
found a mutual inclination between the two orbital planes of $40.2\pm
0.4^{\circ}$.  Again, taking into account that this quantity changes
with time, their value compares favorably with our value of $40.84\pm
0.12^{\circ}$.  Regarding the long-term orbital dynamics,
Borkovits~et~al.\ \cite{Borkovits_2016} found an apsidal period of 68
years and a nodal period of 110 years.  Their apsidal period is
somewhat different than our value of $46.7\pm 0.3$ years.  On the
other hand, we find a nodal period of $107.7\pm 0.3$ years, which is
in reasonably good agreement with their~value.

KIC 7668648 is a 2 + 1 system like KIC 10319590, but is much more
compact (the period ratio is $P_{\rm out}/P_{\rm bin}=7.5$ at the
reference time of BJD 2,454,750, compared to 21.4 for \mbox{KIC
  10319590}).  Also, KIC 7668648 has extra eclipse events, therefore
making the modeling more challenging.
Borkovits~et~al.\ \cite{Borkovits_2016} found a total mass for the
binary of $1.6\pm 0.5\,M_{\odot}$, which agrees with our value of
$1.64\pm 0.01\,M_{\odot}$.  They found a tertiary mass of $0.27\pm
0.08\,M_{\odot}$, which agrees favorably with our value of $0.275\pm
0.003\,M_{\odot}$.  Regarding the long-term orbital dynamics,
Borkovits~et~al.\ \cite{Borkovits_2016} found a mutual inclination
between the two orbital planes of $42\pm 1^{\circ}$.  This value seems
to be too large, given the presence of numerous tertiary eclipse and
occultation events.  We find $I=2.4440\pm 0.0007^{\circ}$ at the
reference time of BJD 2,454,750.  They found an apsidal period of
$54\pm 6$ years, compared to our value of \mbox{$25.826\pm 0.005$
  years}.  They quote a nodal period for the binary of 25 years with
no uncertainty.  We find a nodal period for the binary of 19.7601
years with a small formal uncertainty of 0.0005 years (e.g.,\ around 4
h).

Comparisons for KIC 5255552 are more fraught since it is actually a 2
+ 2 system and not a 2 + 1 system.  Using a model that assumed a 2 + 1
system, Borkovits~et~al.\ \cite{Borkovits_2016} found a binary mass of
$1.7\pm 0.2\,M_{\odot}$, which compares favorably with our value of
$1.70\pm 0.02\,M_{\odot}$.  They found a tertiary mass of $0.7\pm
0.1\,M_{\odot}$.  In~our case, we find a total mass for binary \#2 of
$0.99\pm 0.01\,M_{\odot}$.

For EPIC 2202, Rappaport~et~al.\ \cite{Rappaport_2017} attempted a few
different techniques to measure the masses and radii of the component
stars, for~example, by modeling each binary separately and attempting
to model them jointly.  Their mass measurements had relative
uncertainties on the order of 10\%, similar to what we find here.
Their mass measurements and our mass measurements agree at the
$\approx$$1\sigma$ level.  Their radius measurements also had relative
uncertainties of around 10\%, compared to $\approx$$5\%$ for the
present work.  Based on dynamical simulations, they concluded the
outer orbital period is likely between about 300 and 500 days.
The~posterior sample for this quantity generated by the extensive MCMC
analysis has peaks near 500 days and 1000 days, with~a tail that
extends out to around \mbox{3000 days.}  As~Rappaport~et~al.\ pointed
out, a~few more radial velocity measurements and/or a few more eclipse
observations would greatly constrain the properties of the
outer~orbit.

\section{Summary}\label{SummarySec}

We presented full photodynamical models of the 2 + 1 systems KIC
7668648 and \mbox{KIC 10319590}.  The stellar masses and radii in KIC
7668648 are tightly constrained.  Based on an isochrone analysis,
the~system is very old and must be somewhat metal-poor.  The~system is
compact ($P_{\rm out}/P_{\rm bin}=7.5$), and the two orbital planes
are within a few degrees of being coplanar.  KIC 10319590 does not
have extra eclipse events, so the stellar masses and radii are less
well constrained.  The~mutual inclination of the two orbital planes
oscillates around $\approx$$40^{\circ}$, consistent with a triple
system that underwent Kozai cycles with tidal friction.  KIC 5255552
turns out to be a 2 + 2 system where one of the binaries currently
does not eclipse.  We presented here, for~the first time,
a~comprehensive model that explains the eclipse timing variations of
the eclipsing binary and all the extra eclipse events.  All three
orbital planes are within $\approx$$4^{\circ}$ of being coplanar.
Finally, we performed an extensive MCMC analysis of EPIC 2202 to
constrain the geometrical properties of the system.  Using the
presently available data, the~mutual inclination between the orbital
planes of the two eclipsing binaries is only weakly constrained, where
there is a hint of a mutual inclination of around $10^{\circ}$.
On~the other hand, the posterior distribution for the mutual
inclination between the orbital plane of the shorter period eclipsing
binary and the outer orbit has two distinct peaks near $30^{\circ}$
and $150^{\circ}$, with~values between $50^{\circ}$ and $100^{\circ}$
ruled out.  Further investigation will be needed to see if this
particular bimodal distribution is the result of a type of Kozai
cycles with a tidal friction process that operates in 2 + 1 systems.
Finally, additional radial velocity measurements and additional
eclipse observations would improve the constraints on the various
fitting and derived parameters for all four systems, but~perhaps none
more than those of EPIC~2202.
\vspace{6pt}

\funding{This research was funded by NASA grants NNX14AB91G,
09-KEPLER09-0061, 20-ADAP20-0137, 20-TESS20-0037, NNH20ZDA001N-TESS,
NNH22ZDA001N-XRP, and~NNX13AI76G; National Science Foundation grants
AST-1109928, AST-1617004, and~AST-02206814.}

\institutionalreview{Not applicable.}

\informedconsent{Not applicable.}

\dataavailability{The {\em Kepler} and TESS
data are available at the Mikulski Archive for
Space Telescopes (MAST), see \url{https://archive.stsci.edu}.}

\acknowledgments{{\em~Kepler} was selected as the 10th mission of the
Discovery Program. Funding for this mission is provided by the NASA
Science Mission Directorate. The~{\em Kepler} data were obtained
from the Mikulski Archive for Space Telescopes (MAST). The~Space
Telescope Science Institute is operated by the Association of
Universities for Research in Astronomy, Inc., under NASA contract
NAS5-26555. Support for MAST for non-HST data is provided by the
NASA Office of Space Science via grant NXX09AF08G and by other
grants and contracts.  This paper includes data collected by the
{\em TESS} mission.  Funding for the {\em TESS} mission is provided
by the NASA's Science Mission Directorate.  J.A.O. acknowledges the
valuable contributions of D. R. Short to the continuous development
of the {\tt ELC} code.  Useful discussions with W. F. Welsh,
V. B. Kostov, G. Torres, D. Latham, D. C. Fabrycky, and~members of
the {\em Kepler} Eclipsing Binary Stars Working Group are also
appreciated.}

\conflictsofinterest{The authors declare no conflict of~interest.}

\abbreviations{Abbreviations}{
The following abbreviations are used in this manuscript:\\

\noindent 
\begin{tabular}{@{}ll}
BF & Broadening Function \\
CPOC & Common-Period Observed minus Computed \\
DE-MCMC & Differential Evolution Monte Carlo Markov Chain \\
EB  & Eclipsing Binary \\
{\tt ELC} & Eclipsing Light Curve (code) \\
EPIC  &  Ecliptic Plane Input Catalog  \\
ETV  & Eclipse Timing Variation \\
FFI & Full Frame Image \\
GR & General Relativity \\
IRAF & Image Reduction and Analysis Facility \\
KIC & Kepler Input Catalog \\
MCMC & Monte Carlo Markov Chain \\
MIST & MESA Isochrones and Stellar Tracks \\
NASA & National Aeronautics and Space Administration \\
TESS & Transiting Exoplanet Survey Satellite \\
\end{tabular}
}
\printendnotes

\appendixtitles{yes} 
\appendixstart
\appendix

\section[\appendixname~\thesection]{Eclipse Time Measurements}
\label{appendix_times}

We present tables of the measured eclipse and transit times for
the {\em Kepler} systems.

\begin{table}[H] 
\caption{Eclipse times for KIC~5255552.\label{5255_times}}
\newcolumntype{L}{>{\centering\arraybackslash}X}

\begin{adjustwidth}{-\extralength}{0cm}
\centering 
\setlength{\tabcolsep}{7mm} \resizebox{\linewidth}{!}{\begin{tabular}{llrllr}
\toprule
\textbf{Type} & \textbf{Cycle Number} &  \textbf{Time}  &  
\textbf{Type} & \textbf{Cycle Number} &  \textbf{Time}\\
              &                       &   \textbf{(BJD $-$2,455,000)} &  
              &                       &   \textbf{(BJD $-$2,455,000)}\\
\midrule
        p  &    0 &  $ -29.411314\pm0.00014$  &        s  &    0 &  $ -11.214690\pm0.00021$  \\
        p  &    1 &  $   3.039710\pm0.00018$  &        s  &    1 &  $  21.255632\pm0.00018$  \\
        p  &    2 &  $  35.504448\pm0.00017$  &        s  &    2 &  $  53.705795\pm0.00013$  \\
        p  &    3 &  $  67.971947\pm0.00019$  &        s  &    3 &  $  86.148163\pm0.00022$  \\
        p  &    5 &  $ 132.901825\pm0.00011$  &        s  &    4 &  $ 118.586273\pm0.00014$  \\
        p  &    6 &  $ 165.361801\pm0.00013$  &        s  &    5 &  $ 151.023422\pm0.00015$  \\
        p  &    7 &  $ 197.819473\pm0.00011$  &        s  &    7 &  $ 215.898682\pm0.00014$  \\
        p  &    9 &  $ 262.726929\pm0.00013$  &        s  &    8 &  $ 248.337631\pm0.00017$  \\
        p  &   13 &  $ 392.515259\pm0.00014$  &        s  &   12 &  $ 378.108704\pm0.00052$  \\
        p  &   14 &  $ 424.956268\pm0.00012$  &        s  &   13 &  $ 410.558746\pm0.00021$  \\
        p  &   15 &  $ 457.394623\pm0.00010$  &        s  &   14 &  $ 443.011627\pm0.00014$  \\
        p  &   16 &  $ 489.829254\pm0.00010$  &        s  &   15 &  $ 475.468689\pm0.00020$  \\
        p  &   17 &  $ 522.259521\pm0.00013$  &        s  &   16 &  $ 507.932953\pm0.00018$  \\
        p  &   19 &  $ 587.107361\pm0.00020$  &        s  &   17 &  $ 540.406860\pm0.00012$  \\
        p  &   20 &  $ 619.530273\pm0.00021$  &        s  &   18 &  $ 572.893616\pm0.00017$  \\
        p  &   24 &  $ 749.666687\pm0.00012$  &        s  &   19 &  $ 605.397888\pm0.00011$  \\
        p  &   25 &  $ 782.074097\pm0.00018$  &        s  &   24 &  $ 767.776367\pm0.00015$  \\
        p  &   26 &  $ 814.455322\pm0.00015$  &        s  &   25 &  $ 800.421936\pm0.00023$  \\
        p  &   27 &  $ 846.882751\pm0.00012$  &        s  &   26 &  $ 832.992493\pm0.00020$  \\
        p  &   28 &  $ 879.339661\pm0.00026$  &        s  &   28 &  $ 897.962219\pm0.00014$  \\
        p  &   29 &  $ 911.807190\pm0.00013$  &        s  &   30 &  $ 962.854309\pm0.00011$  \\
        p  &   30 &  $ 944.276001\pm0.00019$  &        s  &   36 &  $1157.493286\pm0.00015$  \\
        p  &   31 &  $ 976.742859\pm0.00013$  &        s  &   37 &  $1189.936401\pm0.00022$  \\
        p  &   32 &  $1009.207153\pm0.00011$  &        s  &   38 &  $1222.381470\pm0.00017$  \\
        p  &   37 &  $1171.488892\pm0.00009$  &        s  &   39 &  $1254.828613\pm0.00022$  \\
        p  &   38 &  $1203.938110\pm0.00014$  &        s  &   40 &  $1287.278687\pm0.00012$  \\
        p  &   39 &  $1236.385498\pm0.00014$  &        s  &   42 &  $1352.192139\pm0.00016$  \\
        p  &   41 &  $1301.272339\pm0.00012$  &        s  &   43 &  $1384.659180\pm0.00017$  \\
        p  &   42 &  $1333.711548\pm0.00010$  &        s  &  114 &  $3690.162354\pm0.00206$  \\
        p  &   43 &  $1366.146851\pm0.00012$  &        s  &  124 &  $4014.639404\pm0.00240$  \\
        p  &  115 &  $3703.093994\pm0.00114$  &        s  &  137 &  $4437.044434\pm0.00259$  \\
        p  &  125 &  $4027.528564\pm0.00167$  &        s  &  147 &  $4761.431152\pm0.00330$  \\
        p  &  137 &  $4417.155762\pm0.00130$  &        s  &  148 &  $4793.882324\pm0.00326$  \\
        p  &  148 &  $4774.140625\pm0.00159$  &                   &      & \\
\bottomrule
\end{tabular}}
\end{adjustwidth}
\end{table}
\unskip

\begin{table}[H] 
\caption{Eclipse times for KIC~7668648.\label{7668_times}}
\newcolumntype{L}{>{\centering\arraybackslash}X}
\begin{adjustwidth}{-\extralength}{0cm}
\centering 
\setlength{\tabcolsep}{7mm} \resizebox{\linewidth}{!}{\begin{tabular}{llrllr}
\toprule
\textbf{Type} & \textbf{Cycle Number} &  \textbf{Time}  &  
\textbf{Type} & \textbf{Cycle Number} &  \textbf{Time}\\
              &                       &   \textbf{(BJD $-$2,455,000)} &  
              &                       &   \textbf{(BJD $-$2,455,000)}\\
\midrule
              p  &    0 &  $  -8.849283\pm0.00015$  &                  s  &    0 &  $ -22.791008\pm0.00011$  \\
              p  &    1 &  $  18.947523\pm0.00012$  &                  s  &    1 &  $   5.016356\pm0.00017$  \\
              p  &    2 &  $  46.745491\pm0.00013$  &                  s  &    2 &  $  32.788330\pm0.00046$  \\
              p  &    3 &  $  74.541428\pm0.00017$  &                  s  &    3 &  $  60.560638\pm0.00013$  \\
              p  &    4 &  $ 102.374245\pm0.00012$  &                  s  &    5 &  $ 116.412834\pm0.00016$  \\
              p  &    5 &  $ 130.222137\pm0.00015$  &                  s  &    6 &  $ 144.038223\pm0.00010$  \\
              p  &    6 &  $ 158.083542\pm0.00010$  &                  s  &    7 &  $ 171.991974\pm0.00014$  \\
              p  &    7 &  $ 185.892426\pm0.00021$  &                  s  &    8 &  $ 199.814865\pm0.00016$  \\
              p  &    8 &  $ 213.679688\pm0.00010$  &                  s  &    9 &  $ 227.583847\pm0.00009$  \\
              p  &    9 &  $ 241.469406\pm0.00010$  &                  s  &   10 &  $ 255.343399\pm0.00012$  \\
              p  &   10 &  $ 269.258972\pm0.00016$  &                  s  &   11 &  $ 283.137451\pm0.00008$  \\
              p  &   11 &  $ 297.058563\pm0.00009$  &                  s  &   12 &  $ 311.096863\pm0.00007$  \\
              p  &   12 &  $ 324.938477\pm0.00008$  &                  s  &   13 &  $ 338.802063\pm0.00007$  \\

           p  &   13 &  $ 352.757568\pm0.00021$  &                  s  &   14 &  $ 366.749939\pm0.00011$  \\
              p  &   14 &  $ 380.611359\pm0.00009$  &                  s  &   15 &  $ 394.620392\pm0.00010$  \\

   p  &   15 &  $ 408.409363\pm0.00012$  &                  s  &   16 &  $ 422.411163\pm0.00009$  \\
	\bottomrule
	\end{tabular}}
	\end{adjustwidth} 
	 \end{table}

\begin{table}[H]\ContinuedFloat 
\caption{{\em Cont.}}
\newcolumntype{L}{>{\centering\arraybackslash}X}
\begin{adjustwidth}{-\extralength}{0cm}
\centering 
\setlength{\tabcolsep}{7mm} \resizebox{\linewidth}{!}{\begin{tabular}{llrllr}
\toprule
\textbf{Type} & \textbf{Cycle Number} &  \textbf{Time}  &  
\textbf{Type} & \textbf{Cycle Number} &  \textbf{Time}\\
              &                       &   \textbf{(BJD $-$2,455,000)} &  
              &                       &   \textbf{(BJD $-$2,455,000)}\\
\midrule

              p  &   16 &  $ 436.207672\pm0.00012$  &                  s  &   17 &  $ 450.182678\pm0.00014$  \\
              p  &   17 &  $ 464.008606\pm0.00007$  &                  s  &   18 &  $ 477.970276\pm0.00017$  \\
              p  &   18 &  $ 491.815094\pm0.00010$  &                  s  &   19 &  $ 505.868103\pm0.00012$  \\
              p  &   19 &  $ 519.690918\pm0.00005$  &                  s  &   20 &  $ 533.790527\pm0.00007$  \\
              p  &   20 &  $ 547.470276\pm0.00014$  &                  s  &   22 &  $ 589.509644\pm0.00009$  \\
              p  &   21 &  $ 575.364380\pm0.00010$  &                  s  &   23 &  $ 617.311890\pm0.00006$  \\
              p  &   22 &  $ 603.157898\pm0.00007$  &                  s  &   24 &  $ 645.080750\pm0.00005$  \\
              p  &   23 &  $ 630.947449\pm0.00009$  &                  s  &   25 &  $ 672.852173\pm0.00006$  \\
              p  &   24 &  $ 658.743652\pm0.00006$  &                  s  &   26 &  $ 700.682007\pm0.00500$  \\
              p  &   25 &  $ 686.545532\pm0.00014$  &                  s  &   27 &  $ 728.691284\pm0.00007$  \\
              p  &   26 &  $ 714.385437\pm0.00006$  &                  s  &   28 &  $ 756.320923\pm0.00005$  \\
              p  &   27 &  $ 742.181885\pm0.00005$  &                  s  &   29 &  $ 784.279114\pm0.00006$  \\
              p  &   29 &  $ 797.893372\pm0.00009$  &                  s  &   30 &  $ 812.098755\pm0.00014$  \\
              p  &   30 &  $ 825.675110\pm0.00007$  &                  s  &   31 &  $ 839.867493\pm0.00007$  \\
              p  &   31 &  $ 853.464966\pm0.00008$  &                  s  &   32 &  $ 867.629028\pm0.00006$  \\
              p  &   32 &  $ 881.260437\pm0.00006$  &                  s  &   33 &  $ 895.427856\pm0.00007$  \\
              p  &   33 &  $ 909.074890\pm0.00008$  &                  s  &   34 &  $ 923.383911\pm0.00007$  \\
              p  &   34 &  $ 936.919189\pm0.00006$  &                  s  &   36 &  $ 979.043884\pm0.00500$  \\
              p  &   35 &  $ 964.780884\pm0.00006$  &                  s  &   37 &  $1006.909119\pm0.00008$  \\
              p  &   36 &  $ 992.636536\pm0.00006$  &                  s  &   38 &  $1034.698730\pm0.00008$  \\
              p  &   37 &  $1020.426086\pm0.00005$  &                  s  &   39 &  $1062.471436\pm0.00010$  \\
              p  &   39 &  $1076.028564\pm0.00005$  &                  s  &   40 &  $1090.262817\pm0.00010$  \\
              p  &   41 &  $1131.713257\pm0.00005$  &                  s  &   41 &  $1118.161987\pm0.00016$  \\
              p  &   42 &  $1159.486938\pm0.00007$  &                  s  &   42 &  $1146.063354\pm0.00008$  \\
              p  &   43 &  $1187.399292\pm0.00016$  &                  s  &   43 &  $1173.883545\pm0.00012$  \\
              p  &   44 &  $1215.179565\pm0.00004$  &                  s  &   44 &  $1201.789551\pm0.00010$  \\
              p  &   45 &  $1242.964355\pm0.00006$  &                  s  &   45 &  $1229.585815\pm0.00008$  \\
              p  &   46 &  $1270.761475\pm0.00005$  &                  s  &   46 &  $1257.352783\pm0.00026$  \\
              p  &   47 &  $1298.570435\pm0.00007$  &                  s  &   47 &  $1285.124146\pm0.00007$  \\
              p  &   48 &  $1326.412354\pm0.00006$  &                  s  &   49 &  $1340.941895\pm0.00008$  \\
              p  &   49 &  $1354.167358\pm0.00008$  &                  s  &   50 &  $1368.583496\pm0.00009$  \\
              p  &   50 &  $1382.125854\pm0.00008$  &                  s  &   51 &  $1396.541382\pm0.00006$  \\
              p  &   51 &  $1409.919189\pm0.00004$  &                  s  &  160 &  $4429.248535\pm0.00451$  \\
              p  &  160 &  $4443.082031\pm0.00326$  &                  s  &  172 &  $4763.061035\pm0.00656$  \\
              p  &  171 &  $4749.165039\pm0.00589$  &                  s  &  174 &  $4818.671387\pm0.00263$  \\
              p  &  173 &  $4804.764648\pm0.00223$  &                     &      &                           \\
\bottomrule
\end{tabular}}
\end{adjustwidth}
\end{table}
\unskip

\begin{table}[H] 
\caption{Eclipse times for KIC~10319590.\label{1031_times}}
\newcolumntype{L}{>{\centering\arraybackslash}X}
\begin{adjustwidth}{-\extralength}{0cm}
\centering 
\setlength{\tabcolsep}{7mm} \resizebox{\linewidth}{!}{\begin{tabular}{llrllr}
\toprule
\textbf{Type} & \textbf{Cycle Number} &  \textbf{Time}  &  
\textbf{Type} & \textbf{Cycle Number} &  \textbf{Time}\\
              &                       &   \textbf{(BJD $-$2,455,000)} &  
              &                       &   \textbf{(BJD $-$2,455,000)}\\
\midrule
              p  &    1 &  $ -34.290009\pm0.00005$  &                  s  &    1 &  $ -23.491531\pm0.00048$  \\
              p  &    2 &  $ -12.955812\pm0.00007$  &                  s  &    3 &  $  19.159035\pm0.00029$  \\
              p  &    3 &  $   8.376382\pm0.00006$  &                  s  &    4 &  $  40.480873\pm0.00034$  \\
              p  &    4 &  $  29.704445\pm0.00006$  &                  s  &    5 &  $  61.798950\pm0.00019$  \\
              p  &    5 &  $  51.027229\pm0.00007$  &                  s  &    6 &  $  83.115196\pm0.00022$  \\
              p  &    6 &  $  72.345612\pm0.00006$  &                  s  &    7 &  $ 104.429291\pm0.00146$  \\
              p  &    7 &  $  93.659721\pm0.00005$  &                  s  &    8 &  $ 125.743103\pm0.00021$  \\
              p  &    8 &  $ 114.971283\pm0.00005$  &                  s  &    9 &  $ 147.056442\pm0.00026$  \\
              p  &    9 &  $ 136.281631\pm0.00005$  &                  s  &   10 &  $ 168.371246\pm0.00027$  \\
              p  &   10 &  $ 157.592392\pm0.00010$  &                  s  &   11 &  $ 189.689651\pm0.00032$  \\
              p  &   11 &  $ 178.906586\pm0.00010$  &                  s  &   12 &  $ 211.012634\pm0.00039$  \\
              p  &   12 &  $ 200.226532\pm0.00013$  &                  s  &   14 &  $ 253.675369\pm0.00037$  \\
              p  &   13 &  $ 221.554733\pm0.00012$  &                  s  &   15 &  $ 275.008789\pm0.00046$  \\
              p  &   14 &  $ 242.891785\pm0.00010$  &                  s  &   16 &  $ 296.333344\pm0.00034$  \\
              p  &   15 &  $ 264.234070\pm0.00010$  &                  s  &   17 &  $ 317.645599\pm0.00046$  \\
              p  &   16 &  $ 285.572205\pm0.00010$  &                  s  &   18 &  $ 338.950592\pm0.00087$  \\
              p  &   17 &  $ 306.895721\pm0.00011$  &                  s  &   19 &  $ 360.263062\pm0.00262$  \\
              p  &   18 &  $ 328.203033\pm0.00013$  &                     &      &      \\
              p  &   19 &  $ 349.504608\pm0.00029$  &                     &      &      \\
\bottomrule
\end{tabular}}
\end{adjustwidth}
\end{table}
\unskip

\section[\appendixname~\thesection]{Radial Velocity Measurements}\label{appb}

We present tables of radial velocity measurements for the {\em Kepler}
systems.  

\begin{table}[H] 
\caption{Radial velocities for KIC~5255552.\label{5255_RVtab}}
\newcolumntype{C}{>{\centering\arraybackslash}X}
\begin{tabularx}{\textwidth}{CCC}
\toprule
\textbf{Time} & \textbf{Primary Radial} &  \textbf{Secondary Radial}\\
\textbf{(BJD $-$2,455,000)} & \textbf{Velocity (km s$^{-1}$)} &  
       \textbf{Velocity (km s$^{-1}$)}\\
\midrule
~736.80885 & $ -21.898 \pm  0.586  $  &  $-64.927\pm 6.138$  \\
~739.77839 & $  -2.379 \pm  0.561  $  &  $-85.521\pm 6.543$  \\
~741.70109 & $  -2.217 \pm  0.661  $  &  $-77.311\pm 9.253$  \\
~742.89303 & $  -5.261 \pm  1.033  $  &  $-83.080\pm 8.484$  \\
1083.89084 & $ -56.240 \pm  0.918  $  &  $8.796\pm 3.550$ \\
1086.76300 & $ -53.508 \pm  0.807  $  &  $17.654\pm 6.857$ \\
1457.77006 & $  -6.904 \pm  0.499  $  &  $-80.306\pm 6.382$  \\
1460.80769 & $ -19.814 \pm  0.983  $  &  $-59.924\pm  12.409$ \\
\bottomrule
\end{tabularx}
\end{table}
\unskip

\begin{table}[H] 
\caption{Radial velocities for KIC~7668648.\label{7668_RVtab}}
\newcolumntype{C}{>{\centering\arraybackslash}X}
\begin{tabularx}{\textwidth}{CCC}
\toprule
\textbf{Time} & \textbf{Primary Radial} &  \textbf{Secondary Radial}\\
\textbf{(BJD $-$2,455,000)} & \textbf{Velocity (km s$^{-1}$)} &  
       \textbf{Velocity (km s$^{-1}$)}\\
\midrule
1084.69763 &  $-46.135 \pm  1.3035$  &  $35.165\pm  1.100$  \\
1085.86778 &  $-39.870 \pm  1.3035$  &  $28.939\pm  1.100$  \\
1086.78690 &  $-34.708 \pm  1.3035$  &  $25.093\pm  1.100$  \\
1088.72642 &  $-20.533 \pm  1.3035$  &  $8.065\pm   1.100$  \\
1092.86019 &  $ 16.776 \pm  1.3035$  &  $-33.987\pm 1.100$  \\
1441.91578 &  $-31.129 \pm  1.3035$  &  $27.645\pm  1.100$  \\
1459.86573 &  $ 38.920 \pm  1.3035$  &  $-43.766\pm 1.100$  \\
\bottomrule
\end{tabularx}
\end{table}
\unskip

\begin{table}[H] 
\caption{Radial velocities for KIC~10319590.\label{1031_RVtab}}
\newcolumntype{C}{>{\centering\arraybackslash}X}
\begin{tabularx}{\textwidth}{CCC}
\toprule
\textbf{Time} & \textbf{Primary Radial} &  \textbf{Tertiary Radial}\\
\textbf{(BJD $-$2,455,000)} & \textbf{Velocity (km s$^{-1}$)} &  
       \textbf{Velocity (km s$^{-1}$)}\\
\midrule
~740.907195  & $-54.739 \pm 0.539 $  & $-6.312 \pm 0.935$  \\
~743.677381  & $-28.326 \pm 1.094 $  & $-4.196 \pm 1.905$  \\
~797.905536  & $-42.982 \pm 1.155 $  & $4.626 \pm  1.770$  \\
1087.850896  & $ 18.385 \pm 0.896 $  & $-41.135 \pm 1.343$  \\
1089.650195  & $ 29.673 \pm 1.698 $  & $-41.073 \pm 1.848$  \\
1092.884236  & $ 21.784 \pm 0.998 $  & $-40.136 \pm 1.578$  \\
\bottomrule
\end{tabularx}
\end{table}
\unskip

\section[\appendixname~\thesection]{Initial Conditions for Best-fitting 
Solutions}\label{appc}

We present the Keplerian elements and Cartesian coordinates for the
best-fitting models for KIC 5255552, KIC 7668648, KIC 10919590, and
EPIC 2202.  For~the 2 + 1 systems, ``orbit 1'' is the inner binary,
and~``orbit 2'' is the orbit of the third star.  For the 2 + 2
systems, ``orbit 1'' is binary \#1, ``orbit 3'' is binary \#2,
and~``orbit 2'' is the outer~orbit.

\begin{table}[H]
\footnotesize
\caption{Initial Conditions for KIC~5255552.\label{K5255_init}}
\newcolumntype{s}{>{\hsize=8pt}X}
\begin{adjustwidth}{-\extralength}{0cm}
\centering 
\setlength{\tabcolsep}{2mm} \resizebox{\linewidth}{!}{\begin{tabular}{rrrrr}
\toprule
\textbf{Parameter $^1$} & 
\textbf{\quad Orbit 1} &  
\textbf{\quad Orbit 2} &
\textbf{\quad Orbit 3} &
\textbf{\quad } \\
\midrule
Period (days) & $ 3.24542950562181574E+01$ & $ 8.78469020616712442E+02$ & $ 3.37428742722387511E+01$ & \\
$e\cos\omega$ & $ 8.43858230779453017E-02$ & $-3.12538841564988568E-01$ & $-1.58095818133700217E-01$ & \\
$e\sin\omega$ & $-2.01248140714242796E-01$ & $-2.59142794081377525E-01$ & $ 3.34654593003067269E-02$ & \\
$i$ (rad) & $ 1.55299257745799957E+00$ & $ 1.56894197073380948E+00$ & $ 1.50355302433959626E+00$ & \\
$\Omega$ (rad) & $ 0.00000000000000000E+00$ & $-6.51169685358067053E-03$ & $-1.83064645017791232E-02$ & \\
$T_{\rm conj}$ (days) $^3$ & $-1.91803437861466108E+02$ & $-3.44453952814519084E+02$ & $-1.99477465106001489E+02$ & \\
$a$ (AU) & $ 2.37479041791028272E-01$ & $ 2.49589708685808942E+00$ & $ 2.03780725713174682E-01$ \\
$e$ & $ 2.18224153744451643E-01$ & $ 4.05999402968881706E-01$ & $ 1.61598962489069714E-01$ \\
$\omega$ (deg) & $ 2.92748904100436107E+02$ & $ 2.19663956770694256E+02$ & $ 1.68048152218808156E+02$ \\
true anomaly (deg) & $ 7.85310580756028571E+01$ & $ 2.97842906513808543E+02$ & $ 2.75857964744907576E+02$ \\
mean anomaly (deg) & $ 5.49784964150570232E+01$ & $ 3.33014529305190138E+02$ & $ 2.93975537578175476E+02$ \\
mean longitude (deg) & $ 3.71279962176038964E+02$ & $ 5.17133770537323926E+02$ & $ 4.42857233809957734E+02$ \\
$i$ (deg) & $ 8.89799203034869635E+01$ & $ 8.98937532239852004E+01$ & $ 8.61472425687896077E+01$ \\
$\Omega$ (deg) & $ 0.00000000000000000E+00$ & $-3.73092747178789985E-01$ & $-1.04888315375800500E+00$ \\

\midrule
\textbf{Parameter $^2$} & 
\textbf{Body 1} &  
\textbf{Body 2} &
\textbf{Body 3} &
\textbf{Body 4} \\
\midrule
Mass ($M_{\odot}$) & $ 9.51079756261333897E-01$ & $ 7.45320608115522254E-01$ & $ 4.83662201051754115E-01$ & $ 5.07904696565614766E-01$ \\
$x$ (AU) & $ 5.03794600189741093E-01$ & $ 7.16371646803592110E-01$ & $-1.03242840572052463E+00$ & $-1.01146692543154515E+00$ \\
$y$ (AU) & $-4.67899141880373948E-03$ & $-3.92415626491691635E-03$ & $ 9.51320489373764025E-04$ & $ 1.36142288835155618E-02$ \\
$z$ (AU) & $-2.65912518155657585E-01$ & $-2.23519459049125374E-01$ & $ 3.23849086237075312E-01$ & $ 5.17546130244308955E-01$ \\
$v_x$ (AU day$^{-1}$) & $ 7.72677494158415293E-04$ & $ 1.03862059292030755E-03$ & $ 1.87193786576186974E-02$ & $-2.07968920350985187E-02$ \\
$v_y$ (AU day$^{-1}$) & $-3.81742407801961489E-04$ & $ 5.11551762917300469E-04$ & $-3.20197759944435109E-04$ & $ 2.69075988896128851E-04$ \\
$v_z$ (AU day$^{-1}$) & $-1.31321442880617898E-02$ & $ 3.70370514413945195E-02$ & $-1.42226781415548623E-02$ & $-1.62152257211280043E-02$ \\
Radius (AU) & $ 4.31875482895648349E-03$ & $ 3.21323426147340080E-03$ & $ 2.15643774488338390E-03$ & $ 2.20870367009257584E-03$ \\
Flux & $ 8.32487087169697521E-02$ & $ 1.66201989798861194E-01$ & $ 9.69194422462402519E-03$ & $ 2.01976985218469164E-02$ \\
\bottomrule
\end{tabular}}
\end{adjustwidth}
\noindent{\footnotesize{\textsuperscript{1} Jacobian instantaneous (Keplerian) elements, reference time BJD 2,454,800. 
\textsuperscript{2} Barycentric Cartesian~coordinates. 
\textsuperscript{3} Times are relative to BJD 2,455,000.000.}}
\end{table}
\unskip

\begin{table}[H]
\footnotesize
\caption{Initial Conditions for KIC~7668648.\label{K7668_init}}
\newcolumntype{s}{>{\hsize=8pt}X}
\begin{adjustwidth}{-\extralength}{0cm}
\centering 
\setlength{\tabcolsep}{5mm} \resizebox{\linewidth}{!}{\begin{tabular}{rrrr}
\toprule
\textbf{Parameter $^1$} & 
\textbf{\quad Orbit 1} &  
\textbf{\quad Orbit 2} &
\textbf{\quad } \\
\midrule
Period (days)& $ 2.77962612105870654E+01$ & $ 2.08004956298134800E+02$ & \\ 
$e\cos\omega$ & $-9.55846192066458003E-03$ & $-1.97615688206508128E-01$ & \\ 
$e\sin\omega$ & $-4.80326253974505957E-02$ & $ 4.78101764400388901E-02$ & \\ 
$i$ (rad) & $ 1.59687600813315744E+00$ & $ 1.55570960109599232E+00$ & \\ 
$\Omega$ (rad) & $ 0.00000000000000000E+00$ & $ 1.11706928536536854E-02$ & \\ 
$T_{\rm conj}$ (days) $^3$  & $-9.25043773987229940E+01$ & $-1.05812755307676596E+02$ & \\ 
$a$ (AU) & $ 2.11445341441081708E-01$ & $ 8.51839426583633652E-01$ & \\
$e$ & $ 4.89744555545068710E-02$ & $ 2.03316927963609340E-01$ & \\
$\omega$ (deg) & $ 2.58745205235794003E+02$ & $ 1.66399476850124302E+02$ & \\
true anomaly (deg) & $ 3.08300008972628746E+02$ & $ 7.76649936028934889E+01$ & \\
mean anomaly (deg) & $ 3.12604885366790086E+02$ & $ 5.57687841621907268E+01$ & \\
mean longitude (deg) & $ 5.67045214208422749E+02$ & $ 2.44704504007769089E+02$ & \\
$i$ (deg) & $ 9.14942556717284390E+01$ & $ 8.91355942907812278E+01$ & \\
$\Omega$ (deg) & $ 0.00000000000000000E+00$ & $ 6.40033554751305966E-01$ & \\
\midrule

\textbf{Parameter $^2$} & 
\textbf{Body 1} &  
\textbf{Body 2} &
\textbf{Body 3} \\
\midrule
Mass ($M_{\odot}$) & $ 8.36254587234367519E-01$ & $ 7.96114827571525296E-01$ & $ 2.73629716446342652E-01$ \\ 
$x$ (AU) & $ 1.38046915200118941E-01$ & $-4.42902170681904253E-02$ & $-2.93032016429504605E-01$ \\
$y$ (AU) & $ 8.89288648314162135E-04$ & $ 3.31668563947641257E-03$ & $-1.23675687401901668E-02$ \\
$z$ (AU) & $ 1.46412807097381470E-01$ & $ 5.33577347000446373E-02$ & $-6.02702321469965385E-01$ \\
$v_x$ (AU day$^{-1}$) & $-1.49455150373738761E-02$ & $ 9.11165287670818828E-03$ & $ 1.91658041307088030E-02$ \\
$v_y$ (AU day$^{-1}$) & $-5.47600719167812516E-04$ & $ 5.75731527397032412E-04$ & $-1.51588902566718357E-06$ \\
$v_z$ (AU day$^{-1}$) & $ 2.33976025887882716E-02$ & $-1.96657133463214485E-02$ & $-1.42900652610584780E-02$ \\
Radius (AU) & $ 4.67389123020501077E-03$ & $ 4.07536158371365510E-03$ & $ 1.33457234260197784E-03$ \\
Flux & $ 1.06031175675393025E-01$ & $ 3.99465803917630147E-01$ & $ 2.22166066646409038E-03$ \\
\bottomrule
\end{tabular}}
\end{adjustwidth}
\noindent{\footnotesize{\textsuperscript{1} Jacobian instantaneous (Keplerian) elements, reference time BJD 2,454,800. 
\textsuperscript{2} Barycentric Cartesian~coordinates.  
\textsuperscript{3} Times are relative to BJD 2,455,000.000.}}
\end{table}
\unskip

\begin{table}[H]
\footnotesize
\caption{Initial Conditions for KIC~10319590.\label{K1031_init}}
\newcolumntype{s}{>{\hsize=8pt}X}
\begin{adjustwidth}{-\extralength}{0cm}
\centering 
\setlength{\tabcolsep}{5mm} \resizebox{\linewidth}{!}{\begin{tabular}{rrrr}
\toprule
\textbf{Parameter $^1$} & 
\textbf{\quad Orbit 1} &  
\textbf{\quad Orbit 2} &
\textbf{\quad } \\
\midrule
Period (days)& $ 2.12654307685400603E+01$ & $ 4.55628563049740364E+02$ & \\ 
$e\cos\omega$ & $ 1.37393071317266097E-02$ & $-9.77426355365841049E-02$ & \\ 
$e\sin\omega$ & $ 1.83096050717336187E-02$ & $ 9.95037109808404596E-02$ & \\ 
$i$ (rad) & $ 1.56834038252700858E+00$ & $ 1.82802557518344755E+00$ & \\ 
$\Omega$ (rad) & $ 0.00000000000000000E+00$ & $ 6.69848081316824828E-01$ & \\ 
$T_{\rm conj}$ (days) $^3$ & $-2.04870001610389039E+02$ & $-1.89207395376568286E+02$ & \\ 
$a$ (AU) & $ 1.84123641627113394E-01$ & $ 1.65016102995011926E+00$ & \\
$e$ & $ 2.28912690417747873E-02$ & $ 1.39479788143644284E-01$ & \\
$\omega$ (deg) & $ 5.31158789150222930E+01$ & $ 1.34488459425773385E+02$ & \\
true anomaly (deg) & $ 1.20064569527552720E+02$ & $ 3.05076421815502783E+02$ & \\
mean anomaly (deg) & $ 1.17774811252421031E+02$ & $ 3.17383628595700600E+02$ & \\
mean longitude (deg) & $ 1.73180448442574999E+02$ & $ 4.77944349215666193E+02$ & \\
$i$ (deg) & $ 8.98592847587306665E+01$ & $ 1.04738150299986302E+02$ & \\
$\Omega$ (deg) & $ 0.00000000000000000E+00$ & $ 3.83794679743900318E+01$ & \\
\midrule
\textbf{Parameter $^2$} & 
\textbf{Body 1} &  
\textbf{Body 2} &
\textbf{Body 3} \\
\midrule
Mass ($M_{\odot}$) & $ 1.09884220076050898E+00$ & $ 7.42680819654872537E-01$ & $ 1.04618946026211956E+00$ \\ 
$x$ (AU) & $-8.68108251627838085E-02$ & $-2.71655796775709668E-01$ & $ 2.84025942978475276E-01$ \\
$y$ (AU) & $ 4.53910059223668616E-02$ & $ 4.54452955085508578E-02$ & $-7.99366704949047852E-02$ \\
$z$ (AU) & $-5.25084361661361854E-01$ & $-5.02979024116999240E-01$ & $ 9.08571310989086389E-01$ \\
$v_x$ (AU day$^{-1}$) & $ 9.96617711333108204E-03$ & $ 2.50829149845738621E-03$ & $-1.22483703622658230E-02$ \\
$v_y$ (AU day$^{-1}$) & $ 5.78918622397648058E-03$ & $ 5.65832462960103207E-03$ & $-1.00973405926929067E-02$ \\
$v_z$ (AU day$^{-1}$) & $ 2.08177360350398112E-02$ & $-3.24657740926403010E-02$ & $ 1.18171792129101970E-03$ \\
Radius (AU) & $ 7.37788983874515438E-03$ & $ 3.33573068311745265E-03$ & $ 6.51472936693705054E-03$ \\
Flux & $ 2.66430315149299768E-01$ & $ 4.11181697270141394E-02$ & $ 4.11351649247532369E-01$ \\
\bottomrule
\end{tabular}}
\end{adjustwidth}
\noindent{\footnotesize{\textsuperscript{1} Jacobian instantaneous (Keplerian) elements, reference time BJD 2,454,750. 
\textsuperscript{2} Barycentric Cartesian~coordinates.  
\textsuperscript{3} Times are relative to BJD 2,455,000.000.}}
\end{table}
\unskip

\begin{table}[H]
\footnotesize
\caption{Initial Conditions for EPIC~2202.\label{EPIC01_init}}

\begin{adjustwidth}{-\extralength}{0cm}
\centering 
\setlength{\tabcolsep}{2mm} \resizebox{\linewidth}{!}{\begin{tabular}{rrrrr}
\toprule
\textbf{Parameter $^1$} & 
\textbf{\quad Orbit 1} &  
\textbf{\quad Orbit 2} &
\textbf{\quad Orbit 3} &
\textbf{\quad } \\
\midrule
Period (days) & $ 1.32573806476518747E+01$ & $ 5.32131371045080868E+02$ & $ 1.43863756806189400E+01$ & \\
$e\cos\omega$ & $-5.94598448262661228E-02$ & $-8.55018881705337103E-02$ & $-3.38389937075848010E-02$ & \\
$e\sin\omega$ & $-1.02380498222947716E-01$ & $-8.59084462583555031E-02$ & $-4.99696389496548515E-02$ & \\
$i$ (rad) & $ 1.57854219985552158E+00$ & $ 1.57578348719679817E+00$ & $ 1.56662070691968336E+00$ & \\
$\Omega$ (rad) & $ 0.00000000000000000E+00$ & $ 2.55025994661455702E-01$ & $ 5.56347576081097195E-01$ & \\
$T_{\rm conj}$ (days) $^3$  & $ 2.40184925569052348E+03$ & $ 2.75795386232665851E+03$ & $ 2.40299016521348676E+03$ & \\
$a$ (AU) & $ 1.13242296211668092E-01$ & $ 1.58329708898028576E+00$ & $ 1.05998419826898130E-01$ \\
$e$ & $ 1.18394423699526213E-01$ & $ 1.21205750768067833E-01$ & $ 6.03493356376100104E-02$ \\
$\omega$ (deg) & $ 2.39853130379876433E+02$ & $ 2.25135896233769216E+02$ & $ 2.35894493170920470E+02$ \\
true anomaly (deg) & $ 2.43169713250248947E+02$ & $ 3.42563864782194742E+02$ & $ 2.46611228663296032E+02$ \\
mean anomaly (deg) & $ 2.55755507769099324E+02$ & $ 3.46389156126581895E+02$ & $ 2.53071089587544577E+02$ \\
mean longitude (deg) & $ 4.83022843630125351E+02$ & $ 5.82311674176191218E+02$ & $ 5.14382089885996834E+02$ \\
$i$ (deg) & $ 9.04438058350178977E+01$ & $ 9.02857432427837239E+01$ & $ 8.97607546042993363E+01$ \\
$\Omega$ (deg) & $ 0.00000000000000000E+00$ & $ 1.46119131602272763E+01$ & $ 3.18763680517803394E+01$ \\
\midrule
\textbf{Parameter $^2$} & 
\textbf{Body 1} &  
\textbf{Body 2} &
\textbf{Body 3} &
\textbf{Body 4} \\
\midrule
Mass ($M_{\odot}$) & $ 6.00160305304624941E-01$ & $ 5.02158972671820969E-01$ & $ 3.88288635827891637E-01$ & $ 3.79412587545848390E-01$ \\
$x$ (AU) & $ 5.21452165213335594E-01$ & $ 4.57168023040150640E-01$ & $-6.82185797329151411E-01$ & $-7.31765186374565557E-01$ \\
$y$ (AU) & $ 1.27283067586334703E-01$ & $ 1.26516988662700669E-01$ & $-1.67244814593741076E-01$ & $-1.97628195027042264E-01$ \\
$z$ (AU) & $ 2.21788796535345500E-01$ & $ 3.20688377818036885E-01$ & $-4.28249446259482247E-01$ & $-3.36997213876308988E-01$ \\
$v_x$ (AU day$^{-1}$) & $ 1.40127468254323644E-02$ & $-2.57720323650417228E-02$ & $ 2.13167100137257731E-02$ & $-9.87124173345826916E-03$ \\
$v_y$ (AU day$^{-1}$) & $-1.22574482288896483E-03$ & $-9.72692399585586560E-04$ & $ 1.12442212517640926E-02$ & $-8.28099496615305845E-03$ \\
$v_z$ (AU day$^{-1}$) & $ 2.23888690613121069E-02$ & $-1.02797998384343212E-02$ & $ 2.31437406675261397E-03$ & $-2.41780642377567787E-02$ \\
Radius (AU) & $ 2.27966651955714217E-03$ & $ 1.76488575141570364E-03$ & $ 1.86695843377417407E-03$ & $ 1.67223634562931240E-03$ \\
Flux & $-2.03408906342277401E-01$ & $ 1.71702881281671521E-01$ & $ 2.74715511695112879E-01$ & $ 2.24976169379865309E-01$ \\
\bottomrule
\end{tabular}}
\end{adjustwidth}
\noindent{\footnotesize{\textsuperscript{1} Jacobian instantaneous (Keplerian) elements, reference timáe BJD 2,457,390. 
\textsuperscript{2} Barycentric Cartáeá©sian~ácoordinates.  
\textsuperscript{3} Times are relative to BJD 2,455,000.000.}}
\end{table}

\newpage

\begin{adjustwidth}{-\extralength}{0cm}

\reftitle{References}

\PublishersNote{}
\end{adjustwidth}
\end{document}